\documentclass[journal]{IEEEtran}
\usepackage{epsfig}
\usepackage{epstopdf}
\usepackage{amssymb}
\usepackage{amsmath}
\usepackage[version=3]{mhchem}
\usepackage{graphicx}
\usepackage{multirow}
\usepackage{wrapfig}
\usepackage{color}
\usepackage{colortbl}
\usepackage{soul}
\usepackage[noadjust]{cite}
\usepackage{subfigure}
\usepackage{enumerate}
\usepackage{cite}
\usepackage{paralist}
\usepackage[normalem]{ulem}
\usepackage{mathtools}
\usepackage{tabularx,booktabs,caption}
\usepackage[flushleft]{threeparttable}
\newcolumntype{C}{>{\centering\arraybackslash}X} 
\DeclarePairedDelimiter\floor{\lfloor}{\rfloor}

\newcommand{\p}{\mathrm{p}}
\newcommand{\E}{\mathrm{E}}
\newcommand{\Var}{\mathrm{Var}}
\def\mathclap#1{\text{\hbox to 0pt{\hss$\mathsurround=0pt#1$\hss}}}
\makeatletter
\newcommand{\vastt}{\bBigg@{3}}
\newcommand{\vast}{\bBigg@{4}}
\newcommand{\Vast}{\bBigg@{5}}
\makeatother

\DeclareMathOperator\erfc{erfc}

\begin{document}

\title{Transmitter and Receiver Architectures for Molecular Communications: A Survey on Physical Design with Modulation, Coding and Detection Techniques}

\author{Murat Kuscu,~\IEEEmembership{Student Member,~IEEE,}
        Ergin Dinc,~\IEEEmembership{Member,~IEEE,}
        Bilgesu A. Bilgin,~\IEEEmembership{Member,~IEEE,}
        Hamideh Ramezani,~\IEEEmembership{Student Member,~IEEE,}
        Ozgur B. Akan,~\IEEEmembership{Fellow,~IEEE,}
\thanks{The authors are with the Internet of Everything (IoE) Group, Electrical Engineering Division, Department of Engineering, University of Cambridge, UK, CB3 0FA (e-mail: \{mk959, ed502, bab46, hr404, oba21\}@cam.ac.uk).}
\thanks{This work was supported in part by the ERC projects MINERVA (ERC-2013-CoG \#616922), and MINERGRACE (ERC-2017-PoC \#780645).}}

\IEEEpeerreviewmaketitle
\maketitle

\begin{abstract}
 Inspired by Nature, molecular communications (MC), i.e., use of molecules to encode, transmit and receive information, stands as the most promising communication paradigm to realize nanonetworks. Even though there has been extensive theoretical research towards nanoscale MC, there are no examples of implemented nanoscale MC networks. The main reason for this lies in the peculiarities of nanoscale physics, challenges in nanoscale fabrication and highly stochastic nature of biochemical domain of envisioned nanonetwork applications. This mandates developing novel device architectures and communication methods compatible with MC constraints. To that end, various transmitter and receiver designs for MC have been proposed in literature together with numerable modulation, coding and detection techniques. However, these works fall into domains of a very wide spectrum of disciplines, including but not limited to information and communication theory, quantum physics, materials science, nanofabrication, physiology and synthetic biology. Therefore, we believe it is imperative for the progress of the field that, an organized exposition of cumulative knowledge on subject matter be compiled. Thus, to fill this gap, in this comprehensive survey we review the existing literature on transmitter and receiver architectures towards realizing MC amongst nanomaterial-based nanomachines and/or biological entities, and provide a complete overview of modulation, coding and detection techniques employed for MC. Moreover, we identify the most significant shortcomings and challenges in all these research areas, and propose potential solutions to overcome some of them.
\end{abstract}
\begin{IEEEkeywords}
	Molecular communications, Nanonetworks, Internet of Nano Things, Transmitter, Receiver,  Modulation, Coding, Detection
\end{IEEEkeywords}

\section{Introduction}

\begin{figure*}[!ht]
	\centering
	\includegraphics[width=18cm]{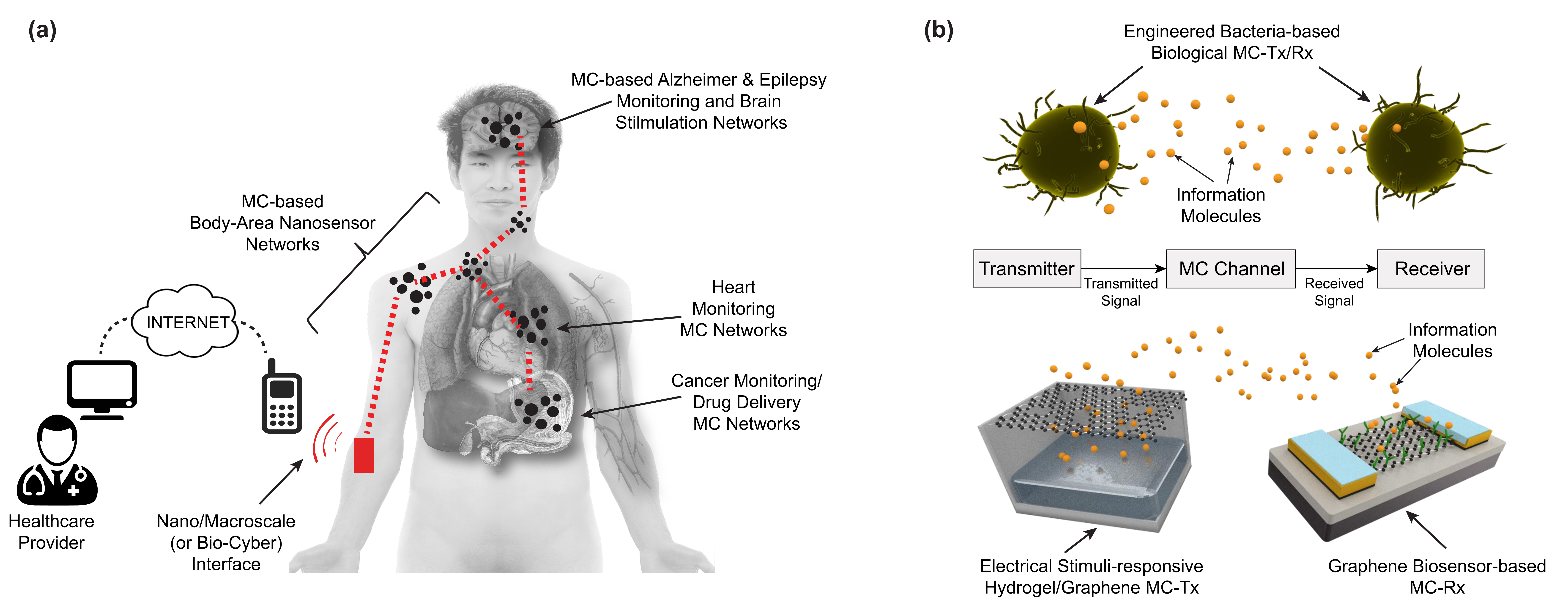}
	\caption{(a) An intrabody continuous healthcare application of IoNT enabled by MC nanonetworks \cite{akyildiz2015internet}. (b) Components of an MC system with biological and nanomaterial-based MC-Tx/Rx design approaches, which are reviewed in Section \ref{sec:Tx} and Section \ref{sec:Rx}. }
	\label{fig:IoNT_MCTxRx}
\end{figure*}

\IEEEPARstart{M}{olecular} communications (MC) is a bio-inspired communication method that uses molecules for encoding, transmitting and receiving information, in the same way that living cells communicate \cite{akyildiz2015internet}. Since it is inherently bio-compatible, energy-efficient and robust in physiological conditions, it has emerged as the most promising method to realize nanonetworks and Internet of Nano Things (IoNT), i.e., artificial networks of nanoscale functional units, such as nano-biosensors and engineered bacteria, integrated with the Internet infrastructure \cite{akyildiz2010internet, akyildiz2015internet}. In that respect, MC is promising for novel applications especially towards information and communication technology (ICT)-based early diagnosis and treatment of diseases, such as continuous health monitoring, smart drug delivery, artificial organs, and lab-on-a-chips \cite{akan2017fundamentals, felicetti2016applications} (see Fig. \ref{fig:IoNT_MCTxRx}(a)). It bears a significant potential as an alternative to conventional wireless communication in environments where the latter may fail, e.g., intrabody medium \cite{atakan2012body} and confined channels like pipe networks \cite{farsad2016comprehensive, qiu2014molecular}.  

Discrete nature of information carrying agents, i.e., molecules, and peculiarities arising from the nanophysical and biochemical processes, and computational and energy-based limitations of communicating nanomachines give rise to novel challenges that cannot be always tackled by the conventional ICT methods, which are mostly tailored for electromagnetic means of communications with macroscale components. This requires rethinking of the existing ICT tools and devising new ones for MC in the light of envisioned applications and considering especially available material and fabrication technologies, which basically set the limitations. 

MC has been extensively studied from various aspects over the last decade. The research efforts are mainly centered around developing information theoretical models of MC channels \cite{pierobon2013capacity, atakan2010deterministic}, devising modulation and detection techniques \cite{kuran2011modulation, kilinc2013receiver}, and system-theoretical modeling of MC applications \cite{chahibi2017molecular, mosayebi2018early}. For the physical design of MC transmitter (Tx) and receiver (Rx), mainly two approaches have been envisioned: biological architectures based on engineered bacteria enabled by synthetic biology, and nanomaterial-based architectures, as shown in Fig. \ref{fig:IoNT_MCTxRx}(b) \cite{akyildiz2015internet}. However, none of these approaches could be realized yet, and thus, there is no implementation of any artificial micro/nanoscale MC system to date. As a result, the overall MC literature mostly relies on assumptions isolating the MC channel from the physical processes regarding the transceiving operations, leading to a plethora of ICT techniques, feasibility and performance of which could not be validated. 

Our objective in this paper is to tackle this discrepancy between the theory and practice by providing a comprehensive account of the recent proposals for the physical design of MC-Tx/Rx and the state-of-the-art in the theoretical MC research covering modulation, coding and detection techniques. We provide an overview of the opportunities and challenges regarding the implementation of Tx/Rx and corresponding ICT techniques which are to be built on these devices. Although we mostly focus on diffusion-based MC in this review, we also partly cover other MC configuration, such as MC with flow, microfluidic MC, bacteria conjugation-based MC, and molecular-motor powered MC. We first investigate the fundamental requirements for the physical design of micro/nanoscale MC-Tx and MC-Rx, e.g., those regarding the energy and molecule consumption, computational complexity and operating conditions. In light of these requirements, both of the design approaches previously proposed for MC-Tx/Rx, i.e., artificial Tx/Rx designs enabled by the discovery of new nanomaterials, e.g., graphene, and biological Tx/Rx architectures based on the engineering tools provided by synthetic biology, are covered.

The literature on physical design of micro/nanoscale artificial MC-Tx is almost nonexistent, except for a few approaches focusing on standalone drug delivery systems \cite{murdan2003electro} and microfluidic neural interfaces with chemical stimulation capabilities \cite{scott2013microfluidic,jonsson2016bioelectronic}, which, however, do not address the communication theoretical challenges of MC. Therefore, regarding the MC-Tx architecture, we mostly investigate and propose several potential research directions towards nanomaterial-based implementation of MC-Tx based on recent advancements in nanotechnology, novel materials and microfluidics. We will particularly focus on the utilization of nanoporous graphene membranes in microfluidic settings to control the release of information molecules through electric field. For the design of artificial MC-Rx, we will focus on nanomaterial-based biosensing approaches, as a biosensor and an MC-Rx have similar functionality, i.e., both of them aim to detect the physical characteristics of molecules in an environment. Previously in \cite{kuscu2016physical}, the structure and performance of available biosensing approaches, i.e., electrical, optical and chemical sensing, are studied to evaluate their feasibility as an artificial MC receiver. It is shown that field effect transistor (FET)-based biosensors (bioFETs) satisfy the requirements of an MC receiver, i.e., nanoscale dimension, in-situ, label free, continuous and selective detection of information carriers. Thus, in this paper, we mostly focus on the operation principles of bioFETs and their main design parameters that must be selected to optimize performance of the device as an MC-Rx.  We review the possible types of bioreceptors, e.g., natural receptor proteins, aptamer/DNAs, which act as the interface between the MC channel and the receiver. We also study available options for the nanomaterials, e.g., silicon nanowire (SiNW) and graphene, used as the transducer channel, and evaluate their feasibility and potential to be a part of MC receiver in terms of mechanical, electrical and chemical properties and associated fabrication challenges. We then provide an overview of studies focusing on the modeling and performance evaluation of bioFETs from MC perspective. 

Regarding biological architectures, there are more opportunities enabled by the synthetic biology tools in engineering bacteria with artificial MC-Tx/Rx functionalities. Biological device architectures, in general, have the advantage of biocompatibility, a necessary condition for many MC applications within healthcare and Internet of Bio-Nano Things (IoBNT) paradigm \cite{akyildiz2015internet}, as well as availability of already existing molecular machinery for sensing, transmitting and propagation at micro/nanoscale, all of which are prone to be utilized for MC nanonetworking. In terms of Tx architectures, biological-based approaches also prove advantageous thanks to the capability of harnessing transmitter molecules or synthesizing them from readily available molecules, which resolves the reservoir problem. Biological architectures considered in this review include the use of genetically engineered flagellated bacteria, e.g., \textit{E. coli} and viruses as encapsulated carriers of information, e.g., RNA or DNA, bacteria with engineered gene circuits for protein-based communication and $Ca^{++}$ circuits for hormonal communication, as well as other mechanisms such as microtubule networks and biological-nanomaterial hybrid approaches. For biological receivers, synthetic biology tools allow performing complex digital and analog computations for detection \cite{daniel2013synthetic}, integrate computation and memory \cite{purcell2014synthetic} and enable continuously keep track of individual receptor states, as naturally done by living cells. In this paper, we provide a review of available studies on the use of genetic circuits for implementing MC receiver architectures. The major challenge in use of genetic circuits for implementing MC-Tx/Rx functionalities in engineered bacteria arises from the fact that information transmission in biological cells is through molecules and biochemical reactions. This results in nonlinear input-output behaviors with system-evolution-dependent stochastic effects that are needed to be comprehensively studied to evaluate the performance of the device. This paper will present a brief account of these challenges and opportunities regarding the use of biological architectures. 

We also provide an overview of the communication techniques proposed for MC. In particular, we will cover modulation, coding and detection methods. Modulation techniques in MC fundamentally differ from that in conventional EM communications, as the modulated entities, i.e., molecules, are discrete in nature and the developed techniques should be robust against highly time-varying characteristics of the MC channel, as well as inherently slow nature of the propagation mechanisms. Exploiting the observable characteristics of molecules, researchers have proposed to encode information into the concentration, e.g., concentration shift keying (CSK) \cite{kuran2012interference}, molecule type, e.g., molecule shift keying (MoSK) \cite{kuran2011modulation}, and molecule release time, e.g., release time shift keying (RTSK) \cite{srinivas2012molecular}. However, the proposed modulation schemes generally suffer from low data rates as it is not practical to use modulation schemes with high number of symbols due to inter-symbol interference (ISI), detection sensitivity and selectivity problems. Therefore, in this paper, we also investigate two new research directions that can potentially alleviate these problems: (1) utilization of DNA/RNA as carrier molecules such that nucleotide strands differing in physical properties, e.g., length, dumbbell hairpins, short sequence motifs/labels and orientation, can enable transmitting high number of symbols that can be selectively received, (2) encoding large amount of information into the base sequences of DNAs, i.e., Nucleotide Shift-Keying (NSK) \cite{organick2018random}, to boost the data rate up to the order of Mbps such that MC can compete with traditional wireless communication systems. 

Another fundamental aspect of MC is coding. In a communication system coding is done at two stages, source coding and channel coding. Source coding is related to the structure of information to be transmitted, which has so far not been discussed in MC literature, and is not covered in this review. On the other hand, various channel coding schemes have been applied to MC channels with the hope of error correction. However, conventional channel coding schemes developed for classical EM-wave based communications are incompatible with nanoscale MC in many aspects. To start with, channel characteristics are very different, where diffusive channels suffer extremely from ISI, which causes high bit error rates (BERs). Novel codes that specifically handle with ISI need to be developed for mitigation of these errors. Moreover, in general, energy is a scarce resource for communicating agents in a nanonetwork, requiring implementation of low complexity codes with low computational cost, which renders the application of state-of-the-art codes, such as Turbo codes \cite{berrou1993near}, unfeasible. Consequently, many researchers have focused on simpler block codes, such as Hamming codes \cite{hamming1950error}, whereas some have devised novel channel codes specifically for MC. In this review, we give a comprehensive account of these channel codes.

Detection is by far the most studied aspect of MC in the literature, but there are still many open issues mainly because of the unavailability of any MC-Rx implementation that can validate the developed methods. This lack of physical models for realistic devices has led the researchers to make simplifying assumptions about the sampling process, receiver geometry and channel and reception noise. Depending on the type of these assumptions, we investigate MC detection methods under two categories: detection with passive/absorbing receivers and detection with reactive receivers. We provide a qualitative comparison of these methods in terms of considered channel and receiver characteristics, complexity, type of required channel state information (CSI), and performance.  

In summary, this paper provides design guidelines for the physical implementation of MC-Tx/Rxs, an overview of the state-of-the-art MC methods on transceiving molecular messages, and a detailed account of the challenges and future research directions. We believe that this comprehensive review will help researchers to overcome the major bottleneck resulting from the long-standing discrepancy between theory and practice in MC, which has, so far, severely impeded the innovation in this field linked with huge societal and economic impact.

The remainder of this paper is organized as follows. In Section \ref{sec:Tx}, we investigate the design requirements of an MC-Tx, and review the physical design options for nanomaterial-based and biological MC-Tx architectures with an overview of state-of-the-art MC modulation and coding techniques. We focus on the opportunities for the physical design of an MC-Rx in Section \ref{sec:Tx}, where we also provide a comprehensive review of the existing MC detection schemes for passive, absorbing and reactive receivers. In Section \ref{sec:Challenges}, we outline the challenges and future research directions towards the implementation of MC Tx/Rxs, and the development of the corresponding coding, modulation, and detection techniques. Finally, we conclude the paper in Section \ref{sec:Conclusion}.

\section{Molecular Communication Transmitter}
\label{sec:Tx}
MC transmitter (MC-Tx) encodes information in a physical property of molecules such as concentration, type, ratio, order or release time, and releases information molecules (IMs) accordingly. To this aim, information to be transmitted is required to be mapped to a sequence of bits through source and channel coding to represent the information with less number of bits, and to introduce additional bits to the information with the purpose of providing error correction, respectively. Then, the modulator unit encodes the information in a property of molecules and control the release of IMs according to a predetermined modulation scheme. Finally, a power source and IM generator/container are required to provide energy and IM molecules for MC-Tx. The interconnection of these components in an MC-Tx is illustrated in Fig. \ref{fig:MCTX}. In this section, we first discuss the requirements of an MC-Tx and investigate the utilization of different IMs. Then, we review the available approaches in physical design of MC-Txs, which can be categorized into two main groups, (i) nanomaterial-based artificial MC-Tx and (ii) biological MC-Tx based on synthetic biology.

\subsection{Design Requirements for MC-Tx}
\label{sec:TxRequirements}

\subsubsection{\textbf{Miniaturization}}
Many novel applications promised by MC impose size restrictions on their enabling devices, requiring them to be micro/nanoscale. Despite the avalanching progress in nanofabrication of bioelectronics devices over the last few decades \cite{daksh2016recent}, fabrication of fully functional nanomachines capable of networking with each other and their surroundings in order to accomplish a desired task still evades us \cite{abbasi2016nano}.
Moreover, with miniaturization the surface area to volume ratio increases, causing surface charges to become dominant in molecular interactions. This causes the behavior of molecules passing through openings with one characteristic dimension less than $100$nm to be significantly different than that observed in larger dimensions \cite{schoch2008transport}. Consequently, as the release aperture of many considered molecular transmitter architectures are commonly in nanoscale, peculiarities arising from this phenomenon need to be taken into account in transmitter design.

\subsubsection{\textbf{Molecule Reservoir}}
The size restriction introduces, besides the need for energy harvesting due to infeasibility of deploying a battery unit, which is a problem for the whole nanomachine in general, another very important problem for any transmitter architecture, namely, limited resources of IMs \cite{akyildiz2008nanonetworks}. Typically, a transmitter module would contain reservoirs, where the IMs are stored to be released. However, at dimensions in question, without any replenishment these reservoirs will inevitably deplete, rendering the nanomachine functionally useless in terms of MC. Moreover, the replenishment rate of transmitter reservoirs has direct effect on achievable communication rates \cite{huang2017capacity,bafghi2018diffusion}. Possible theoretically proposed scenarios for reservoir replenishment include local synthesis of IMs, e.g., use of genetically engineered bacteria whose genes are regulated to produce desired transmitter proteins \cite{unluturk2015genetically}, as well as, employment of transmitter harvesting methods \cite{bilgin2017fast}. Corresponding technological breakthroughs necessary for realization of these approaches are emerging with advancements in the relevant fields of genetic engineering \cite{pinero2015engineered}, and materials engineering \cite{haefner2017chemically,di2018molecularly}.

\subsubsection{\textbf{Biocompatibility}}
Most profound applications of nanodevices with MC networking capabilities, such as, neural prosthetics, tissue engineering, targeted drug delivery, BMIs, immune system enhancement, etc., require the implantation of corresponding enabling devices into biological tissue \cite{katz2014implantable}. Combined with increased toxicity of materials at nanoscale \cite{sharifi2012toxicity}, the issue of biocompatibility of these devices stands out as one of the most challenging research issues to be addressed. Moreover, corrosion of implants inside body is another commonly known issue \cite{asghari2017biodegradable} restricting device durability, which due to increased surface to volume ratio at nanoscale has an effectively amplified effect on nanoscale implants compared to their larger versions. In addition to biochemical toxicity and durability, flexibility of implanted devices is another important aspect of biocompatibility, as mechanical mismatch between soft biological tissue and the implanted device triggers inflammation followed by scar tissue formation \cite{anderson1993mechanisms}, which eventually restricts the implant to carry out its intended task. Accordingly, the last decade has witnessed extensive research on novel biocompatible materials such as polymers \cite{asghari2017biodegradable}, soft organic electronics \cite{someya2016rise}, dendrimers, and hydrogels \cite{nguyen2017biocompatible}. It is imperative to note that, all issues of biocompatibility mentioned here specifically apply to nanomaterial-based device architectures, and they can be almost completely avoided by the choice of a biological architecture, e.g., genetically engineered organisms \cite{akyildiz2015internet}.

\subsubsection{\textbf{Transmission Performance}}
Performance of a molecular transmitter should also be evaluated by how much control it has over the transmission of molecules. Important performance metrics include off-state leakage, transmission rate precision, resolution and range, as well as transmission delay. Off-state leakage, or unwanted leakage of molecules from transmitter, from MC perspective, degrades communication significantly by contributing to background channel noise. On the other hand, it is unacceptable for some niche applications, such as targeted drug delivery \cite{allen2004drug}, where leakage of drug molecules would cause undesired toxicity to the body. During operation, control over rate of transmission is essential. For instance, in neural communications information is encoded, among other fashions, in the number of neurotransmitters released at a chemical synapse, and accordingly transmitter architectures envisioned to stimulate neurons via release of neurotransmitters, e.g., glutamate \cite{simon2009organic}, will need to encode information in neurotransmitter concentrations via very precise transmissions. In this respect, transmission rate precision, resolution and range are all variables that are determinant in the capacity of communication with the recipient neuron. Finally, transmission delay is in general a quantity to be minimized, as it contributes to degradation in communication rates. Moreover, for instance in neural communications, where information is also encoded in frequency of signals, there is an upper bound on permissible transmission delay of a transmitter in order to satisfy a given lower bound on maximum frequency transmittable.

\subsection{Physical Design of MC-Tx}
\label{sec:TxPhysicalDesign}
\begin{figure}[!t]
	\centering
	\includegraphics[width=0.49\textwidth]{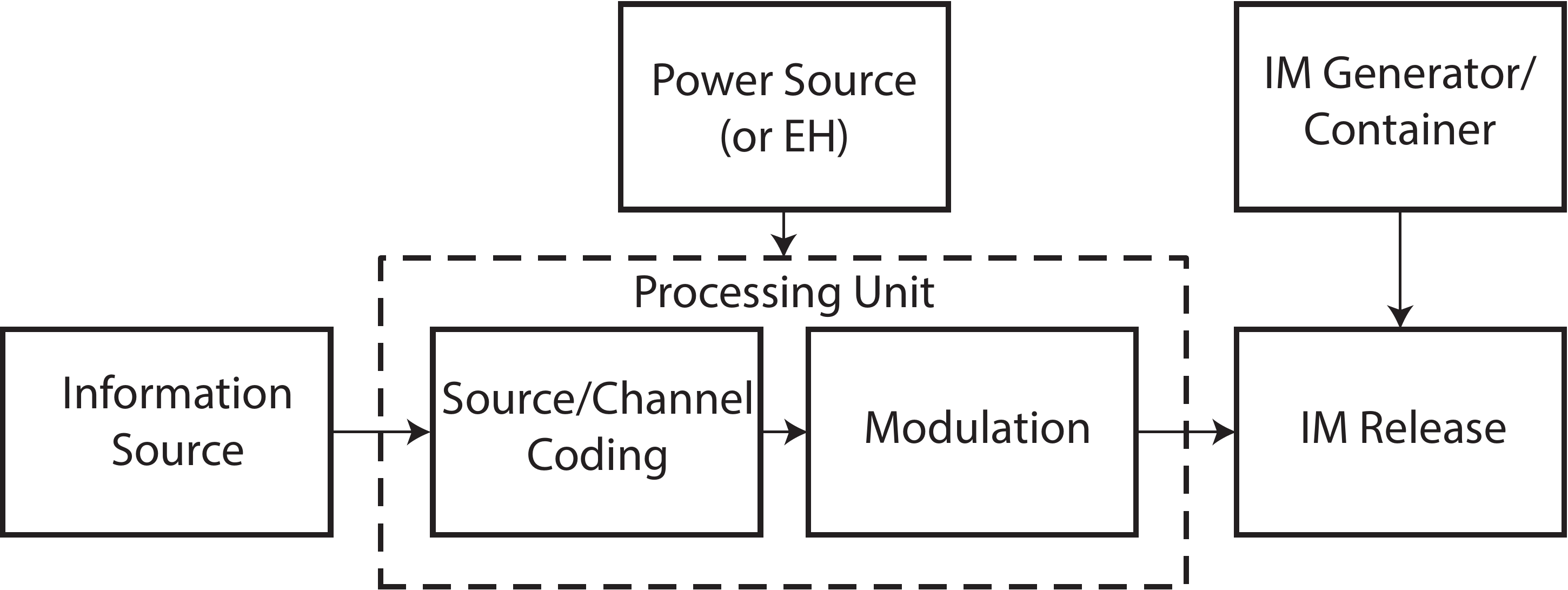}
	\caption{Physical architecture of an MC transmitter.}
	\label{fig:MCTX}
\end{figure}

MC uses molecules for information transfer unlike the traditional communication systems that rely on electromagnetic (EM) and acoustic waves. Therefore, the physical design of MC-Tx significantly differs from the traditional transmitters. The key components of an MC-Tx is illustrated in Figure \ref{fig:MCTX}. The first element in MC-Tx is the information source that represents either a bit-wise information generated by a nanomachine or a biological signal from living cells such as neurons and cardiomyocytes.

The next building block in an MC-Tx is the processing unit, which performs coding, modulation and control of IM release. This block may not appear in biological systems utilizing MC in a basic way such as hormonal communication inside human body, where the information is transferred via a single molecule type with on-off keying. For this purpose, MC-Tx processing unit can operate via chemical pathways \cite{farsad2016comprehensive} or through biocompatible micro-scale processors, or via synthetic genetic circuits \cite{moon2012genetic}. In case of biological Tx architectures, processing units performing logic gates and memory functions in the form of synthetic genetic circuits can be embedded into cells \cite{moon2012genetic, siuti2013synthetic}.

Inside the processing unit, the first step is to represent analog or digital information with the least number of bits possible through source coding. Then, channel coding block introduces extra bits to make data transmission more robust against error-prone MC channels. After the coding step, digital information is transformed into an analog molecular signal according to a predetermined modulation scheme. There are various modulation schemes proposed for MC by using molecular concentration, type of molecules or time of molecule release. The detailed discussion of modulation techniques for MC can be found in Section \ref{sec:TxModulation}. The processing unit controls IM release according to the data being transmitted and the modulation scheme. IM molecules are provided from IM generator/container, which can be realized as either a reservoir containing genetically engineered bacteria to produce IM molecules on demand, or a container with limited supply of IM molecules.

There are two types of molecule release mechanisms \cite{bossert1963analysis}: instant release, where certain number of molecules (in mol) are released to the medium at the same time, and continuous release, where the molecules are released with a constant rate over a certain period of time (mol/s). Instant release can be modeled as an impulse symbol, while continuous release can be considered as a pulse wave. According to application requirements and channel conditions, one of the release mechanisms can be more favorable. Bit-wise communications, for example, require rapid fade-out of IMs from the channel as these molecules will cause ISI for the next symbol, thus, instant release is more promising. In addition, alarm signals for warning plants and other insects can favor continuous release to increase detection probability and reliability. 

MC-Tx architecture also shows dependency on propagation channel. There are three main propagation channels that can be exploited by MC: free diffusion, flow-assisted propagation, and motor-powered propagation \cite{farsad2016comprehensive,cobo2010bacteria}. Diffusion is the movement of molecules from high concentration to low concentration with random steps. This process requires no external energy and the molecules travel with the thermal energy of the medium, e.g., liquid or air. Diffusion process can be modeled with the Fick's law, i.e., a partial differential equation calculating the changes in molecular concentration in a medium, as 
\begin{equation}
\frac{\partial c}{\partial t}=D \nabla^2 c, 
\end{equation}
where $c$ is the molecule concentration, and $D$ is the diffusion coefficient, which depends on the temperature, pressure and viscosity of the medium. Calcium signaling between cells and propagation of neurotransmitters in synaptic cleft are good examples of biological systems utilizing diffusion for molecular information transfer. Diffusion-based MC provides short range information transfer in the order of micrometers \cite{farsad2016comprehensive}. 

Second type of MC channel is flow-assisted propagation, in which molecules propagate via both diffusion and flow, such as flow in the blood stream and air flow created by wind. Flow-assisted propagation can be modeled with the advection-diffusion equation, which is represented as
\begin{equation}
\frac{\partial c}{\partial t}+\nabla .  (\textbf{v}c)=D \nabla^2 c, 
\end{equation}
where $\textbf{v}$ is the velocity vector for the flow in the medium. The diffusion case, flow-assisted propagation does not require any external energy source, and uses the thermal energy that exists in the propagation medium. Communication range of flow-assisted propagation can reach up to several meters as in hormonal communications in the blood stream and communications via pheromone between plants \cite{farsad2016comprehensive}.

Lastly, motor-powered propagation includes molecular motors and bacteria with flagella, e.g., \textit{E. coli}. Molecular motor-powered microtubules, also known as walkway-based architectures \cite{pierobon2010physical,walsh2008hybrid}, use predefined microtubules with motor proteins, which carry IMs from one end to other end of the tube. Motor proteins such as Kinesis consume external energy in the form of adenosine triphosphate (ATP) to move IMs in microtubules with the diameters of 20-30nm \cite{hiyama2010biomolecular, enomoto2011design}. This motion can be mathematically represented as \cite{farsad2016comprehensive}
\begin{equation}
x_i=x_{i-1}+v_{avg} \Delta t,
\end{equation}
where $x$ is the location of the molecule inside the microtubule, $v_{avg}$ is the average speed of molecules moving via molecular motors, and $\Delta t$ is the time step. Molecular motors enable medium range molecular communication up to a millimeter \cite{pierobon2010physical}. In addition, \cite{cobo2010bacteria,gregori2011physical} propose a nanonetwork architecture in which engineered bacteria with flagella is utilized to carry IMs for medium range MC ($\mu$m-mm). As in the molecular motor case, bacteria consume energy for movement via flagella. 
\begin{figure*}[!t]
	\centering
	\includegraphics[width=0.95\textwidth]{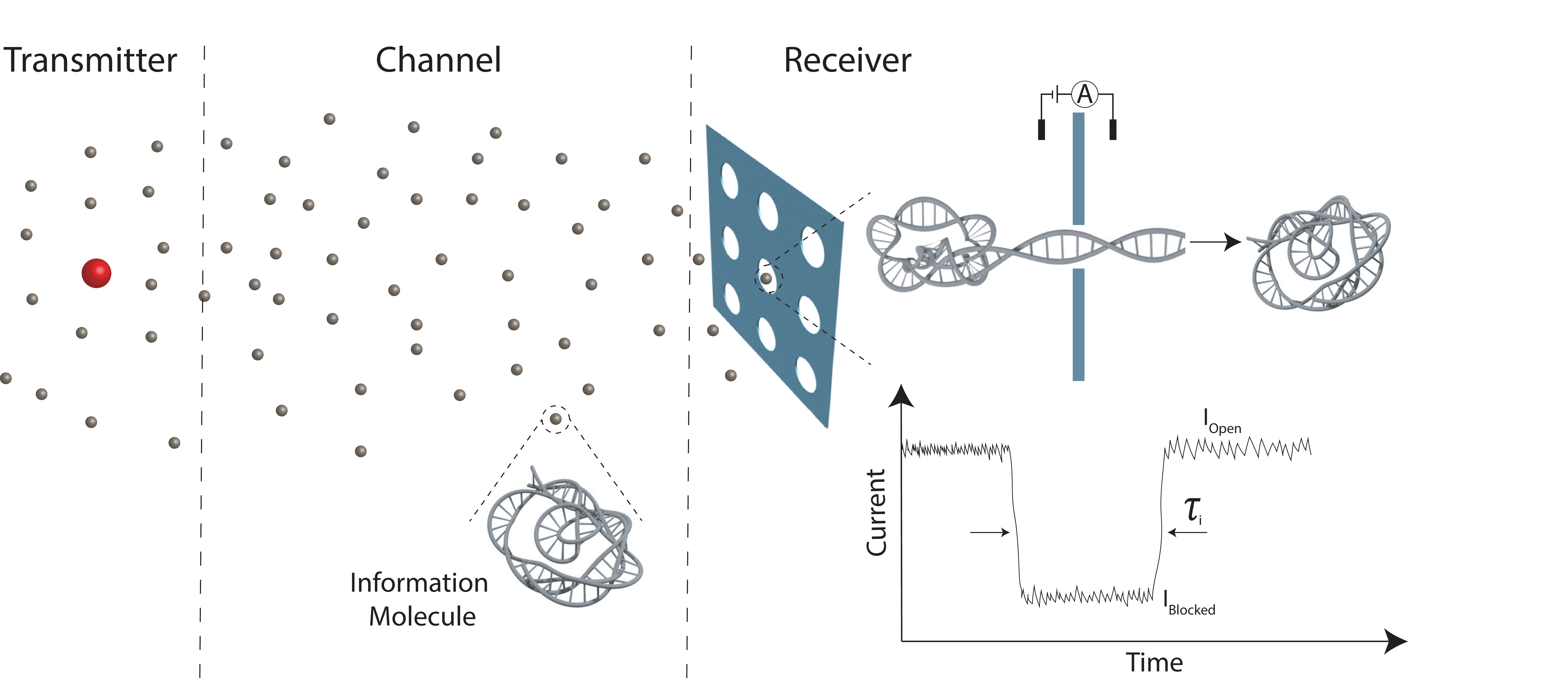}
	\caption{System model for the DNA-based MC.}
	\label{fig:DNA}
\end{figure*}

The processing unit in MC-Tx may require a power source for performing coding and modulation operations especially in the nanomaterial-based MC-Txs. Battery-powered devices suffer from limited-lifetime and random battery depletions, i.e., unexpected exhaustion of sensor battery due to hardware or software failure, that significantly affect the reliability of systems that rely on MC. Therefore, MC-Tx unit is required to operate as a self-sustaining battery-less device by harvesting energy from its surrounding. For this purpose, energy harvesting (EH) from various sources, such as solar, mechanical, chemical, can be utilized in MC-Tx architectures. Hybrid energy harvesting techniques can be employed as well to increase the output power reliability \cite{akan2018internet}. Concerning the intrabody and body area applications, human body stands as a vast source of energy \cite{dagdeviren2017energy}, which has been exploited to power biomedical devices and implants in many ways, e.g., thermoelectric EH from body heat \cite{leonov2013thermoelectric}, vibrational EH from heartbeats \cite{amin2012powering}, and biochemical EH from perspiration \cite{jia2013epidermal}. Nevertheless, design and implementation of EH methods at micro/nanoscales for MC still stand as important open research issues in the literature. 

\subsubsection{\textbf{Information Carrying Molecules}}
\label{sec:TxInformationMolecules}
Reliable and robust MC necessitates chemically stable IMs that can be selectively received at the receiver. In addition, environmental factors, such as molecular deformation due to enzymes and changes in pH can have severe effects on IMs \cite{hiyama2010molecular} and degrade the performance of MC. Next, we provide a discussion on several types of IMs that can be utilized for MC.

\paragraph {\textbf{Nucleic Acids}} Nucleic acids, i.e., deoxyribonucleic acid (DNA) and ribonucleic acid (RNA), are promising candidates as IMs by being biocompatible and chemically stable especially in in-body applications. In the nature, DNA is carrying information one generation to another. RNA is utilized as messenger molecules for intercellular communication in plants and animals to catalyze biological reactions such as localized protein synthesis and neuronal growth \cite{donnelly2010subcellular, mittelbrunn2012intercellular}. In a similar manner, nucleic acids can be exploited in MC-Tx to carry information. DNA has a double stranded structure whereas RNA is single stranded molecule often folded on to itself, and the existence of hydroxyl groups in RNA makes it less stable to hydrolysis compared to DNA. Hence, DNA can be expected to outperform RNA as an information carrying molecule. 

Recent advancements in DNA/RNA sequencing and synthesis techniques have enabled DNA-encoded MC \cite{bell2016digitally,chen2017ionic, slonkina2003polymer}. For information transmission, communication symbols can be realized with DNA/RNA strands having different properties, i.e., length \cite{bell2016translocation}, dumbbell hairpins \cite{bell2016digitally,bell2016direct}, short sequence motifs/labels \cite{chen2017ionic} and orientation \cite{butler2006determination}. For information detection, solid-state \cite{chen2017ionic,dekker2007solid} and DNA-origami \cite{hernandez2014dna} based nanopores can be utilized to distinguish information symbols based on the properties of DNA/RNA strands by examining the ionic current characteristics during \textit{translocation}, i.e., while DNA/RNA strands pass through the nanopores as illustrated in Fig. \ref{fig:DNA}. The utilization of nanopores for DNA symbol detection also enables the miniaturization of MC-capable devices towards the realization of IoNT. According to \cite{bell2016digitally}, 3-bit barcode-coded DNA strands with dumbbell hairpins can be detected through nanopores with 94\% accuracy. In \cite{butler2006determination}, four different RNA molecules having different orientations are translocated with more than 90\% accuracy while passing through transmembrane protein nanopores. Nanopore-based detection can also pave the way for detecting cancer biomarkers from RNA molecules for early detection of cancers as in \cite{wang2011nanopore}. However, this requires high sensitivity and selectivity.

Translocation time of DNA/RNA molecules through nanopores depends on the voltage, concentration and length of the DNA/RNA symbols, and the translocation of symbols can take from a few ms up to hundred ms time frames \cite{bell2016digitally,bell2016direct}. Considering the slow diffusion channel in MC, transmission/detection of DNA-encoded symbols do not introduce a bottleneck and multiple detections can be performed during each symbol transmission. Therefore, the utilization of DNA/RNA strands is promising for high capacity communication between nanomachines as the number of symbols in the modulation scheme can be increased by exploiting multiple properties of DNA/RNA at the same time. Hence, the utilization of DNA as information carrying molecule paves the way for high capacity links between nanomachines by enabling higher number of molecules that can be selectively received at the Rx.

In addition, information can be encoded into the base sequences of DNAs, which is also known as nucleotide shift-keying (NSK). In \cite{organick2018random}, 35 distinct data files over 200MB were encoded and stored by using more than 13 million nucleotides. More importantly, this work proposes a method for reading the stored data in DNA sequences using a random-access approach. Thanks to high information density of DNA, application of NSK in DNA-based MC may boost the typically low data rates of MC up to the extend of competing with traditional wireless communication standards. In NSK, information carrying DNA and RNA strands can be placed into bacteria and viruses. In \cite{cobo2010bacteria}, bacteria based nanonetworks, where bacteria are utilized as a carrier of IMs, have been proposed and analytically analyzed. In \cite{shipman2017crispr}, a digital movie is encoded into DNA sequences and these strands are placed into bacteria. However, DNA/RNA reading/writing speed and cost at the moment limit the utilization of NSK in a practical system. Theoretical and experimental investigation of DNA/RNA-encoded MC stands as a significant open research issue in the MC literature.

\paragraph{\textbf{Elemental Ions}} Elemental ion concentrations, e.g., $Na^{+}$, $K^{+}$, $Ca^{++}$, dictate many processes in biological systems. For instance, in a neuron at steady-state higher $Na^{+}$ and $K^+$ ions are maintained via active transmembrane ion channels in extra- and intracellular medium, respectively, and an action potential is instigated by a surge of $Na^{+}$ ions into the cell upon activation of transmembrane receptors by neurotransmitters from other neurons. On the other hand, $Ca^{++}$ ions are utilized in many complex signaling mechanisms in biology including neuronal transmission in an excitatory synapse, exocytosis, cellular motility, apoptosis and transcription \cite{clapham2007calcium}. This renders elemental ions one of the viable ways to communicate with living organisms for possible applications. Correspondingly, many transmitter device \cite{isaksson2007electronic} and nanonetwork architectures \cite{nakano2005molecular} based on elemental ion signaling have been proposed in literature.

\paragraph{\textbf{Neurotransmitters}} Neural interfacing is one of the hottest contemporary research topics with a diverse range of applications. In this respect, stimulation of neurons using their own language, i.e., neurotransmitters, stands out as the best practice. Various transmitter architectures for the most common neurotransmitters, e.g., glutamate, $\gamma$-amino butyric acid (GABA), aspartate and acetylcholine have been reported in literature \cite{simon2009organic,uguz2017microfluidic,jonsson2016chemical}.

\paragraph{\textbf{Proteins}} Proteins comprise the basic building blocks of all mechanisms in life. They are synthesized by cells from amino acids utilizing information encoded inside DNA, and regulate nearly all processes within the cell, including the protein synthesis process itself, so that they form a self regulatory network. Proteins are commonly used as IMs by biological systems both in intracellular pathways, e.g., enzymes within vesicles between the endoplasmic reticulum and the Golgi apparatus \cite{lee2004bi}, and intercellular pathways, e.g., hormones within exosomes, vesicles secreted from a multitude of cell types via exocytosis \cite{denzer2000exosome}. Artificial transmission of proteins have been long considered within the context of applications such as protein therapy \cite{schwarze1999vivo}. From MC point of view, use of proteins as IMs by genetically engineered bacteria is regarded as one of the possible biological MC architectures, where engineered genetic circuits have been proposed to implement various logic gates and operations necessary for networking \cite{walsh2010synthetic}.

\paragraph{\textbf{Other Molecules}} Many other types of molecules have been used or considered as IMs in literature. These include synthetic pharmaceuticals \cite{chourasia2003pharmaceutical}, therapeutic nanoparticles \cite{cho2008therapeutic}, as well as organic hydrofluorocarbons \cite{kuran2012interference} and isomers \cite{kim2013novel}.

\subsubsection{\textbf{Nanomaterial-based MC-Tx Architectures}}
\label{sec:TxNanomaterial}

In the literature, MC-Tx is generally assumed to be an ideal point source capable of perfectly transmitting molecular messages encoded in the number, type or release time of molecules to the channel instantly or continuously, neglecting the stochasticity in the molecule generation process, and the effect of Tx geometry and channel feedback. Despite of the various studies investigating MC, physical implementation of MC-Tx stands as an important open research issue especially in micro/nanoscale. However, there exist some works on macroscale demonstration of MC. In \cite{farsad2013tabletop}, the authors implemented a macroscale MC system with an electronically controlled spray as MC-Tx, which is capable of releasing alcohol, and alcohol metal-oxide sensor as MC-Rx. According to this experiment, the macroscale MC setup achieves 0.2bps with 2m communication range. Since there are almost no micro/nanoscale implementation of MC-Tx, we investigate and propose MC-Tx architectures by exploiting recent advancements in nanotechnology, novel materials and microfluidics. 

As discussed in Section \ref{sec:TxRequirements}, design of MC-Tx in micro/nanoscale is an extremely challenging task including various requirements such as biocompability, miniaturization and lifetime of the device. For reducing the dimensions of MC-Tx architectures to microscales, microfluidics and microfluidic droplet technologies are promising. Inside droplets, IMs can be transmitted in a precisely controlled manner, such that even logic gates can be implemented with microfluidic chips as demonstrated in \cite{prakash2007microfluidic}. Feasibility of utilizing droplets for communication purposes has been suggested in \cite{fuerstman2007coding}. In addition, microfluidic chips can be fabricated by using Polydimethylsiloxane (PDMS), which is a biocompatible polymer \cite{peterson2005poly}. The authors of \cite{farsad2012chip} theoretically compare achievable data rates of passive transport, i.e., diffusive channel, and active transport, i.e., flow-assisted channel via external pressure or molecular motors, in a microfluidic environment. According to this study, active transport improves achievable data rates thanks to faster movement of IMs from Tx to Rx compared to passive transport. A new modulation scheme based on distance between droplets has been introduced in \cite{de2013communications} by exploiting hydrodynamic microfluidic effects. 
\begin{figure*}[!t]
	\centering
	\includegraphics[width=0.95\textwidth]{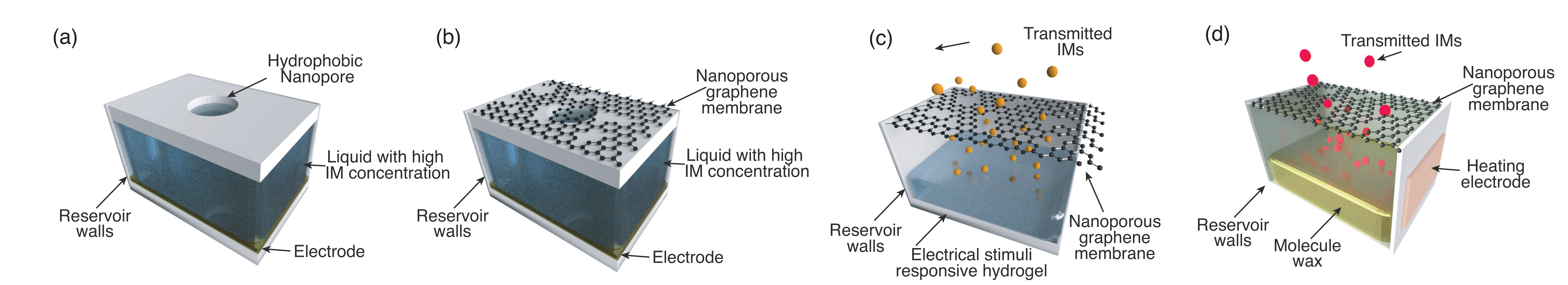}
	\caption{MC-Tx Architectures (a) microfluidic-based MC-Tx with hydrophobic nanopore, (b) microfluidic-based MC-Tx with hydrophobic nanopore and nanoporous graphene membrane, (c) MC-Tx with thin film hydrogels with nanoporous graphene membrane, (d) MC-Tx with molecule wax and nanoporous graphene membrane.}
	\label{fig_mctx_nano}
\end{figure*}

Although it is not originally proposed as MC-Tx, microfluidic neural interfaces with chemical stimulation capabilities, i.e., devices that release neurotransmitters such as glutamate or $\gamma$-aminobutyric acid (GABA) to stimulate or inhibiting neural signals \cite{scott2013microfluidic,iezzi2009microfluidic,jonsson2016bioelectronic}, operates same as MC-Tx. Therefore, the studies on neural interfaces with chemical stimulation capabilities can be considered as baseline for designing micro/nanoscale MC-Tx. However, leakage of IM molecules while there is no signal transmission stands as a significant challenge for microfluidic based MC-Txs. It is not possible to eliminate the problem, but there are some possible solutions to reduce or control the amount of leakage. Hydrophobic nanopores can be utilized as shown in Figure \ref{fig_mctx_nano}(a) such that the liquid inside the container can be separated from the medium when there is no pressure inside the MC-Tx. This solution has two drawbacks. Since there is a need for external pressure source, it is hard to design a practical stand-alone MC-Tx with this setup. Secondly, this solution does not completely eliminate the IM leakage; hence, the authors of \cite{jones2016can} suggest the utilization of porous membranes and electrical control of fluids to further improve MC-Tx against IM leakage.

Nanoporous graphene membranes can provide molecule selectivity \cite{walker2017extrinsic} by adjusting pore size depending on the size of IM. In addition, graphene is also proven to be biocompatible as demonstrated in \cite{fabbro2016graphene}. If the pore size of the graphene membrane can be adjusted in a way that IMs can barely pass through, negatively charging the graphene membrane can decrease the pore size with the additional electrons such that some level of control can be introduced to reduce lM leakage. In addition, an electrode plate placed at the bottom of the container can be utilized to pull and push charged IMs as illustrated in Figure \ref{fig_mctx_nano}(b). This way, an electric field (E-field) can be generated in the liquid containing charged IMs and the direction of E-field can be utilized to ease or harden the release of charged IMs.  

MC-Tx can be also realized with electrical stimuli-responsive thin film hydrogels by performing E-field modulated release and uptake of IMs \cite{murdan2003electro} as in Figure \ref{fig_mctx_nano}(c). Hydrogels, widely utilized in smart drug delivery, are bio-compatible polymers that can host molecules, and reversibly swell/deswell in water solution upon application of a stimuli, resulting in release/uptake of molecules. This architecture can be further improved via a porous graphene controlling its micro-environment. E-field stimuli can be generated by two electrodes placed at the opposite walls of the reservoir. Refilling the IM reservoirs is another significant challenge in MC-Tx design, which can be solved with the replenishable drug delivery methods, e.g., oligodeoxynucleotides (ODN) modification \cite{brudno2014refilling}, click chemistry \cite{brudno2018replenishable}, and refill lines \cite{whyte2018sustained}.

Up to this point, we consider the design of MC-Tx only in liquid medium. For airborne MC, we propose an MC-Tx architecture consisting of a molecular reservoir sealed by a porous graphene membrane, accommodating a wax layer containing IMs as in Figure \ref{fig_mctx_nano}(d). The rest of the reservoir is filled with water, in which the IMs dissolve. Upon application of heat via E-field through conducting walls of the reservoir, the module sweats IM-rich water through membrane pores, and IMs become airborne upon evaporation, and provide a continuous release of IMs.

\subsubsection{\textbf{Biological MC-Tx Architectures}}
\label{sec:TxBiological}
An alternative approach to nanomaterial-based architectures for MC, which in most cases are inspired by their biological counterparts, resides in rewiring the already established molecular machinery of biological realm to engineer biological nanomachines that network via MC to accomplish specific tasks. At the cost of increased complexity, this approach has various advantages over nanomaterial-based designs including inherent biocompatibility and already integrated production, transport and transmission modules for a wealthy selection of IMs and architectures.
	
\paragraph{\textbf{Bacterial Conjugation-based Transmission}} 
\textit{Bacterial conjugation} is one of the lateral gene transfer processes between two bacteria. More specifically, some \textit{plasmids} inside bacteria, mostly circular small double-stranded DNA molecules physically distinct from the main bacterial DNA, can replicate and transfer itself into a new bacteria \cite{hanahan1983studies}. These plasmids encode the conjugative "sex" pilus, which upon receiving a right molecular stimulus from a neighboring bacteria is translated. The produced pilus extends out of the donor cell, attaches to the recipient cell, and then retracts to get the two cells in contact with their intracellular media joined through the pilus hole. Single strand DNA (ssDNA) of the plasmid is then transferred from the donor cell to the recipient cell. utilizing bacterial conjugation as a means of information transmission for MC necessitates at least partial control over bacteria behavior, which is achieved by means of genetically engineering the bacteria.
			
At the core of all genetical engineering schemes lies the concept of gene regulation via a process known as \textit{RNA interference} (RNAi), which has provided us with the means of manipulating gene expressions in targeted cells or bacteria, paving the way for genetical engineered bacteria based MC architectures \cite{hannon2002rna}. RNAi is observed to serve as a mediator of inter-kingdom MC \cite{mellies2010interkingdom}. In particular, it is shown that short hairpin RNA (shRNA) expressing bacteria elicit RNAi in mammals \cite{xiang2006short}, rendering bacteria mediated RNAi-based diagnosis and therapeutics a promising prospect \cite{riglar2018engineering}. In this respect, recently \cite{mckay2018platform} proposed a platform of genetically engineered bacteria as vehicles for localized delivery of therapeutics towards applications for Crohn's disease. From MC perspective, genetically engineered bacteria have been envisioned to be utilized in various ways with different transmitter architectures. Below, we collate various novel research directions in biological transmitter architectures for MC enabled by the recent advancements in genetic engineering.
			
It is important to note that, conjugative pili are typically few microns in length, which from MC point of view dramatically decreases the transmission radius. Accordingly, \cite{cobo2010bacteria} has proposed engineered bacteria with flagella, e.g., \textit{E. coli} as the carriers of information, i.e., sequenced DNA strands, which establish a communication link between two nanomachine nodes that can interface and exchange DNA strands with the bacteria. The sequenced DNA message resides within a \textit{plasmid}. The behavior of bacteria is controlled via chemotaxis by release of attractants from the nodes, and is regulated by encoded active regions on the plasmid. To avoid interference with behavior regulation the message section of the plasmid is inactivated, i.e., it is not expressed, which can be achieved by avoiding \textit{consensus promoter} sequences within the message. Consensus promoters are necessary for the RNA polymerase to attach to the DNA and start transcription. The message section of the plasmid also contains the destination address, which renders message relaying across nodes possible. The authors develop a simulator for flagellated bacteria propagation, which they combine with analytical models for biological processes involved to obtain the end-to-end delay and capacity of the proposed MC channel for model networking tasks. In \cite{gregori2011physical}, utilization of flagellated bacteria for medium range ($\mu$m-mm) MC networks is suggested, where authors present a physical channel characterization of the setup together with a simulator based on it. \cite{sugranes2012capacity} extends the model in \cite{cobo2010bacteria} by accounting for mutations in the bacteria population, and also considering asynchronous mode of operation of the nodes. In \cite{balasubramaniam2012opportunistic}, the authors also consider a similar setup utilizing bacterial conjugation, but employ opportunistic routing, where opportunistic conjugation between bacteria is allowed in contrast to strict bacteria-node conjugation assumed in \cite{cobo2010bacteria}. However, it requires node labeling of bacteria, and additional attractant release by bacteria to facilitate bacteria-bacteria contact for opportunistic routing. The authors extend this work with \cite{balasubramaniam2013multi} by additionally assuming the nanomachine nodes as capable of releasing antibiotics that kill bacteria with useless or no content to decrease noise levels by avoiding overpopulation. Moreover, they allow multiple number of plasmids per bacterium, which, contrary to their previous work and \cite{cobo2010bacteria}, allows simultaneous communication between multiple source and destination nodes over the same network. Later, \cite{balasubramaniam2014exploiting} considers the design of the nanomachine node in more detail, and \cite{petrov2014analytical} derives an analytical model to estimate the successful delivery rates within proposed framework. \cite{petrov2015incorporating} analyzes via simulations the end-to-end delay and reliability of conjugation based transmission within the same scenario for varying system parameters such as quantity of bacteria, area of deployment and number of target sites. An important factor that restricts reliable successful delivery in conjugation based bacterial networks is incomplete DNA transfer due fragile nature of the process, the effect of which is pronounced over multiple transfers. To mitigate losses due to this effect, which always results in a loss of information from the tail part of DNA, \cite{petrov2014forward} devises a \textit{Forward-Reverse Coding} (FRC) scheme, where messages are encoded in both directions and sent simultaneously over the same channel via dedicated sets of bacteria, to equally distribute losses to both ends of DNA. The performance of FRC is later compared with a \textit{cyclic shift coding} (CSC) scheme, in which DNA message is partitioned into $N$ smaller blocks and is cyclically shifted to create $N$ versions of the same content starting with corresponding blocks, and similar to FRC transmitted simultaneously via dedicated bacteria populations \cite{petrov2016performance}. Expectedly, CSC provided higher link probability than FRC, which outperformed straight encoding. In a very similar spirit, \cite{furubayashi2016packet} considers packet fragmentation at transmitter and reassembly at receiver, which has the additional advantage of faster propagation due to diminished size of carriers. They observe that this protocol increases reliability over diffusive channels, however, it degrades communication when there is strong enough drift, which can be attributed to the heightened diffusion rates.
			
The MC transmitter module of the architecture based on bacterial conjugation outlined above, indeed, is composed of several submodules, i.e., the transmitter module of the nanomachine node, the bacteria itself via chemotaxis and the pilus-based sexual conjugation module, which is a reflection of the trademark high complexity involved with biological architectures. In return, the reward is the achievement of unprecedented data rates via MC thanks to the high information density of DNA.

\paragraph{\textbf{Virus-based Transmission}}
Viruses have initially been studied as infectious agents and as tools for investigating cell biology, however, their use as templates for transferring genetic materials to cells has provided us with the means of genetically engineering bacteria \cite{palese1996negative}, and unlocked a novel treatment technique in medicine, e.g., gene therapy \cite{dunbar2018gene}.
A virus is a bio-molecular complex that carries DNA (or RNA), which is packed inside a protein shell, called \textit{capsid}, that is enveloped by a lipid layer. The capsid has (at least) an entrance to its interior, through which the nucleic payload is packed via a ring ATPase motor protein \cite{yu2010mechanochemistry}. Located near the entrance are functional groups that facilitate docking on a cell by acting as ligands to receptors on it. Once docked, the nucleic content is injected into the cell, which triggers the cell's production line to produce more of virus' constituent parts and DNA. These self-organize into fully structured viruses, which finally burst out of the host cell to target new ones.
			
From MC perspective, viral vectors are considered to be one of the possible solutions in transmitting DNA between nanonetworking agents. The concept is similar to bacterial conjugation based transmission, as both encode information into transmitted DNA, but, in contrast viral vectors are immotile and their propagation is dictated by passive diffusion. However, they are comparatively smaller than bacteria, so that more of them can be deployed in a given channel, and they diffuse faster. Moreover, the ligand-receptor docking mechanism, which serves as a header for receptor/cell specific long range targeting, enables the possibility of design of very complex and large scale networking schemes with applications in gene therapy \cite{felicetti2016applications}. Furthermore, considering high information density of DNA, virus based MC stands out as one of the MC protocols viable to support high data rates. Yet, models of MC networks utilizing viral vectors are still very few. In particular, \cite{walsh2010synthetic} proposes utilization of engineered cells as platforms for devices and sensors to interface to nanonetworks. These engineered cells are assumed to be capable of virus production and excretion to facilitate desired networking, where a modular approach is presented for modeling of the genetic circuitry involved in the modulation of viral expression based on incoming extracellular signaling. Based on the model developed in \cite{walsh2010synthetic}, \cite{walsh2013reliability1} and \cite{walsh2013reliability2} investigate viral MC networking between nanomachines that communicate with each other in a diffusive medium via DNA messages transmitted within viruses, where the former analyses the reliability of a multi-path topology, and the latter derives reliability and delay in multi-hop relay networks.

\paragraph{\textbf{Genetic Circuit Regulated Protein Transmission}}
We have so far investigated biological transmitter architectures, where the message to be conveyed has been encoded in sequences of nucleotides, e.g., DNA or RNA. Yet, the most abundant form of intercellular MC interaction in nature occurs via proteins, whose expression levels are determined by the genetic circuitry and metabolic state within the cells. Typically, proteins that are produced within a cell via transcriptional processes are either excreted out via specialized transmembrane protein channels, or packed into vesicles and transported to extracellular medium via exocytosis. As a result of exocytosis, either the vesicle is transported out wholly, e.g., an exosome \cite{mathivanan2010exosomes}, or it fuses with the cell membrane and only the contents are spewed out. Proteins on the membranes of exosomes provide addressing via ligand-receptor interactions with membrane proteins of recipient cell. Upon a match, the membranes of exosome and the recipient cell merge, and exosome contents enter recipient cell. This establishes a one-to-one MC link between two cells. In case the contents are merely spewed out to the external medium, they diffuse around contributing to the overall concentrations within the extracellular medium. This corresponds to local message broadcast, and it has been long observed that populations of bacteria regulate their behavior according to resulting local molecule concentrations resulting from these broadcast messages, referred to as the phenomenon of \textit{quorum sensing}.
			
This mode of communication, even though lacking the information density of nucleotide chains, and therefore supporting lower data rates, are commonly employed by nature as MC schemes. They are considerably more energy efficient compared to DNA transmission schemes, which justifies their use by nature as signaling agents for comparatively simple nanonetworking tasks. In this respect, \cite{abadal2011automata} proposes quorum sensing as a means to achieve synchronization amongst bacteria nodes of a nanonetwork, where they model a bacterium as a finite state automaton, and \cite{abadal2012quorum} investigates quorum sensing as a tool for power amplification of MC signals increasing the range of transmitted signals. In contrast, \cite{nakano2012molecular} considers exosome secretion as a possible means for realization of nanonetworks composed of large number of bio-nanomachines. \cite{unluturk2015genetically} presents a detailed model of engineered genetic circuitry based on mass action laws that regulate gene expressions, which in turn dictate transmitter protein production. Their work stands as a basis for future engineered genetic circuitry based protein transmission.
\begin{table*}[!t]\scriptsize
	\centering
	\begin{threeparttable}
		\centering
		\caption{Comparison Matrix for MC Modulation Schemes}
		\label{tab:modulation}
		\begin{tabular}{llllll}
			\toprule	
			\textbf{Description}   & \textbf{Encoding Mechanism} & \textbf{ISI Reduction} & \textbf{\# of Molecule Type}&  \textbf{\# of Symbols} \\
			\midrule
			On-off keying (OOK) \cite{mahfuz2010characterization}           & Concentration & No            & $1$ &  $2$  \\
			Concentration shift keying (CSK) \cite{kuran2012interference}    & Concentration    & No        & $1$ ($b$ Concentration levels) &     $2^b$       \\
			Pulse amplitude modulation (PAM) \cite{garralda2011diffusion} & Concentration    & No        & $1$ ($b$ Concentration levels) &     $2^b$\\
			Molecule shift keying (MoSK) \cite{kuran2011modulation}          & Molecule type      & Moderate  & $k$     &  $k$     \\
			Depleted MoSK (D-MoSK) \cite{kabir2015d}           & Molecule type & No    & $k$          &  $2^k$      \\
			Isomer-based ratio shift keying (IRSK) \cite{kim2013novel}      & Molecule ratio& Moderate & $k$        &$k$             \\
			Release time shift keying (RTSK) \cite{murin2018exploiting,murin2016communication,srinivas2012molecular} & Release timing& No & $1$  ($b$ Timing intervals)         & $2^b$    \\
			Molecular array-based communication (MARCO) \cite{atakan2012nanoscale} & Molecule order& High     & $k$           & $2^k$\\
			\bottomrule
		\end{tabular}%
	\end{threeparttable}	
\end{table*}%

\paragraph{\textbf{Enzyme Regulated $Ca^{++}$ Circuits}}		
$Ca^{++}$ ions are utilized in many complex signaling mechanisms in biology including neuronal transmission in an excitatory synapse, exocytosis, cellular motility, apoptosis and transcription \cite{clapham2007calcium}. Moreover, they play a major role in intracellular and intercellular signal transduction pathways, where the information is encoded into local $Ca^{++}$ concentration waves under the control of enzymatic processes. Intracellular organelles including mitochondria and endoplasmic reticulum accumulate excess $Ca^{++}$ ions, and serve as $Ca^{++}$ storages. Certain cellular events, such as an extracellular signal generated by a toxin, trigger enzymatic processes that release bound $Ca^{++}$ from organelles effectively increasing local cytosolic $Ca^{++}$ concentrations \cite{brini2013intracellular}. These local concentrations can propagate like waves within cells, and can be injected to adjacent cells through transmembrane protein gap junction channels called \textit{connexins} \cite{cotrina1998connexins}.
			
Establishing controlled MC using this inherent signaling mechanism was first suggested by \cite{nakano2007molecular}, where authors consider communicating information between two nanomachines over a densely packed array of cells that are interconnected via connexins. The authors simulate MC within this channel to analyze system parameter dependent behavior of intercellular signal propagation and its failure. They also report on experiments relating to so-called \textit{cell wires}, where an array of gap junction transfected cells are confined in a wire configuration and signals along the wire are propagated via $Ca^{++}$ ions. The communication was characterized as limited range and slow speed. Later, $Ca^{++}$ relay signaling over 1 cell thick cell wires were investigated in \cite{walsh2010synthetic} and \cite{nakano2010design} via dedicated simulations, where the former explored amplitude and frequency modulation characteristics and the latter aimed at understanding the communication capacity of the channel under stochastic effects. Simulations to determine the effects of tissue deformation on $Ca^{++}$ propagation and the capacity of MC between two nanomachines embedded within two-dimensional cell wires with thickness of multiple cells were carried out in \cite{barros2014transmission}. As part of their study, authors propose various transmission protocols and compare their performance in terms of achieved rates. In a later study \cite{barros2014using}, authors employ $Ca^{++}$ signaling nanomachines embedded within deformable cell arrays to infer deformation status of the array as model for tissue deformation detection. This is achieved by estimating the distance between nanomachines from observed information metrics coupled with strategic placements of nanomachines. Motivated by tissue health inference via embedded nanomachines, the authors of \cite{barros2015comparative} identify three categories of cells that employ $Ca^{++}$ signaling, namely \textit{excitable}, \textit{non-excitable} and \textit{hybrid}, which respectively model muscle cells, epithelium cells and astrocytes, and model $Ca^{++}$ communication behavior within channels comprised of these cells. We refer the reader to \cite{barros2017ca2+} for a more detailed discussion of existing literature, theoretical models, experiments, applications and future directions in this field of MC.

\paragraph{\textbf{Other Biological Architectures}}
In addition to the biological transmission architectures described above, molecular motors sliding on cytoskeletal protein structures, e.g., microtubules, and carrying cargo, i.e., vesicles, between cells  is another option that has been considered by the MC community. In particular, \cite{moore2006design} and \cite{enomoto2006molecular} both describe a high level architecture design for MC over such channels. The comparison of active, i.e., molecular motors on microtubules, and passive, i.e., diffusion, vesicle exchange among cells shows that, active transport is a better option for intercellular MC in case of low number of available vesicles, and passive transport can support higher rates when large numbers of vesicles are available\cite{eckford2010microchannel}. Two design options to form a microtubule nanonetwork in a self-organizing manner, i.e., via a polymerization/depolymerization process and molecular motor assisted organization, are proposed in \cite{enomoto2011design}. A complementary approach to molecular motor based microtubular MC is presented in \cite{farsad2015design}, where an on-chip MC testbed design based on kinesin molecular motors is presented. In this approach, instead of molecular motors gliding over microtubules as carriers of molecules, microtubules are the carriers of molecules, e.g., ssDNA, gliding over a kinesin covered substrate.
			
Other transmitter approaches that involve the use of biological entities are  biological-nanomaterial hybrid approaches. A promising approach is to utilize IM production mechanisms of bacteria in nanomaterial based transmitter architectures. In this direction, \cite{sankaran2018optoregulated} reports on an optogenetically controlled living hydrogel, that is, a permeable hydrogel matrix embedded with bacteria from an endotoxin-free \textit{E. coli} strain, which release IMs, i.e., anti-microbial and anti-tumoral drug deoxyviolacein, in a light-regulated manner. The hydrogel matrix is permeable to deoxyviolacein, however spatially restricts movement of the bacteria. This hybrid approach proposes a solution the the reservoir problem of nanomaterial based architectures. Moreover, to cover a variety of IMs simultaneously this approach can be built upon by utilizing many engineered bacteria, and controlling their states via external stimulus to control their molecular output \cite{lindemann2016engineering}.

\subsection{Modulation Techniques for MC}
\label{sec:TxModulation}
In MC, several modulation schemes have been proposed to encode information into concentration, molecule type, and molecule release time as in Table \ref{tab:modulation}. The first and simplest modulation method that was proposed for MC is {\em on-off keying (OOK)}, in which certain number of molecules are released for high logic and no molecule is released to represent low logic \cite{mahfuz2010characterization}. In a similar manner by using a single molecule, {\em concentration shift keying (CSK)} that is analogous to amplitude shift keying (ASK) in traditional wireless channels is introduced in order to increase the number of symbols in the modulation scheme by encoding information into concentration levels \cite{kuran2012interference}. In \cite{garralda2011diffusion}, a similar modulation scheme based on the concentration levels is proposed and named as \textit{pulse amplitude modulation (PAM)}, which uses pulses of continuous IM release instead of instantaneous release as in CSK. The number of concentration levels that can be exploited significantly depends on molecule type and channel characteristics as ISI at MC-Rx can be a limiting factor.  

Hitherto, we have discussed modulation techniques that use single type of molecules. However, information can be encoded by using multiple molecules such that each molecule represents different symbol, and $k$ information symbols can be represented with $2^k$ different types of molecules in {\em molecule shift keying (MoSK)} \cite{kuran2011modulation}. That is, 1-bit MoSK requires two molecules to encode bit-0 with molecule A and bit-1 with molecule B. The modulation of each molecule in MoSK is based on other modulation techniques such as OOK. MoSK can achieve higher capacity, but the main limiting factor for this modulation type is the number of molecules that can be selectively received. The authors of \cite{kabir2015d} further improve MoSK by representing low logic with no molecule release and enabling simultaneous release of different type of molecules such that $2^k$ symbols can be represented with $k$ molecules, which is named as depleted MoSK (D-MoSK). That is, 1-bit D-MoSK is equivalent to OOK. 2-bit D-MoSK requires only two molecules, and four distinct symbols can be encoded with these molecules (molecule A and molecule B), such as N (00), A (01), B (10), and AB (11), where N represents no molecule release. Furthermore, the authors of \cite{kim2013novel} propose the utilization of isomers, i.e., the molecules having the same atoms in a different orientation, and a new modulation scheme, named as {\em isomer-based ratio shift keying (IRSK)}, in which information is encoded into the ratio of isomers, i.e., molecule ratio-keying. 

The release time of molecules can be also used to encode information. In \cite{garralda2011diffusion}, the authors propose pulse position modulation in which signaling period is divided into two blocks such that a pulse in the first block means high logic and a pulse in the second block means low logic. More complex modulation schemes-based on release timing, i.e., {\em release time shift keying (RTSK)}, where information is encoded into the time interval between molecule release has been investigated in \cite{murin2017diversity,murin2018exploiting,murin2016communication,srinivas2012molecular}. The channel characteristics in case of RTSK is significantly different than other modulation schemes as additive noise is distributed with inverse Gaussian distribution in the presence of flow in the channel \cite{srinivas2012molecular}, and Levy distribution without any flow \cite{murin2016communication}. 

In MC, ISI is an important performance degrading factor during detection due to random motion of particles in the diffusive channels. The effects of ISI can be compensated by considering ISI-robust modulation schemes. In \cite{movahednasab2016adaptive}, an adaptive modulation technique exploiting memory of the channel is utilized to encode information into emission rate of IMs, and this approach makes the channel more robust against ISI by adaptive control of the number of released molecules. In addition, the order of molecules can be also used for information encoding as in {\em molecular array-based communication (MARCO)} \cite{atakan2012nanoscale}. In this approach, different types of molecules are released consecutively to transmit symbols, and by assuming perfect molecular selectivity at the transmitter, the effects of ISI can be reduced. 

The modulation schemes based on concentration, molecule type/ratio/order, and molecule release time offer limited number of symbols. Therefore, MC suffers from low data rates by considering limited number of symbols and slow diffusive propagation. To tackle this problem, large amount of information can be encoded into the base sequences of DNAs, i.e., {\em Nucleotide Shift-Keying (NSK)}. For this purpose, information can be encoded directly using nucleotides with an error coding algorithm such as Reed Solomon (RS) block codes \cite{blawat2016forward}, or an alphabet can be generated out of DNA sequences (100-150bp per letter) to encode information. The latter approach can yield higher performance in terms of BER considering the complexity and size of MC-Tx and MC-Rx architectures. Although identification of base pairs with nanopores can be performed with relatively low costs and high speeds \cite{clarke2009continuous,derrington2010nanopore}, there is yet no practical system to write DNA sequences with a microscale device. Therefore, future technological advancements towards low-cost and practical synthesis/sequencing of DNA are imperative for MC communications with high data rates, e.g., on the order of Mbps.

\subsection{Coding Techniques for MC}
\label{sec:TxCoding}
Encoding of information before transmission is classically done for two reasons, \textit{source coding} is done for statistically efficient representation of data form a discrete input source, and \textit{channel coding} is done to control errors that occur due to channel noise via introducing redundant bits. Source coding practices are independent of channel characteristics, and as a result they do not differ for MC with respect to traditional communications from ICT point of view. For this reason, we do not cover source coding in this review. On the other hand, MC channels are typically diffusive, a process which has slow and omnidirectional propagation. As a consequence IMs quickly accumulate in the channel after a series of transmissions, rendering MC extremely noisy and susceptible to ISI. Moreover, as coding has a computational burden on both transmitter and receiver ends, and energy is a scarce resource at nanoscale, energy efficiency of employed channel codes is also a crucially important aspect. This calls for utilization lower complexity block codes, such as simple parity codes or cyclic codes, e.g., Hamming codes \cite{hamming1950error}, instead of the state-of-the-art high complexity codes with high computational burden like Turbo codes \cite{berrou1993near}. On the other hand, the noisy nature of MC channels and over-pronounced effects of ISI renders channel codes developed for conventional EM communications ill-adjusted for MC, calling for invention of novel coding techniques specifically tailored for MC. Table \ref{tab:Coding} enlists the channel coding practices so far employed in MC literature. Below, we summarize these works and highlight their contributions, where we focus more on codes that have been specifically designed for MC.
	
	\begin{table*}[!t]\scriptsize
		\begin{threeparttable}
			\centering
			\caption{Comparison Matrix for MC Channel Codes}
		    \label{tab:Coding}%
			\begin{tabular}{llllllllll}
				\toprule	
				\multirow{2}{*}{\textbf{Code}} & \multirow{2}{*}{\textbf{Mod.}} & {\textbf{Transmission/}} & \textbf{Channel Model} & \multirow{2}{*}{\textbf{Detection}} & \multirow{2}{*}{\textbf{Reception}} & \multirow{2}{*}{\textbf{C.D}.} & \textbf{\# of Mol.} & \multirow{2}{*}{\textbf{Coding Rate ($\frac{k}{n}$)}} & \multirow{2}{*}{\textbf{E.C.}}\\
				& & \textbf{\# of Mol.} & \textbf{Prop./Dim./B.N.} & & & & \textbf{Types} & & \\
				\midrule
				ISI-free code \cite{shih2012channel,shih2013channel} & MoSK & Reg./Single & Diff.+Drift/1d/No & Abs. & Imm. & No & 2 & $\frac{2}{4},\frac{2}{5},\frac{4}{7},\frac{3}{8},\frac{3}{11},\frac{4}{23},\frac{5}{47}$ & No\\
				MoCo-based code \cite{ko2012new} & OOK & Reg./Single & Diff.+Drift/1d/Yes & Abs. & Add.Thr.$=$$1$ & Yes & 1 & $2/4$ & No\\
				Hamming codes \cite{leeson2012forward,bai2014minimum}& OOK & Reg./Many & Diff./3d/No & Non-Abs. & Add.Thr. & No & 1 & $\frac{2^m-m-1}{2^m-1}$, $m=3,4,5$ & Yes\\
				SOCC \cite{lu2015self} & OOK & Reg./Many & Diff./3d/No & Non-Abs. & Add.Thr. & No & 1 & $k/(k+1),$ $k=1,2$ & Yes\\
				SOCC \cite{lu2017energy} & OOK & Reg./Many & Diff./3d/No & Abs. & Add.Thr. & No & 1 & $k/(k+1),$ $k=1,2$ & Yes\\
				EG-LDPC \cite{lu2015comparison} & OOK & Reg./Many & Diff./3d/No & Abs. & Add.Thr. & No & 1 & $\frac{4^s-3^s}{4^s-1}$, $s=2,3,4$ & Yes\\
				Convolutional code \cite{mahfuz2013performance} & PAM & Reg./Many & Diff./3d/No & Abs. & Add.Thr. & No & 1 & $1/2$ & No\\
				SPC code \cite{marcone2018parity} & OOK & Reg./Many & Diff./3d/Yes & Reactive & LLR & No & 2 & $2/3$ & No\\
				\bottomrule
			\end{tabular}%
			\begin{tablenotes}
				\scriptsize
				\item Abbreviations | Abs.: Absorbing -- Add.Thr.: Additive with threshold -- B.N.: Background Noise -- C.D.: Channel Dependent -- Diff.: Diffusion -- Dim.: Dimension -- E.C.: Energy Considered -- Imm.: Immediate -- Mod.: Modulation -- Mol.: Molecule -- Non-Abs.: Non-absorbing -- Prop.: Propagation -- Reg.: Regular
			\end{tablenotes}
		\end{threeparttable}
	\end{table*}%
	
\begin{figure}[!t]
	\centering
	\includegraphics[width=\linewidth]{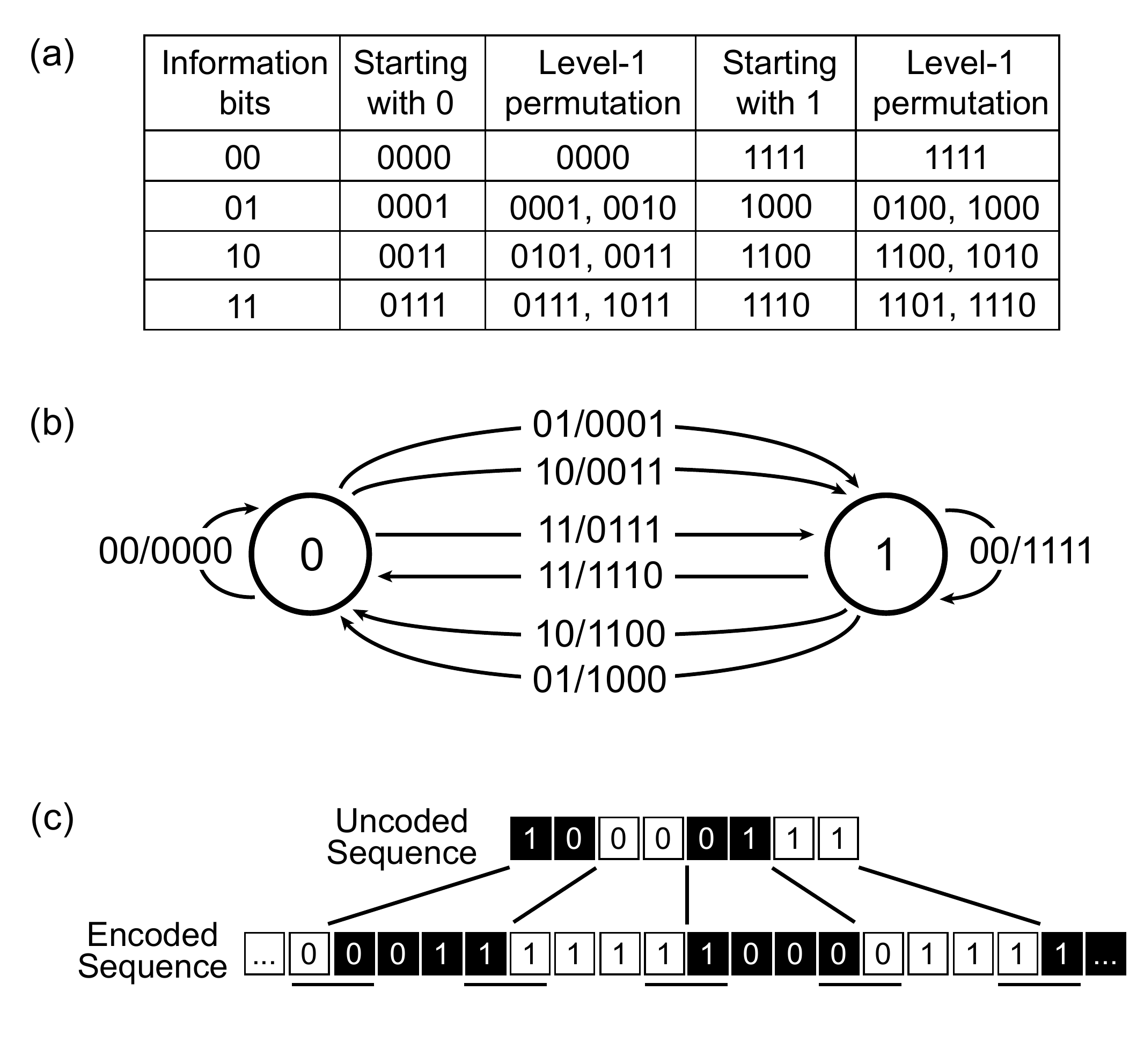}
	\caption{(a) Employed codewords for two states, e.g., starting with 0 or 1, together with their level-1 permutations, (b) the state transition diagram, and (c) an encoding example for the ISI-free $(4,2,1)$ code. Note the same bits at the boundaries of contiguous words. \cite{shih2012channel}}
	\label{fig:Coding}
\end{figure}

The first MC specific code in literature aimed at mitigating the effects of ISI in MC, as it is the main source of high BERs. In \cite{shih2012channel} authors introduce the \textit{ISI-free} coding scheme under MoSK modulation, where two distinguishable molecules encode for bit 0 and bit 1, respectively. The receiver is \textit{absorbing}, i.e., detects everything that hits, and it immediately receives bit 0 or bit 1 upon detection depending on the type of detected molecule. The authors work with the example of a $(4,2,1)$ ISI-free code, where an $(n,k,l)$ ISI-free code is an $(n,k)$ block code, i.e., maps $k$-bit information into $n$-bit codewords, and is error-free provided that there are no more than level-$l$ crossovers. Here, crossover is the phenomenon of late detection of a molecule belonging to previously transmitted symbols, and a level-$l$ crossover means that the detected molecule was transmitted $l$ symbols ago. ISI-free code is a fixed code, in that, it is invariant with respect to change in channel parameters. The (4,2,1) code implemented in \cite{shih2012channel} is based on the idea of finding a codebook with codewords, whose level-1 permutation sets, i.e., possible detection sequences under maximum level-1 crossover assumption, are disjoint. However, level-1 permutation sets of codewords depend on values of neighboring bits at the boundary of contiguous codewords, where if they are same, crossovers between contiguous codewords do not contribute new elements to the level-1 permutation set, making finding codewords with disjoint permutation sets easier. To achieve this, authors devise a two state encoder architecture, whose codeword assignments and state diagram are illustrated in Fig. \ref{fig:Coding} together with an example encoding. Note that contiguous codewords have neighboring bits always same. The receiver decodes the information bits from the codeword received by adding the number of 1's modulo $n=4$, and converting the result to binary, which is a fairly simple decoding rule, and therefore favorable for MC. The authors also compare the ISI-free $(4,2,1)$ code with convolutional and repetition codes, and verify comparable BER performance with much less computational burden. In their following work \cite{shih2013channel}, authors extend the ISI-free $(n,k,l)$ codes to account for higher level crossovers, namely, for levels $l=2,3,4,5$. Furthermore, they introduce the ISI-free $(n,k,l,s)$ codes, in which the codewords have at least $l$ final and $s$ initial identical bits, instead of the symmetric at least $l$ identical bits at both ends of ISI-free $(n,k,l)$ codes, and show that they significantly outperform $(n,k,l)$ codes under similar computational burdens. In essence, the motivation for $(n,k,l,s)$ codes comes from the asymmetry of inter-codeword error probabilities arising from crossover at the start of a codeword and at the end of a codeword. More specifically, if one compares the probability of a given molecule having level-$l$ crossover forwards via arriving later than the $l$ molecules released after it, to the probability of having level-$l$ crossover backwards via the preceding $l$ molecules arriving after the given molecule, one finds latter to fall far more rapidly with increasing $l$. Thus, to reduce computational burden $s$ is typically chosen lower than $l$ signifying the low probability of backwards crossover errors. Finally, the authors demonstrate that ISI-free $(n,k,l,s)$ codes can deliver better BERs than convolutional codes with less computational resources. As its weaknesses, the work considers a very simple one dimensional diffusive channel model with positive drift velocity, which lacks many phenomena diffusive MC enjoys in three dimensions. In particular, transmitted IMs are doomed to hit the receiver, which is very different from the three dimensional case, where there is always the probability that no molecules will reach the receiver. The extent of this simplification reveals itself in the assumption that the transmitter releases a single molecule per symbol, which, thanks to the drift in channel and the absorbing nature of receiver, is always detected.

MC-adapted version of classical Hamming codes were introduced in \cite{ko2012new}, where the traditional Hamming distance metric on the codeword space, given by the number of bit differences between two binary codewords, is replaced by the so-called \textit{molecular coding distance function} (MoCo). In its essence, MoCo is defined in terms of the negative logarithms of probabilities $Pr(\{x\rightarrow y\})$ of receiving codeword $y$ when $x$ was transmitted, and the code aims at generating a codebook with maximal minimum pairwise MoCo distance between constituent codewords. MoCo distance is not a metric, as it is not symmetric, and the triangular inequality is not verified by the authors. This work, too, considers a one dimensional diffusive channel with drift and an absorbing receiver, however, in contrast to \cite{shih2012channel,shih2013channel}, it uses synchronized time slotted OOK modulation scheme with only a single type of IM and the receiver is additive with a threshold equal to 1, i.e., it counts the number of hits in a period and claims high logic reception with a single hit. In case of $(4,2)$ block codes, the authors demonstrate that the code generated using MoCo performs superior to Hamming code by carrying out an error rate analysis for both codes. However, as shortcomings, MoCo depends on detection probabilities that are sensitive to variations in channel properties, and are in general unknown to Tx and Rx. Moreover, even if adaptive techniques may be envisioned, calculation of MoCo-based codebook (at Tx) and decoding region partition (at Rx) require significant computational resources at Tx and Rx, respectively, and overhead communication would have to be considered for synchronized code updating.
	
The authors of \cite{leeson2012forward} propose classical Hamming codes to introduce error correction in MC, where they consider a three-dimensional diffusive channel with time slotted OOK modulation scheme in a channel with finite memory. The receiver is modeled as a non-absorbing sphere which immediately detects molecules that arrive to it, and reception is decided upon additive count of arriving molecules during transmission period. Hamming codes are error correcting block codes with coding ratio $k/n=(2^m-1)/(2^m-m-1)$, where $m$ is the number of parity check bits, and \cite{leeson2012forward} considers Hamming codes for $m=3,4,5$. Their results show that Hamming codes can deliver coding gains up to $\approx 1.7$dB at transmission distances of 1$\mu$m and for low BERs. Here, coding gain is defined as the gain the code introduces in required number of IMs per transmission to achieve a given BER. At high BERs, i.e., low quantities of transmitted IMs, extra ISI introduced by parity bits overweighs error correction, and uncoded transmission performs better. The authors also incorporate an energy model for transmission, where energy is taken to be proportional to the number of transmitted IMs, and show that the energy required to transmit the extra parity bits causes coded transmission to be energy inefficient at small communication distances, however coding becomes more efficient for larger distances. Later on, over the same channel model a Hamming minimum energy code (MEC) scheme was proposed in \cite{bai2014minimum}. In a trade off of having larger codeword lengths against generating codewords with lower average weights by using more 0-bits, the authors trade between rate and energy efficiency of communication. In subsequent works \cite{lu2015self, lu2017energy}, again over the same channel except with an absorbing receiver in \cite{lu2017energy}, self-orthogonal convolutional codes (SOCCs) are proposed, and their performance against Hamming MECs and uncoded transmissions are investigated with respect to both BER and energy efficiency. Both works conclude that, in nanoscale MC SOCCs have higher coding gains, i.e., they are more energy efficient, compared to uncoded transmission, and to the Hamming MECs when low BER ($10^{-5}$-$10^{-9}$) region. Moreover, SOCCs are also reported to have shorter critical distances than Hamming MECs, where the critical distance is defined as the distance at which extra energy requirements of employing coding are compensated by the coding gain. \cite{lu2017energy} additionally explores the energy budget of nano-to-macro and macro-to-nanomachine MC, and arrive at the conclusion that in MC involving macro-machines the critical distance of the codes decrease. Yet in another work \cite{lu2015comparison}, Hamming codes are evaluated against cyclic 2 dimensional Euclidean geometry low density parity check (EG-LDPC) and cyclic Reed-Muller codes by considering the same channel model as in \cite{lu2017energy}. Again, the comparison of codes is carried out for different MC scenarios involving nano and macro-machines, and it reveals in the case of nano-to-nanomachine MC that, in BER region $10^{-3}-10^{-6}$ Hamming codes with $m=4$ are superior, and at lower BER regions LDPC codes with $s=2$ exhibit the lowest energy cost. Here, $s\geq 2$ is the density parameter in LDPC codes, where coding density increases to 1 monotonically as $s\rightarrow \infty$. Moreover, in macro-to-nano and nano-to-macro MC the results indicate that LDPC codes with $s=2$ and $s=3$ are the best options, respectively.
	
The performance of convolutional coding techniques in diffusive MC systems has been investigated by \cite{mahfuz2013performance} utilizing PAM with $M=1,2,4$ pulse amplitude levels for varying key factors such as transmission rate and communication range ($0.8\mu$m-$1$mm). The findings indicate that, while convolutional coding with high transmission rate and $M=1$, i.e., OOK modulation, does outperform the uncoded transmission in short- and medium-range communications, no coding does better than convolutional codes in long range MC. Furthermore, increase in the number of pulse amplitude levels causes deterioration in achieved BERs, which is attributed to increased ISI, implying that, OOK modulation is better suited to MC than PAM.
	
All the aforementioned works apply various channel coding techniques for error correction in MC, however they do not provide any details into mechanisms of implementation of these codes from device architecture perspective. In \cite{marcone2017biological}, authors devise a molecular single parity check (SPC) encoder with OOK modulation for an MC design based on genetically engineered bacteria, which are assumed to network with each other via signaling molecules, e.g., N-acyl homoserine lactones (AHLs). The implementation of joint encoder-modulator module is achieved via design of genetic circuits that regulate gene expression levels, and the transmission materializes from ensuing biomolecule concentrations dictated by biochemical reactions. Developed SPC encoder, which appends to 2 bits information a parity check bit via  biological XOR gate based on designed genetic circuits, provides an error detection mechanism, however with no correction. Still, the introduced design serves as a basis for genetic circuit based designs of more complex block codes with error correction capabilities. In this paper, there is no evaluation of the proposed coding scheme as the paper considers only the transmitter side of MC. A year later, in \cite{marcone2018parity}, authors extended their work in \cite{marcone2017biological} by introducing biological analog decoder circuit, which computes the a-posteriori log-likelihood ratio (LLR) of transmitted bits from observed transmitter concentrations. LLR is defined as the gain of probability of detection over non-detection in dB. This enabled them to analyze the whole end-to-end MC over a diffusive channel. Via simulations they manage to verify intended operation of designed modulated SPC encoder and the analog decoder, and observe network performance close to an electrical network operating in high noise.

\section{Molecular Communication Receiver}
\label{sec:Rx}
\begin{figure}
	\centering
	\includegraphics[width=\linewidth]{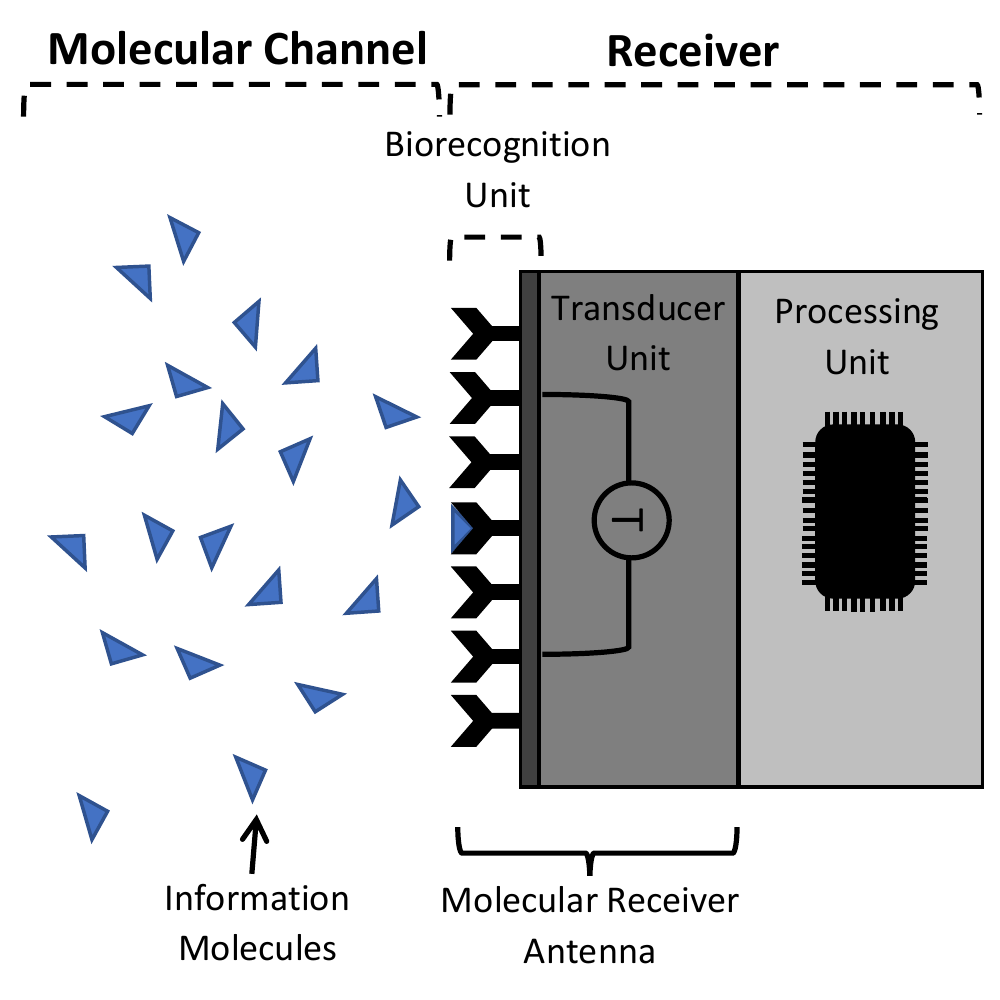}
	\caption{Components of an MC-Rx.}
	\label{fig:MCcomponents}
\end{figure}

The MC receiver (MC-Rx) recognizes the arrival of target molecules to its vicinity and detects the information encoded in a physical property of these molecules, such as concentration, type or release time. To this aim, it requires a molecular receiver antenna that consists of a biorecognition unit followed by a transducer unit. The biorecognition unit, i.e., the interface with the molecular channel, holds a molecular recognition event specifically selective to the information carrying molecules, e.g., it selectively reacts to these target molecules. Then, the transducer unit generates a processable signal, e.g., electrical or biochemical signal, based on this molecular reaction. Finally, a processing unit is needed to detect the transmitted information based on the output of the molecular antenna. The interconnection of these components in an MC-Rx is illustrated in Fig. \ref{fig:MCcomponents} \cite{kuscu2016physical}. Since this structure is fundamentally different from EM communication receivers, it is necessary to thoroughly investigate the receiver architecture specification. To this aim, in this section, we first discuss the requirements of a receiver to be operable in an MC application and the communication theoretical performance metrics that must be taken into consideration while designing the receiver. Then, we review the available approaches in physical design of MC-Rxs, which can be categorized into two main groups, (i) biological receivers based on synthetic gene circuits of engineered bacteria and (ii) nanomaterial-based artificial MC-Rx structures.

\subsection{Design Requirements for MC-Rx}
\label{sec:RxRequirements}
While designing the MC-Rx, its integrability to a mobile nanomachine with limited computational, memory and energy resources, that requires to operate independently in an MC setup, must be taken into consideration. This dictates the following requirements for the functionality and physical design of the receiver \cite{kuscu2016physical}.
\begin{itemize}
	\item {\textbf{In situ operation:}} In-device processing of the molecular message must be one of the specifications of the receiver since it cannot rely on any post-processing of the transduced signals by an external macroscale 	device or a human controller.
	\item \textbf{Label-free detection:} Detection of information carrying molecules, i.e., IMs, must be done based on their intrinsic characteristics, i.e., no additional molecular labeling procedure or preparation stage must be required.
	\item \textbf{Continuous operation:} MC-Rx requires to observe the molecular channel continuously to detect the signal encoded into concentration, type/ratio/order or release time of molecules. Thus, the functionality of molecular antenna and the processing unit should not be interrupted. Since receptors are needed in the biorecognition unit for sensing target molecules, it is important to have re-usable receptors, i.e., they must return to their initial state after signal detection to be ready for the next channel use.
	\item \textbf{Energy efficiency:} Due to limitations of nanomachines, the energy usage of the MC-Rx must be optimized. Additionally, as discussed in Section \ref{sec:TxRequirements}, batteries may not be feasible solutions for long term activity of nanomachines, e.g., as an implanted device. Thus, the receiver may need to be designed with EH units to be energy self-reliance.
	\item \textbf{Biocompatibility and biodurability:} One of the most important MC application areas is the life science, e.g., it promises diagnosis and treatment techniques for diseases caused by dysfunction of intrabody nanonetworks such as neurodegenerative diseases \cite{akan2017fundamentals}. These \textit{in vivo} applications dictate further requirements for the device. Firstly, it must not have any toxic effects on the living system. Moreover, the device needs to be flexible not to cause any injury to the living cells due to mechanical mismatches. Furthermore, there must not be any physiological reactions between device and environment and it must not cause immunological rejection. On the other hand, the physiological environment should not degrade the performance of the device with time.
	\item \textbf{Miniaturization:} Lastly, to be integrated into a nanomachine, the MC-Rx must be built on micro/nanoscale components.
\end{itemize}

\subsection{Communication Theoretical Performance Metrics}
\label{sec:RxMetrics}
In addition to the general performance metrics defined for EM communications, e.g., Signal-to-noise ratio (SNR), Bit error rate (BER), and mutual information, new performance metrics are needed to fully evaluate the functionality of an MC-Rx since molecules are used as IMs in MC. The most important performance metrics are summarized as follows.
\begin{itemize}
	\item \textbf{Limit of detection (LoD):} LoD is a well-known performance metric in biosensing literature, which shows the minimum molecular concentration in the vicinity of the biosensor needed for distinguishing between the existence and the absence of target molecules \cite{armbruster2008limit}. Since the input signal in MC-Rxs is a physical probability of information carriers, LoD corresponds to the sensitivity metric used for evaluating the performance of EM communication receivers, which indicates the minimum input signal power needed to generate a specified SNR at the output of the device. 
	\item \textbf{Selectivity:} This metric is defined based on the relative affinity of the biorecognition unit to the information molecules and interferer molecules \cite{nair2007design}, which can be non-IMs in the medium or other type of IMs in case of MC system with multiple IMs such as MoSK. High selectivity, i.e., very lower probability of interferer-receptor binding compared to target molecule-receptor binding, is needed to uniquely detect the information molecules in the vicinity of the MC-Rx. 
	\item \textbf{Operation range:} The biorecognition unit does not provide infinite range of molecular concentration detection as it has finite receptor density. The response of the device can be divided into two regions, (i) the linear operation, which is the range of molecular concentration in the vicinity of the MC-Rx that does not lead to saturation of receptors, and (ii) the saturation region \cite{kalantar2013sensors}. When the information is encoded into the concentration of molecules, the receiver must work in the linear operation region as the output of molecular antenna provides better information about changes in the molecular concentration. However, for other encoding mechanisms, e.g., MoSK, the receiver can work in both regions. 
	\item \textbf{Molecular sensitivity:} In addition to the aforementioned sensitivity metric that is mapped to LoD for MC-Rxs, a molecular sensitivity can also be defined for an MC-Rx. This metric indicates the smallest difference in the concentration of molecules that can be detected by an MC-Rx \cite{nair2007design}. It is of utmost importance when the information is encoded in molecular concentration. The metric can be defined as the ratio of changes in the output of the molecular receiver antenna to changes in the molecular concentration in the vicinity of the MC-Rx when the device performs in the linear operation region. 
	\item \textbf{Temporal resolution:} This metric is defined to evaluate the speed of sampling the molecular concentration by the receiver. Since the electrical processes are much faster than molecular processes, it is expected that the diffusion and binding kinetics limit the temporal resolution. Thus, biorecognition unit should be realized in transport-limited manner to detect all the messages carried by molecules into the vicinity of the MC-Rx, i.e., the binding kinetics should not be a limiting factor on the sampling rate \cite{kuscu2016physical}.\color{black}
\end{itemize}

\subsection{Physical Design of MC Receiver}
\label{sec:RxPhysicalDesign}
Most of existing studies on the performance of MC ignore the physical design of the receiver and assume that the receiver can perfectly count the number of molecules that (i) enter a reception space with transparent boundaries \cite{meng2014receiver,kilinc2013receiver,noel2014optimal,mosayebi2014receivers}, (ii) hit a 3-D sphere that absorbs molecules \cite{yilmaz20143,damrath2017equivalent,dinc2017theoretical}, or (iii) bind to receptors located on its surface \cite{pierobon2011noise,aminian2015capacity,deng2015modeling}. However, the processes involved in the molecular-to-electrical transduction affect the performance of the receiver, thus, the comprehensive communication theoretical modeling of these processes are required. Available approaches in physical design of MC-Rxs can be categorized into two main groups as follows.

\subsubsection{\textbf{Biological MC-Rx Architectures}}
\label{sec:RxBiological}
Synthetic biology, the engineering of biological networks inside living cells, has seen remarkable
advancements in the last decade, such that it becomes possible to device engineered cells, e.g., bacteria, for use as biological machines, e.g., sensors and actuators, for various applications. Synthetic biology
also stands as a promising means of devising nanoscale biotransceivers for IoNT applications, by
implementing transmission and reception functionalities within living cells by modifying the natural
gene circuits or creating new synthetic ones \cite{unluturk2015genetically}. The technology is already mature enough to allow
performing complex digital computations, e.g., with networks of genetic NAND and NOR gates, as well
as analog computations, such as logarithmically linear addition, ratiometric and power-law computations,
in synthetic cells \cite{daniel2013synthetic}. Synthetic gene networks integrating computation and memory is also proven
feasible \cite{purcell2014synthetic}. More importantly in this context, the technology enables implementing bio-nanomachines
capable of observing individual receptors, as naturally done by living cells; thus, stands as a suitable
domain for practically implementing more information-efficient MC detectors based on the binding state
history of individual receptors, as discussed in Section \ref{sec:RxDetection}.

\begin{figure}
	\centering
	\subfigure[Biological circuit in intracellular environment of an MC-Rx]{\includegraphics[width=\linewidth]{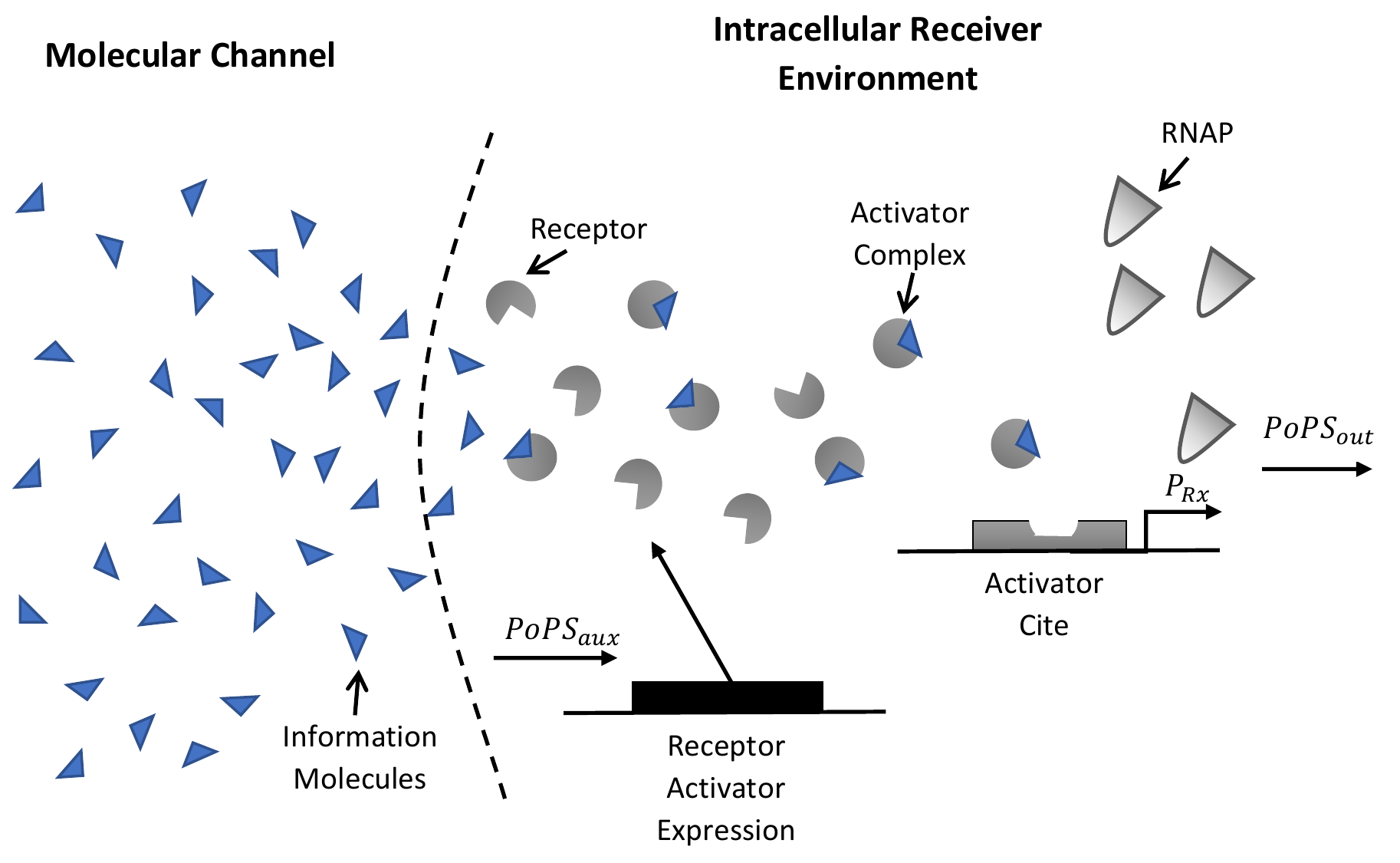}}
	\subfigure[ Functional block diagram]{\includegraphics[width=7cm]{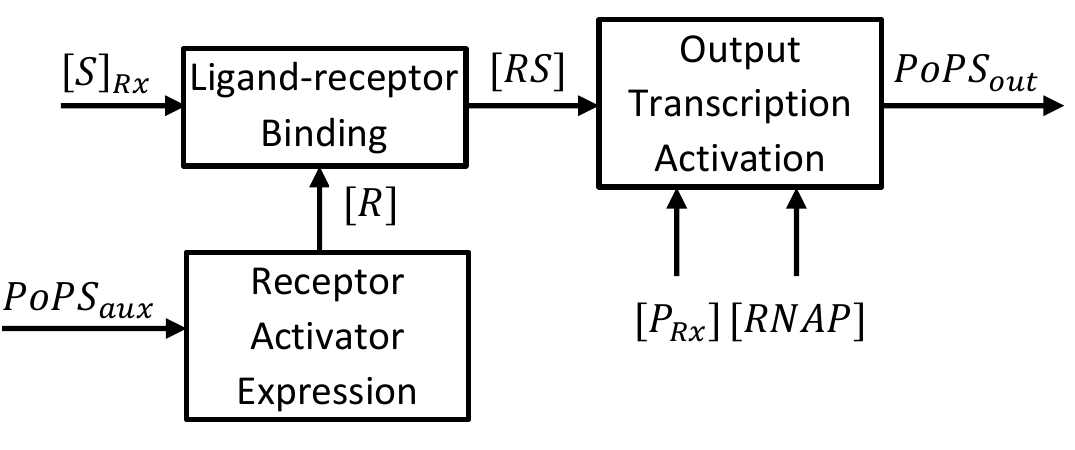}}
	\caption{Biological circuit of an MC-Rx \cite{pierobon2014systems}.}
	\label{fig:geneticCom}
\end{figure} 

In DNAs, gene expression is the process that produces a functional gene product, such as a protein. The rate of this expression can be controlled by binding of another protein to the regulatory sequence of the gene. In biological circuits, activation and repression mechanisms, that regulate the gene expressions, are used to connect DNA genes together, i.e., the gene expression of a DNA generates a protein which can then bind to the regulatory sequence of next DNA to control its expression \cite{myers2016engineering}. In \cite{baker2006engineering}, polymerases per second (PoPS) is defined as the unit of input and output of a biological cell, i.e., the circuit processes a PoPS signal as input and generates another PoPS signal at its output using the aforementioned connections among DNAs. The literature in applying synthetic biology tools to design bacteria-based MC-Rxs is scarce. An MC biotransceiver architecture integrating molecular sensing, transmitting, receiving and processing functions through genetic circuits is introduced in \cite{unluturk2015genetically}. However, the analysis is based on the assumption of linearity and time-invariance of the gene translation networks, and does not provide any insight into the associated noise sources. The necessary biological elements for an MC-Rx are shown in Fig. \ref{fig:geneticCom} \cite{pierobon2014systems}. The process of receiving information is initiated by Receptor Activation Expression, which gets a PoPS auxiliary signal as its input and generates receptor proteins, denoted by $R$, at its output. The incoming signal molecules from molecular channel, i.e., $S_{Rx}$, then bind to these receptors in Ligand-Receptor Binding unit and form activator complexes, called $RS$. Finally, the concentration of $RS$ initiates the Output Transcription Activation, in which RNA polymerase proteins, $RNAP$, bind to to the promoter sequence, $P_{Rx}$ and produce the PoPS output signal, $PoPS_{Out}$. Assuming that the intracellular receiver environment is chemically homogeneous, the transfer function of aforementioned biological circuit is derived in \cite{pierobon2014systems} to provide a system theoretic model for the MC-Rx. Genetic circuits are shown to provide higher efficiency for analog computation compared to digital computations \cite{sarpeshkar2014analog}. Thus, analog computing functionalities of genetic circuits are utilized in \cite{marcone2018parity} to derive an analog parity-check decoder circuit.

The major challenge in use of genetic circuits for implementing MC-Rx arises from the fact that information transmission in biological cells is through molecules and biochemical reactions. This results in nonlinear input-output behaviors with system-evolution-dependent stochastic effects that are needed to be comprehensively studied to evaluate the performance of the device. In \cite{arkus2005mathematical,maini2004using,klipp2006mathematical}, initiative studies on mathematical modeling of cellular signaling are provided. However, it is shown in \cite{harper2018estimating} that the characterization of the communication performance of these systems is not analytically tractable. Additionally, a computational approach is proposed to characterize the information exchange in a bio-circuit-based receiver using the experimental data published in \cite{ramalho2016single}. The results in \cite{harper2018estimating} reveal that the rate of information transfer through biocircuit based systems is extremely limited by the existing noises in these systems. Thus, comprehensive studies are needed to characterize these noises and derive methods to mitigate them. Additionally, existing studies have only focused on the single transmitter-receiver systems, thus, the connection among biological elements for MC with multiple receiver cells remains as an open issue. 

\subsubsection{\textbf{Nanomaterial-based Artificial MC-Rx Architectures}}
\label{sec:RxNanomaterial}
Similar to MC-Rxs, biosensors are also designed to detect the concentration of an analyte in a solution \cite{scheller2001research}. Thus, existing literature on MC-Rx design are focused on analyzing the suitability and performance of available biosensing options for receiving the information in MC paradigm \cite{kuscu2016physical,kuscu2016modeling,farsad2013tabletop,farsad2017novel,unterweger2018experimental}. In this direction, researchers must consider the fundamental differences between a biosensor and a MC-Rx, arising from their different application areas, as stated below.
\begin{itemize}
	\item Biosensors are designed to perform typically in equilibrium condition. However, MC-Rxs must continuously observe the environment and detect the information encoded into a physical property related to the molecules, such as concentration, type/ratio/order, or arrival time.
	\item Biosensors are mostly designed for laboratory applications with macroscale readout devices and human observers to compensate the lack of an integrated processor, which is not applicable for an MC-Rx.
\end{itemize}
Thus, while the biosensing literature provides insights for MC-Rx design, ICT requirements of the device and its performance in an MC paradigm must be considered to reach appropriate solutions.

Among existing biosensing options, the electrical biosensors are mainly under the focus of MC-Rx design \cite{kuscu2016physical}. The remaining options, i.e., optical and mechanical sensing, need macroscale excitation and detection units \cite{borisov2008optical,arlett2011comparative,tamayo2013biosensors} making them inappropriate for an MC-Rx that requires in situ operation. Biocatalytic \cite{wang2008electrochemical}
and affinity-based \cite{rogers2000principles} sensors are two types of electrical biosensors with different molecular recognition methods.\textit{ Biocatalytic recognition} is based on two steps. First, an enzyme, immobilized on the device, binds with the target molecule producing an electroactive specie, such as hydrogen ion. Arrival of this specie near the working electrode of the transducer is then being sensed as it modulates one of electrical characteristics of the device. Glucose and glutamate sensors are examples of the biocatalytic
electrical sensors \cite{wang2008electrochemical,huang2010nanoelectronic}. Alternatively, binding of receptor-ligand pairs on the recognition layer of the sensor is the foundation of affinity-based sensing \cite{rogers2000principles}. 

\textit{Affinity-based sensing}, which is feasible for a a wider range of target molecules, such as receptor proteins, and aptamer/DNAs \cite{poghossian2014label,rogers2000principles}, provides a less complicated sensing scheme compared to biocatalytic-based sensing. For example, in the biocatalytic-based sensing, the impact of the additional products of the reaction between target molecule and enzyme in biocatalytic-based sensing on the performance of the device and the application environment must be analyzed thoroughly. Hence, affinity-based recognition is more appropriate for the design of a general MC system. However, this recognition method is not possible for non-electroactive information-carrying molecules, such as glutamate and acetylcholine, which are highly important neurotransmitters in the mammalian central nervous system \cite{bear2007neuroscience}. Thus, it is essential to study the design of biocatalytic-based electrical biosensors as MC-Rx when these types of information carriers are dictated by the application, e.g., communication with neurons. 

Recent advances in nanotechnology led to the design of field effect transistor (FET)-based biosensors (bioFETs) providing both affinity-based and biocatalytic-based electrical sensing with use of nanowires, nanotubes, organic polymers, and graphene as the transducer unit \cite{shan2018high,xu2017real,fu2017biosensing,huang2010nanoelectronic,liu2007label,patolsky2006nanowire,kim2009enhancement,kim2013highly}. Detection of target molecules by bioFETs is based on the modulation of transducer conductivity as a result of either affinity-based or biocatalytic-based sensing. Simple operation principles together with the extensive literature on FETs established through many
years, electrical controllability of the main device parameters, high-level integrability, and plethora
of optimization options for varying applications make FET-based biosensing technology also a quite
promising approach for electrical MC-Rx. Moreover, these sensors promise label-free, continuous and in situ operation in nanoscale dimensions. Thus, the design of an MC-Rx based on the principles of affinity-based bioFETs is the main approach considered in the literature \cite{kuscu2016physical,kuscu2016modeling}, which will be overviewed in the rest of this section. 

\begin{figure}
	\centering
	\includegraphics[width=\linewidth]{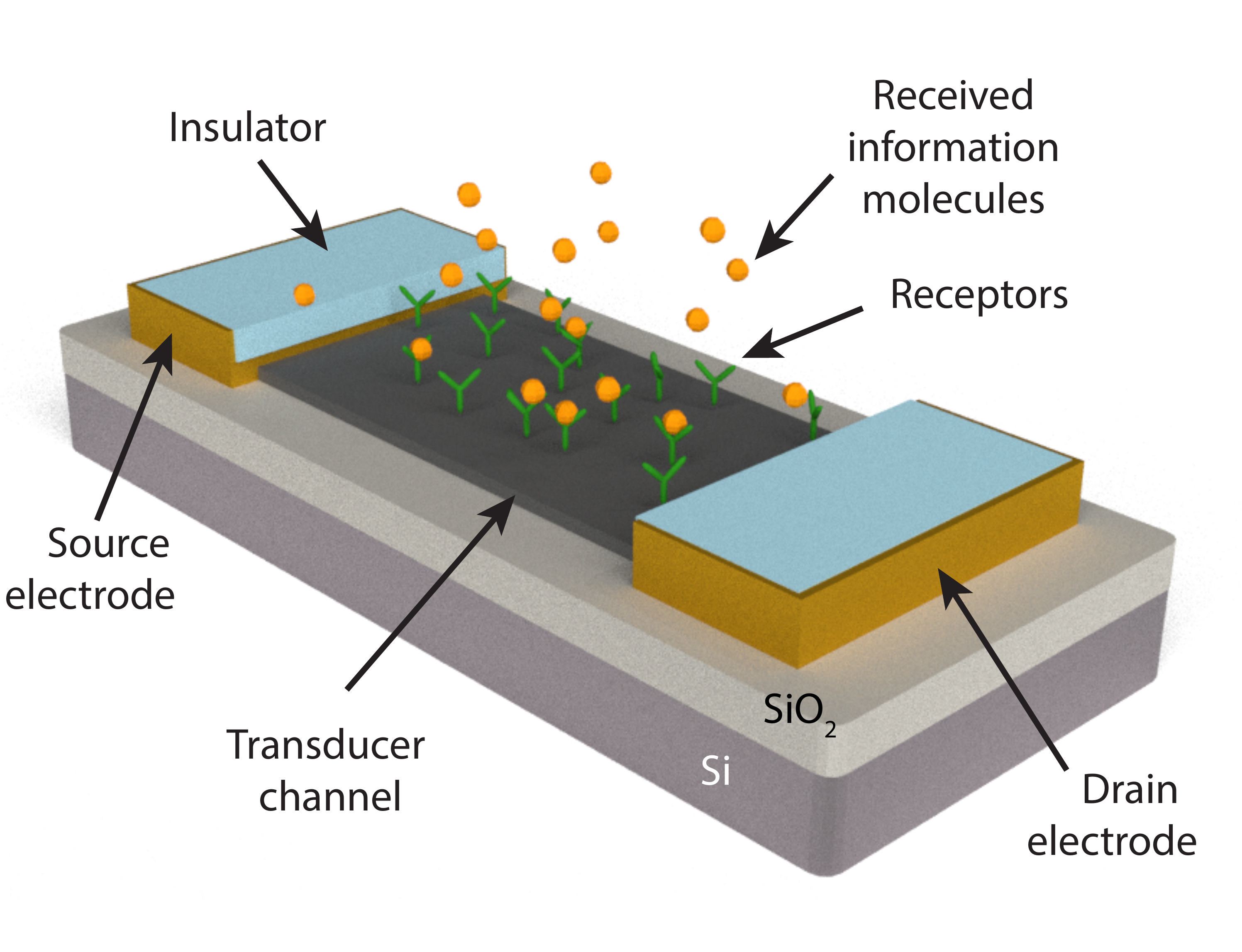}
	\caption{Conceptual design of a bioFET-based MC-Rx.}
	\label{fig:bioFET}
\end{figure}

\paragraph{\textbf{Nanoscale bioFET-based MC-Rx Architectures}}
As shown in Fig. \ref{fig:bioFET}, a bioFET consists of source and drain electrodes and a transducer channel, which is functionalized by ligand receptors in affinity-based sensors \cite{poghossian2014label}. In this type of sensors, the binding of target molecules or analytes to the ligand receptors leads to accumulation or depletion of the
carriers on the semiconductor channel. This modulates the transducer conductivity, which, in turn, alters the flow of current between source and drain. Thus, by fixing the source-to-drain potential, the current flow becomes a function of the analyte density and the amount of analyte charges. Since the ligand receptors are being selected according to the target molecule, i.e., ligands, bioFETs do not require any complicated post-processing such as labeling of molecules. Moreover, the measurement of source-drain current does not need any macroscale readout unit. Thus, bioFETs can be used for direct, label-free, continuous and in situ sensing of the molecules in an environment as a standalone device.

One of significant advantages of bioFETs over other electrical sensors is their wide range of design parameters. A list of FET-based biosensors and their applications in sensing different type of molecules is provided in Table \ref{tab:bioFETs}. In the following, we further describe these vast design options.

\begin{table*}[]
	\centering
	\caption{Design Options, Performance and Applications of bioFETs }
	\begin{tabular}{ c c c c c c}\toprule	
		\begin{tabular}{@{}c@{}}\textbf{Transducer}\\\textbf{channel}\end{tabular}&\textbf{Bioreceptor} &\begin{tabular}{@{}c@{}}\textbf{Limit of}\\\textbf{ detection}\end{tabular} &\textbf{Application}& \textbf{Ref.} \\\midrule
		SWCNT& Acetylcholine(Ach) receptor & 100 pM& ACh detection & \cite{kim2013highly}\\
		SWCNT& Receptor protein (antibody)& 1 ng/ml& Prostate cancer detection& \cite{kim2009enhancement}\\
		SWCNT& Anti Carcinoembryonic antigen (CEA)& $\sim$ 55 pM& CEA detection& \cite{lo2009oriented}\\
		SiNW& Receptor protein (antibody)& $\sim$2 fM& Prostate cancer detection& \cite{patolsky2006nanowire}\\
		SiNW&Estrogen receptors&10 fM& dsDNA detection&\cite{zhang2010highly}\\
		ZnO NW& Anti-Immunoglobulin G(IgG) antibodies& $\sim$0.3 nM& IgG antibodies sensing& \cite{liu2007label}\\
		Graphene& Glucose oxidase (enzyme)& 0.1 mM& Glucose sensor& \cite{huang2010nanoelectronic} \\
		Graphene& Glutamic dehydrogenase(enzyme)& 5 $\mu$M& Glutamate sensor& \cite{huang2010nanoelectronic}\\
		Graphene& Pyrene-linked peptide nucleic acid (pPNA) & 2 pM& DNA sensor& \cite{fu2017biosensing}\\
		Graphene & Single-stranded probe DNA& 10 pM& DNA sensor& \cite{xu2017real}\\
		Graphene&Immunoglobulin E (IgE) aptamers&0.3 nM&IgE protein detection&\cite{ohno2010label}\\
		MoS$_2$& Glucose oxidase (enzyme)& 300 nM& Glucose sensor& \cite{shan2018high}\\\bottomrule
	\end{tabular}
	\label{tab:bioFETs}
\end{table*}

\textbf{Type of bioreceptors:} First important design parameter arises from the type of receptors used in the Biorecognition Unit, which causes the selectivity of receiver for a certain type of molecules that will be used as information carrier in MC paradigm. Among possible receptor types for affinity-based bioFETs, natural receptor proteins and aptamer/DNAs are appropriate ones for an MC-Rx, since their binding to the target molecule is reversible and their size is small enough to be used in a nanomachine \cite{poghossian2014label,rogers2000principles}. As an example, the FET transducer channel is functionalized with natural receptors, e.g., neuroreceptors, to detect taste in bioelectronic tongues \cite{song2014bioelectronic} and odorant in olfactory biosensors \cite{lim2017field}. An advantage of these type of receptors is their biocompatibility that makes them suitable for \textit{in vivo} applications. In addition, use of aptamers, i.e., artificial single-stranded DNAs and RNAs, in the recognition unit of bioFETs provides detectors for a wide range of targets, such as small molecules, proteins, ions, aminoacids and other oligonucleotides
\cite{eissa2017vitro,ohno2010label,xu2017real}. Since an immense number of aptamer-ligand combinations with different affinities exist, it provides a powerful design option to control the selectivity of the MC-Rx. The appropriate aptamer for a target ligand can be find using SELEX process, i.e., searching a large library of DNAs and RNAs to determine a convenient nucleic acid sequence \cite{luzi2003new}. 

\textbf{Material used for transducer channel:}
One of the most important bioFET design parameters is the material used as the transducer channel between source and drain electrodes, which determines
the receiver geometry and affects the electrical noise characteristics of the device. Nanowires (NWs) \cite{ganguly2009functionalized,patolsky2006nanowire}, single walled carbon
nanotubes (SWCNTs), graphene \cite{yang2010carbon}, molybdenum disulfide (MoS2) \cite{sarkar2014mos2}, and organic materials like
conducting polymers \cite{torsi2013organic} are some examples of nanomaterials
suitable for use in a bioFET channel. In the first generation of bioFETs, one dimensional materials such as SWCNT and NW were used as the channel in a bulk form. However, use in the form of single material
or aligned arrays outperformed the bulk channels in terms of sensitivity
and reduced noise \cite{curreli2008real}. Among possible NW materials, such as SnO$_2$, ZnO and In$_2$O$_3$ \cite{senveli2013biosensors}, silicon NW (SiNW) bioFETs have shown high sensitivity, high integration density, high speed sampling and low power consumption \cite{tran2016toward,duan2012quantification,gao2011silicon,shen2011integrating,pei2014device,chen2011silicon}. However, their reliable and cost-effective fabrication is still an important open challenge \cite{zhan2014graphene,xu2017real}. Comprehensive reviews exist on the performance of SiNW bioFETs, their functionality in biomedical applications such as disease diagnostic, their top-down and bottom-up fabrication paradigms and integration within complementary metal-oxide-semiconductor (CMOS) technology \cite{zhang2012silicon,hobbs2012semiconductor,curreli2008real,chen2011silicon,tran2018cmos}.
It is concluded that SiNW bioFETs must be designed according to the requirements arose by applications since defining an ideal characteristic for these devices is difficult. Additionally, both theoretical and experimental studies are needed to find the impact of structure parameters on their functionality. Moreover, fine balancing of important structural factors, such as number of NWs, their doping concentration and length, in SiNW bioFETs design and fabrication remains a challenge, which affects the sensitivity, reliability and stability of the device. SWCNT-based bioFETs offer higher detection sensitivity due to their electrical characteristic, however, these devices also face fabrication challenges, such that their defect free fabrication is the most challenging among all candidates \cite{maehashi2009label}. Note that existence of defects can adversely affect the performance of SWCNT bioFETs in an MC-Rx. Moreover, for \textit{in vivo} applications, the biocompatibility of CNTs and biodurability of functionalized SWCNTs are still under doubt \cite{smart2006biocompatibility,kumar2017carbon,sireesha2018review}. While both NWs and CNTs have one-dimensional structure, use of two-dimensional materials as the transducer channel leads to higher sensitivity, since a planar structure provides higher spatial coverage, more bioreceptors can be functionalized to its surface and all of its surface atoms can closely interact with the bond molecules. Thus, graphene, with its extraordinary electrical, mechanical and chemical characteristics, is a promising alternative for the transducer channel of bioFETs \cite{park2012ultrasensitive,zhan2014graphene}. 

Intrinsic flexibility of graphene provides higher chance of integration into devices with non-planar surfaces which can be more
suitable for the design of nanomachines in an MC application \cite{yang2010carbon}. There is currently tremendous amount of interest in building different configurations of graphene bioFETs, e.g., back-gated \cite{ping2016scalable} or solution-gated \cite{park2012ultrasensitive}, with research showing its superior sensing performance for various analytes, e.g., antigens \cite{gao2016specific}, DNA \cite{manoharan2017simplified}, bacteria \cite{wu2017graphene}, odorant compounds \cite{park2012ultrasensitive}, and glucose \cite{viswanathan2015graphene}.

\begin{figure}
	\centering
	\includegraphics[width=\linewidth]{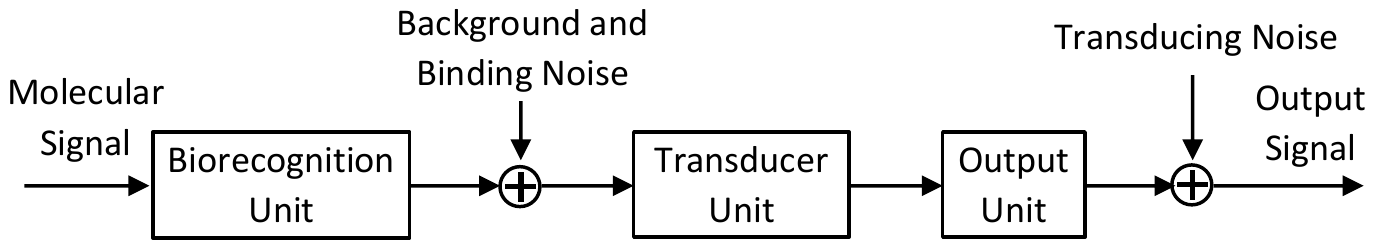}
	\caption{Block diagram of a bioFET-based MC-Rx.}
	\label{fig:bioFETblock}
\end{figure}

General block diagram of a bioFET-based MC-Rx is shown in Fig. \ref{fig:bioFETblock} . The biorecognition Unit is the interface between communication channel and the receiver, thus, it models sensing of the concentration of ligands. The random motion of ligands near the surface of the receiver, which is governed by Brownian motion, leads to fluctuations in the number of bound receptors. This fluctuation can be modeled as a binding noise, which depends on the transmitted signal and can adversely affect the detection of the ligands concentration \cite{hassibi2005biological}. The communication theoretical models of binding noise are presented in \cite{berezhkovskii2013effect,pierobon2011noise}. Background noise, also called biological interference \cite{kuscu2016physical}, is resulted from binding of molecules different from targeted ligands that might exist in the communication channel and show similar affinity for the receptors \cite{hassibi2005biological}. Note that this is different from intersymbol interference and co-channel interference studied in \cite{pierobon2014statistical,pierobon2012intersymbol}. Stochastic binding of ligands to the receptors modulated the conductance of the FET channel, which is modeled by Transducer unit in Fig. \ref{fig:bioFETblock}. This conductance changes is then reflected into the current flowing between source and drain electrodes of FET by Output Unit. The surface potential of the FET channel, thus its source-to-drain current, can be affected by undesirable ionic adsorptions in application with ionic solutions \cite{kuscu2016physical}. Hence, a reference electrode can be used in the solution to stabilize the surface potential \cite{nair2007design}. Finally, the transducing noise shown in Fig. \ref{fig:bioFETblock} covers the impact of noise added to the received signal during transducing operation including thermal noise, caused by thermal fluctuations of charge carriers on the bound ligands, and $1/f$ (flicker) noise, resulted from traps and defects in the FET channel \cite{deen2006noise}. Flicker noise can be the dominant noise source in low frequencies as it increases with decreasing the frequency \cite{kuscu2016physical}. Detailed information on the impact of the aforementioned noise processes on the performance of bioFETs can be found in \cite{deen2006noise, rajan2013performance}.
Moreover, experimental studies are provided in \cite{kutovyi2018origin} on the noise resulted from ion dynamic processes related to ligand-receptor binding events of a liquid-gated SiNW array FET by measuring the noise spectra of device before and after binding of target molecules. 

While the existing biosensing literature can provide insight for the MC-Rx design, there is a need for
investigation of design options according to the communication theoretical requirements of
an MC-Rx. Few studies have focused on evaluating the performance of bioFETs in an MC paradigm. A SiNW bioFET-based MC-Rx is modeled in \cite{kuscu2016physical} based on equilibrium
assumption for the receptor-ligand reaction at the receiver surface. The study provides a circuit model for the transducer unit of the receiver. This work is further extended in \cite{kuscu2016modeling}, where the spatial and temporal correlation effects resulting from finite-rate transport of ligands to the stochastic ligand-receptor binding process are considered to derive the receiver model and its noise statistics. 
In \cite{Khalid2018System}, an MC-Rx consisting of an aerosol sampler, a SiNW bioFET functionalized with antibodies and a detection stage is designed for virus detection. The performance of the receiver is studied by considering the system in steady state. While the receiver model in \cite{Khalid2018System} takes into account the flicker noise and thermal noise, it neglects the interference noise by assuming that the MC-Rx performs in a perfectly sanitized room. 
Moreover, the models used in all of these studies assume
the ligand-receptor binding process in thermal equilibrium and they do not capture well the correlations
resulting from the time-varying ligand concentration occurring in the case of molecular communications. More importantly, these studies only cover SiNW bioFET-receivers, and do not provide much
insight into the performance of other nanomaterials as the transducer channel, such as graphene that promises to provide higher detection sensitivity due to its two-dimensional structure.

Thus, the literature misses stochastic models for nanomaterial-based MC nanoreceiver architectures, which are needed to study the performance of the receiver in MC scenarios, i.e., when the device is exposed to time-varying concentration signals of different types and amplitudes. These models must capture the impacts of receiver geometry, its operation voltage characteristics and all fundamental processes involved in sensing of molecular concentration, such as molecular transport, ligand-receptor binding kinetics, molecular-to-electrical transduction by changes in the conductance of the channel. To provide such a model for graphene-based bioFETs, the major factors that influence the graphene properties must be taken into account. The number of layers is the most dominant factor since electronic band structure, which has a direct impact on the electrical properties of the device, is more complex for graphene with more number of layers \cite{pumera2011graphene,novoselov2005two,castro2007biased,craciun2009trilayer,hibino2009dependence}. Next important parameter is the substrate used in graphene-based bioFET, specially when the number of layers is less than three \cite{wang2008electronic,hibino2009dependence}. The carrier mobility in graphene sheet is reported to be reduced by more than an order of magnitude on SiO$_2$ substrate due to charged impurities in the substrate and remote interfacial phonon scattering \cite{chen2008intrinsic}. On the other hand, it is shown that impacts of substrate can be reduced in suspended single-layer graphene sheet, resulting in higher carrier mobility \cite{bolotin2008ultrahigh}. Moreover, as a result of graphene's large surface area, the impact of impurities on its performance can be substantial \cite{wang2008electronic}. The atomic type, amounts and functional groups on the edges of graphene, which are hard to measure and control, are also among the properties that can result in trial to trial variations in the performance of the fabricated device \cite{neto2009electronic,enoki2007electronic,shenoy2008edge}. Additionally, inherent rippling in graphene sheet, defects and size of sheet also affect the properties of the device \cite{wang2008electronic,geim2007rise,geringer2009intrinsic,brey2008exchange,tung2009high,wu2009synthesis,ni2008reduction,moreno2009ultralong,schedin2007detection}.

\paragraph{\textbf{Other MC-Rx Architectures}}
Few studies exist on the practical MC systems taking into account the physical design of the receiver. In \cite{farsad2013tabletop}, the isopropyl alcohol (rubbing alcohol) is used as information carrier and commercially available metal oxide semiconductor alcohol sensors are used as MC-Rx. This study provides a test-bed for MC with macroscale dimensions, which is later on utilized in \cite{kim2015universal} to estimate its combined channel and receiver model. This test-bed is extended to a molecular multiple-input multiple-output (MIMO) system in \cite{koo2016molecular} to improve the achievable data rate. In \cite{farsad2017novel}, the information is encoded in pH level of the transmitted fluid and a pH probe sensor is used as the MC-Rx. Since use of acids and bases for information
transmission can adversely affect the other processes
in the application environment, such as in the body, magnetic nanoparticles (MNs) are used as information-carrying molecules in microfluidic channels in
\cite{unterweger2018experimental}. In this study, a
bulky susceptometer is used to detect the concentration of MNs and decode the transmitted messages. In addition, performance of MN-based MC where an external magnetic field is employed to attract the MNs to a passive receiver is analyzed in \cite{wicke2018magnetic}. 

However, the focus of the aforementioned studies is using macroscale and commercially available sensors as receiver. Thus, these studies do not contribute to the design of a nanoscale MC-Rx. As discussed in Section \ref{sec:TxInformationMolecules}, recent advancements in DNA/RNA sequencing and synthesis techniques have enabled DNA/RNA-encoded MC \cite{bell2016digitally,chen2017ionic}. For information transmission, communication symbols can be realized with DNA/RNA strands having different properties, i.e., length \cite{bell2016translocation}, dumbbell hairpins \cite{bell2016digitally,bell2016direct}, and short sequence motifs/labels \cite{chen2017ionic}. For information detection, solid-state \cite{chen2017ionic} and DNA-origami \cite{hernandez2014dna} based nanopores can be utilized to distinguish information symbols, i.e., the properties of DNA/RNA strands by examining the current characteristics while DNA/RNA strands pass through the nanopores. As presented in Fig. \ref{fig:DNA}, MC-Rx contains receptor nanopores through which these negatively charged DNA strands pass thanks to the applied potential, and as they do, they obstruct ionic currents that normally flow through. The duration of the current obstruction is proportional to the length of the DNA strand that passes through, which is utilized for selective sensing. The use of nanopores for DNA/RNA symbol detection also enables the miniaturization of MC capable devices towards the realization of IoNT. According to \cite{bell2016digitally}, 3-bit barcode-coded DNA strands with dumbbell hairpins can be detected through nanopores with 94\% accuracy. In \cite{butler2006determination}, four different RNA molecules having different orientations are translocated with more than 90\% accuracy while passing through transmembrane protein nanopores. The time of the translocation event depends on the voltage, concentration and length of the DNA symbols, and the translocation of symbols can take up to a few ms up to hundred ms time frames \cite{bell2016digitally,bell2016direct}. Considering the slow diffusion channel in MC, transmission/detection of DNA/RNA-encoded symbols do not introduce a bottleneck and multiple detections can be performed during each symbol transmission. 

\subsection{Detection Methods for MC}
\label{sec:RxDetection}
Detection is one of the fundamental aspects of communications having tremendous impact over the overall communication performance. The detection of MC signals is particularly interesting due to the peculiarities of the MC channel and communicating nanomachines, which impose severe constraints on the design of detection methods. For example, the limited energy budget and computational capabilities of nanomachines due to their physical design restrict the complexity of the methods. The memory of the diffusion channel causes severe ISI and leads to time-varying channel characteristics with very short coherence time. The stochastic nature of Brownian motion and sampling of discrete message carriers bring about different types of noise, e.g., counting noise and receptor binding noise. The physiological conditions, in which most of the nanonetwork applications are envisioned to operate, imply the abundance of molecules with similar characteristics that can lead to strong molecular interference. These challenges have been addressed in MC to different extents. In this section, we will provide an overview of the state-of-the-art MC detection approaches, along with a discussion on their performances and weaknesses. 

We classify the existing approaches according to the considered channel and received signal models, which reflect the envisioned device architectures that impose different constraints or allow different simplifications over the problem. Accordingly, we divide the detection methods into two main categories: MC detection with passive and absorbing receivers, and MC detection with reactive receivers. 

\subsubsection{\textbf{MC Detection with Passive and Absorbing Receivers}}
\label{sec:RxDetectionPassive}
The nonlinearities and complexity of the MC system often lead researchers to use simplifying assumptions to develop detection methods and analyze their performances. To this end, the intricate relationship between molecular propagation and sampling processes is often neglected. 

\begin{figure}[!t]
	\centering
	\includegraphics[width=0.8\linewidth]{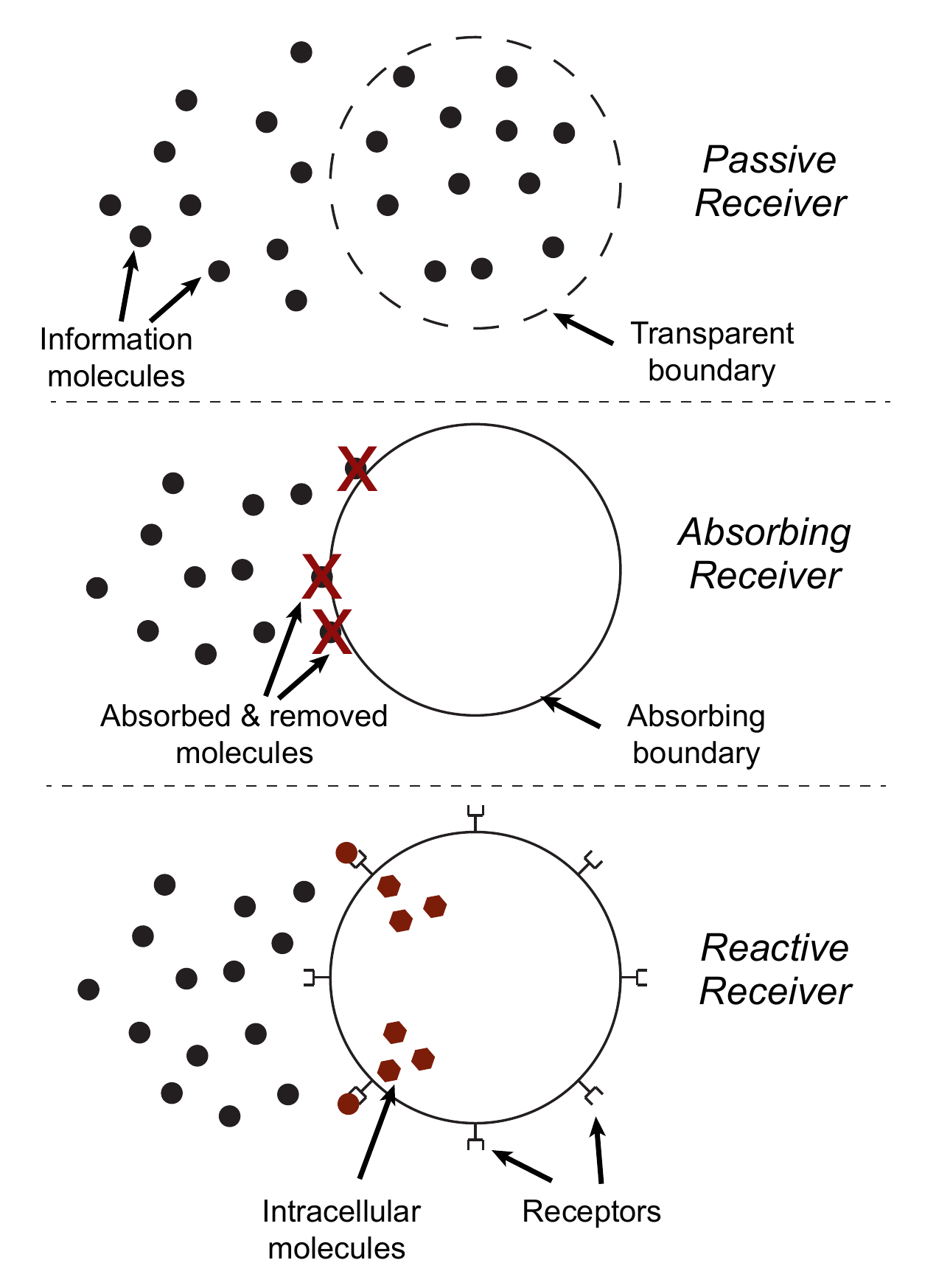}
	\caption{Hypothetical MC-Rx models used for developing detection methods.}
	\label{fig:receiverAssumptions}
\end{figure}

Passive receiver (PA) concept is the most widely used simplifying assumption in MC literature, as it takes the physical sampling process out of the equation, such that researchers can focus only on the transport of molecular messages to the receiver location. Accordingly, the passive receiver is often assumed to be a spherical entity whose membrane is transparent to all kind of molecules, and it is a perfect observer of the number of molecules within its spherical \emph{reception space}, as shown in Fig. \ref{fig:receiverAssumptions} \cite{pierobon2011diffusion}. In the passive receiver approximation, the receiver has no impact on the propagation of molecules in the channel. Passive receivers can also be considered as if they include ligand receptors, which are homogeneously distributed within the reception space with very high concentration and infinitely high rate of binding with ligands, such that every single molecule in the reception space is effectively bound to a receptor at the time of sampling. 

Another modeling approach, i.e., absorbing receiver (AB) concept \cite{yilmaz2014three}, considers receiver as an hypothetical entity, often spherical, which absorbs and degrades every single molecule that hits its surface, as demonstrated in Fig. \ref{fig:receiverAssumptions}. This approach improves the assumption of passive receiver one step further towards a more realistic scenario including a physical interaction between the receiver and the channel. In contrast to passive receiver, absorbing receiver can be considered to have receptors located over the surface. For a perfect absorbing receiver, this means very high concentration of receptors with infinitely high absorption rate, such that every molecule that hits the surface is bound and consumed instantly. 

Physical correspondence of both models is highly questionable. Nevertheless, they are widely utilized in the literature as they provide upper performance limits. However, ignoring the receptor-ligand reactions, which often leads to further intricacies, e.g., receptor saturation, stands as a major drawback of these approaches. 

\paragraph{\textbf{Received Signal Models}}
When constructing the received signal models for diffusion-based MC, the transmitter (Tx) geometry is usually neglected assuming that the Tx is a point source that does not occupy any space. This assumption is deemed valid when the distance between the Tx and Rx are considerably larger than the physical sizes of the devices. Throughout this section, we will mostly focus on OOK modulation, where the Tx performs an impulsive release of a number of molecules to transmit bit-1, and does not send any molecule to transmit bit-0. This is the most widely used modulation scheme in MC detection studies, as it simplifies the problem while capturing the properties of the MC channel. However, we will also briefly review the detection schemes corresponding to other modulation methods, e.g., timing-based modulation and MoSK, throughout the section. 

Molecular propagation in the channel is usually assumed to be only through free diffusion, or through the combination of diffusion and uniform flow (or drift). In both cases, the channel geometry is often neglected and assumed to be unbounded, and molecules are assumed to propagate independently from each other. In some studies addressing passive receivers, researchers consider the existence of enzymes in the channel, which reduce the impact of the ISI by degrading the residual messenger molecules through first-order reaction \cite{noel2014improving}. For a three dimensional free diffusion channel with uniform flow in the presence of degrading enzymes, the number of molecules observed in the spherical reception space of a passive receiver follows non-stationary Poisson process \cite{noel2014unifying, kilinc2013receiver}, i.e.,
\begin{align}
N_{\text{RX}|\text{PA}}(t) \sim \operatorname{Poisson} \left( \lambda_\text{RX}(t) \right),
\end{align}
where the time-varying mean of this process $\lambda_\text{RX}(t)$ can be given by
\begin{align}
\lambda_\text{RX}(t) = \lambda_\text{noise} + Q \sum_{j = 1}^{\floor*{\frac{t}{T_s}+1}} s[i] P_\text{obs}\Bigl(t-(j-i)T_s\Bigr).
\end{align}
The mean depends on the number of transmitted molecules $Q$ to represent bit 1, the symbols transmitted in the current symbol interval as well as in the previous symbol intervals, i.e., $s[i]$, and the length of a symbol interval $T_s$. Most MC studies include an additive stationary noise in their models representing the interfering molecules available in the channel as a result of an independent process in the application environment. These molecules are assumed to be of the same kind with the messenger molecules and their number is represented by a Poisson process and captured by $\lambda_\text{noise}$. Channel response is integrated into the model through the function $P_\text{obs}(t)$, which is the probability of a molecule transmitted at time $t=0$ to be within the sampling space at time $t$. When Tx-Rx distance is considerably large, ligands are typically assumed to be uniformly distributed within the reception space. As a result, the channel response can be written as   
\begin{align}
P_\text{obs}(t) = \frac{V_\text{RX}}{(4\pi D t)^{3/2}} \exp\left( -k C_E t - \frac{|\vec{r}_\text{eff}|^2}{4Dt}\right),
\end{align}
where $V_\text{RX} = \frac{4}{3} \pi d_\text{RX}^3$ is the volume of the spherical receiver with radius $d_\text{RX}$, $D$ is the diffusion coefficient, $C_E$ is the uniform concentration of the degrading enzymes in the channel, $k$ is the rate of enzymatic reaction, and $\vec{r}_\text{eff}$ is the effective Tx-Rx distance vector, which captures the effect of uniform flow \cite{noel2014unifying}. Assuming that the Tx and Rx are located at $\vec{r}_\text{TX} = (0, 0, 0)$, $\vec{r}_\text{RX} = (x_0, 0, 0)$, respectively, and the flow velocity is given by $v_x, v_y, v_z$ in 3D Cartesian coordinates, the magnitude of the effective distance vector can be written as follows
\begin{align}
|\vec{r}_\text{eff}| = \sqrt{(x_0 - v_xt)^2 + (v_y)^ + (v_z)^2}.
\end{align}

For an absorbing receiver, the received signal is usually taken as the number of molecules absorbed by the Rx within a time interval \cite{yilmaz2014three}. For a diffusion channel without flow, the probability density for a molecule emitted at $t = 0$ to be absorbed by a perfectly absorbing receiver of radius $r_r$ and located at a distance $r$ from the Tx at time $t$ is given by
\begin{align}
f_\text{hit}(t) = \frac{r_r}{r} \frac{1}{\sqrt{4 \pi D t}} \frac{r-r_r}{t} \exp\left(- \frac{(r-r_r)^2}{4Dt}\right),
\end{align}
and the cumulative distribution function (CDF) is given by
\begin{align}
F_\text{hit}(t) = \int_{0}^{t}f_\text{hit}(t') dt' = \frac{r_r}{r} \erfc\left[\frac{r-r_r}{\sqrt{4Dt}}\right],
\end{align}
where $\erfc$ is the complementary error function \cite{yilmaz2014three}. The CDF can be used to calculate the probability of a molecule transmitted at time $t=0$ to be absorbed within the $k$th signaling interval, i.e., 
\begin{align}
P_k = F_\text{hit}(kT_s) - F_\text{hit}([k-1]T_s).
\end{align}
When considering multiple independent molecules emitted at the same time, the number of molecules absorbed at the $k$th interval becomes Bernoulli random variable with the success probability of $P_k$. Assuming that the success probability is low enough, Gaussian approximation of the Bernoulli random process can be used to write 
\begin{align}
N_{\text{RX}|\text{AB}}[i] \sim \mathcal{N}\left(\mu[i], \sigma^2[i] \right),
\end{align}
where its signal-dependent mean and variance can be written as a function of current and previously transmitted bits $s[i]$, i.e.,
\begin{align}
\mu[i] &= Q \sum_{i}^{k} P_k s[i-k+1], \\
\sigma^2[i] &= \sigma^2_\text{noise} + Q \sum_{i}^{k} P_k (1-P_k) s[i-k+1].
\end{align}
Note that as in the case of passive receiver, the received signal model includes the contribution of a stationary noise through its variance $\sigma^2_\text{noise}$. Unfortunately, in the literature, there is no analytical model for absorbing receivers in diffusion-based MC channels with uniform flow and degrading enzymes. 

\begin{table*}\scriptsize
	  \begin{threeparttable}
  \centering
	\caption{Comparison Matrix for MC Detectors with Passive and Absorbing Receivers}
	  \label{tab:Detection}%
    \begin{tabular}{llllllllll}
    \toprule	
    \multirow{2}{*}{\textbf{Description}} & \textbf{Channel} & \textbf{Modulation \&} & \textbf{Detector} & \textbf{Rx} & \textbf{Measurement} & \textbf{CSI} & \multirow{2}{*}{\textbf{Complex.}} & \multirow{2}{*}{\textbf{Perf.}} \\
    & \textbf{Characteristics} & \textbf{Transmit Waveform} & \textbf{Type} & \textbf{Type} & \textbf{Method} & \textbf{Req.} &  & \\
    \midrule
    One-shot fixed-threshold detector \cite{meng2014receiver} & Diffusion & CSK/OOK-Impulse & SbS   & PA    & Sampling & Ins. CSI & Low   & Sub. \\
    Weighted sum detector \cite{jamali2017design} & Diff./Flow/En./EI & CSK/OOK-Impulse & SbS   & PA    & Sampling & Ins. CSI & Low   & Opt. \\
    Weighted sum detector \cite{noel2014optimal, noel2014unifying} & Diff./Flow/En./EI & CSK/OOK-Impulse & SbS   & PA    & Sampling & Ins. CSI & Low   & Sub. \\
    Adaptive-threshold detector \cite{shahbazi2018adaptive} & Diff./EI & CSK/OOK-Impulse & SbS   & PA    & Sampling & No CSI & Moderate & Sub. \\
    Adaptive-threshold detector \cite{he2016improving, alshammri2017low, damrath2016low} & Diffusion & CSK/OOK-Impulse & SbS   & AB    & Energy & No CSI & Low   & Sub. \\
    Linear equalizer (MMSE) \cite{kilinc2013receiver} & Diffusion & CSK/OOK-Pulse & Seq.   & PA    & Sampling & Ins. CSI & Very High & Sub. \\
    Nonlinear equalizer (DFE) \cite{kilinc2013receiver} & Diffusion & CSK/OOK-Pulse & Seq.   & PA    & Sampling & Ins. CSI & Very High & Sub. \\
    ML sequence detector with Viterbi \cite{kilinc2013receiver} & Diffusion & CSK/OOK-Pulse & Seq.   & PA    & Sampling & Ins. CSI & Very High & Opt. \\
    ML sequence detector with Viterbi \cite{noel2014optimal} & Diff./Flow/En./EI & CSK/OOK-Impulse & Seq.   & PA    & Sampling & Ins. CSI & Very High & Opt. \\
    ML sequence detector with Viterbi \cite{shahmohammadian2012optimum} & Diffusion & MoSK-Impulse & Seq.   & PA    & Sampling & Ins. CSI & Very High & Opt. \\    
    Near ML sequence det. with RS Viterbi \cite{meng2014receiver} & Diffusion  & CSK/OOK-Impulse & Seq.   & PA    & Sampling & Ins. CSI & High  & Sub. \\    
    Strength-based ML detector \cite{mahfuz2015comprehensive} & Diffusion & CSK/ASK-Impulse & SbS   & PA    & Energy & Ins. CSI & High  & Opt. \\
    Sampling-based ML detector \cite{mahfuz2014comprehensive} & Diffusion & CSK/ASK-Impulse & SbS   & PA    & Sampling & Ins. CSI & High  & Opt. \\
    Derivative-based detector \cite{yan2018derivative} & Diffusion & CSK/OOK-Impulse & SbS   & AB    & Energy & Ins. CSI & Moderate & Sub. \\
    Noncoherent detector \cite{li2016low} & Diffusion & CSK/B-CSK-Pulse & SbS   & PA    & Sampling & No CSI & Low   & Sub. \\
    Local convexity-based noncoherent det. \cite{li2016local} & Diffusion & CSK/OOK-Pulse & SbS   & PA    & Sampling & No CSI & Moderate & Sub. \\
    Noncoherent ML threshold-based det. \cite{jamali2018non} & Diff./EI & CSK/OOK-Impulse & SbS   & PA\&AB & Sampling & Stat. CSI & Low   & Opt. \\
    Noncoherent ML sequence detector \cite{jamali2018non} & Diff./EI & CSK/OOK-Impulse & Seq.   & PA\&AB & Sampling & Stat. CSI & High  & Opt. \\
    Noncoherent decision-feedback detector \cite{jamali2018non} & Diff./EI & CSK/OOK-Impulse & Seq.   & PA\&AB & Sampling & Stat. CSI & Very High & Sub. \\
    Noncoherent blind detector \cite{jamali2018non} & Diff./EI & CSK/OOK-Impulse & SbS   & PA\&AB & Sampling & No CSI & Low   & Sub. \\
    ML sequence detector with CC codes \cite{jamali2018constant} & Diff./EI & CSK/ASK-Impulse & Seq.   & PA  & Sampling & No CSI & Very High & Opt. \\
    Asynchronous fixed-threshold detector \cite{noel2017asynchronous} & Diffusion & CSK/OOK-Impulse & SbS   & PA & Sampling & Ins. CSI & Moderate & Sub. \\
    Asynchronous adaptive-th. det. with DF \cite{noel2017asynchronous} & Diffusion & CSK/OOK-Impulse & SbS   & PA    & Sampling & Ins. CSI & High  & Sub. \\
    Adaptive-threshold detector (Mobile MC) \cite{chang2018adaptive} & Diff./TV & CSK/OOK-Impulse & SbS   & PA    & Sampling & Ins. CSI & Very High & Sub. \\
    Single-sample th. det. (Mobile MC) \cite{ahmadzadeh2017stochastic} & Diff./Flow/TV  & CSK/OOK-Impulse & SbS   & PA    & Sampling & Out. CSI & High  & Sub. \\
    ML sequence detector (Timing Channel) \cite{murin2017time} & Diffusion & Release time-Impulse & Seq.   & AB    & Arrival time & Ins. CSI & Very High & Opt. \\
    Sequence detector with modified Viterbi \cite{murin2017time} & Diffusion & Release time-Impulse & Seq.   & AB    & Arrival time & Ins. CSI & High  & Sub. \\
    Symbol by symbol timing detector \cite{murin2017time} & Diffusion & Release time-Impulse & SbS   & AB    & Arrival time & No CSI & Low   & Sub. \\
    ML detector with FA \& LA times \cite{murin2017optimal, murin2018exploiting} & Diffusion & Release time-Impulse & SbS   & AB    & Arrival time & Ins. CSI & Moderate & Opt. \\
    \bottomrule
    \end{tabular}%
    \begin{tablenotes}
	\scriptsize
	\item Abbreviations | Diff.: Diffusion -- En.: Enzyme -- EI: External Interference -- TV: Time-varying -- SbS: Symbol-by-symbol detector -- Seq.: Sequential detector -- PA: Passive receiver -- AB: Absorbing receiver -- Ins. CSI: Instantaneous CSI -- Stat. CSI: Statistical CSI -- Out. CSI: Outdated CSI -- Sub.: Sub-optimal -- Opt. Optimal -- Complex.: Complexity -- Perf.: Performance -- DF: Decision-Feedback -- FA: First Arrival -- LA: Last Arrival -- CC codes: Constant Composition codes -- RS: Reduced-State
\end{tablenotes}
\end{threeparttable}
\end{table*}%

\paragraph{\textbf{Detection Methods}}
Detection methods for MC in general can be divided into two main categories depending on the method of concentration measurement: sampling-based and energy-based detection. Passive receivers are usually assumed to perform sampling-based detection, which is based on sampling the instantaneous number of molecules inside the reception space at a specific sampling time \cite{mahfuz2014comprehensive}. Absorbing receivers, on the other hand, are typically assumed to utilize energy-based detection, which uses the total number of molecules absorbed by the receiver during a prespecified time interval, that is usually the symbol interval \cite{mahfuz2015comprehensive}. In some studies, passive receivers are also considered to perform energy-based detection through taking multiple independent samples of number of molecules inside the reception space at different time instants during a single symbol interval, and passing them through a linear filter which outputs their weighted sum as the energy of the received molecular signal \cite{noel2014optimal, noel2014unifying}. 

As in conventional wireless communications, detection can be done on symbol-by-symbol or sequential basis. The symbol-by-symbol detection tends to be more practical in terms of complexity, whereas the sequence detectors require the receiver to have a memory to store the previously decoded symbols. Due to the MC channel memory causing a considerable amount of ISI for high data rate communication, the sequence detectors are more frequently studied in the literature. 

Next, we review the existing MC detection techniques developed for passive and absorbing receivers by categorizing them into different areas depending on their most salient characteristics. A comparison matrix for these methods can also be seen in Table \ref{tab:Detection}. 

\textbf{Symbol-by-Symbol (SbS) Detection:} SbS MC detectors in the literature are usually proposed for very low-rate communication scenarios, where the ISI can be neglected, asymptotically included into the received signal model with a stationary mean and variance, or approximated by the weighted sum of ISI contributions of a few previously transmitted symbols. In \cite{meng2014receiver}, a one-shot detector is proposed based on the asymptotic approximation of the ISI assuming that the sum of decreasing ISI contributions of the previously transmitted symbols can be represented by a Gaussian distribution through central limit theorem (CLT) based on Lindeberg's condition. A fixed-threshold detector is proposed maximizing the mutual information between transmitted and decoded symbols. Similarly, in \cite{noel2014optimal, noel2014unifying}, a matched filter in the form of a weighted sum detector is proposed using a different asymptotic ISI approximation as though it results from a continuously emitting source leading to a stationary Poisson distribution of interference molecules inside the reception space. In this scheme, a passive receiver performs energy-based detection taking multiple samples at equally spaced sampling times during a single symbol transmission, and the weights of the samples are adjusted according to the number of molecules expected at the corresponding sampling times. This matched filter is proven to  be optimal in the sense that it maximizes SNR at the receiver. However, the optimal threshold of this detector does not lend itself to a closed-form expression, and thus, it should be numerically obtained through resource intensive search algorithms. Similarly, in \cite{jamali2017design}, considering also the external sources of interference, another linear matched filter is designed maximizing the expected signal-to-interference-plus-noise-ratio (SINR) for SbS detection, and shown to outperform previous schemes especially when the ISI is severe. There are also adaptive-threshold-based SbS detection methods relying on receivers with memory of varying length taking into account only the ISI contribution of a finite number of previously transmitted symbols \cite{mahfuz2014comprehensive, mahfuz2015comprehensive, damrath2016low, he2016improving, alshammri2017low, shahbazi2018adaptive}. In these schemes, the adaptive threshold is updated for each symbol interval using the ISI estimation based on the previously decoded symbols. SbS detection is also considered in \cite{jamali2018non, noel2017asynchronous}, which will be discussed next in the context of noncoherent and asynchronous detection.

\textbf{Sequence Detection and ISI Mitigation:} Optimal sequence detection methods based on Maximum a Posteriori (MAP) and Maximum Likelihood (ML) criteria are proposed in \cite{kilinc2013receiver} for MC with passive receivers. Even though the complexity of the sequence detectors are reduced by applying Viterbi algorithm, it still grows exponentially with increasing channel memory length. To reduce the complexity further, a sub-optimal linear equalizer based on Minimum Mean-Square Error (MMSE) criterion is proposed. To improve the performance of the sub-optimal detection, a nonlinear equalizer, i.e., Decision-Feedback Equalizer (DFE), is also proposed in the same study. DFE is shown to outperform linear equalizers with significantly less complexity than optimal ML and MAP sequence detection methods. Similarly, a near-optimal ML sequence detector employing Viterbi algorithm is proposed in \cite{meng2014receiver}. Another optimal ML sequence detector is proposed in \cite{noel2014optimal} for MC with uniform flow and enzymes that degrade information molecules. 

In addition to the sequence detection methods and equalizers, there are other approaches proposed to overcome the effects of the ISI on detection. For example, in \cite{akdeniz2018optimal}, authors propose to shift the sampling time by increasing the reception delay to reduce the effect of ISI. In \cite{yan2018derivative}, a derivative-based signal detection method is proposed to enable high data rate transmission. The method is based on detecting the incoming messages relying on the derivative of the channel impulse response (CIR).

\textbf{Noncoherent Detection:} Most of the MC detection methods requires the knowledge of the instantaneous CSI in terms of CIR. However, CIR in MC, especially in physiologically relevant conditions, tends to change frequently, rendering the detection methods relying on the exact CIR knowledge useless. Estimating the instantaneous CIR is difficult and requires high computational power. To overcome this problem, researchers propose low-complexity noncoherent detection techniques. For example, in \cite{damrath2016low}, the authors develop a simple detection method for absorbing receivers, which does not require the channel knowledge. In this scheme, the receiver performs a threshold-based detection by comparing the number of absorbed molecules in the current interval to that of the previous symbol interval. The adaptive threshold is updated in every step of detection with the number of molecules absorbed. However, this method performs poorly when a sequence of consecutive bit-1s arrives. Similarly, in \cite{li2016low}, the difference of the accumulated concentration between two adjacent time intervals is exploited for noncoherent detection. In \cite{li2016local}, the local convexity of the diffusion-based channel response is exploited to detect MC signals in a noncoherent manner. A convexity metric is defined as the test statistics, and the corresponding threshold is derived. There are also methods requiring only the statistical CSI rather than the instantaneous CSI \cite{jamali2018non}. 
Additionally, constant-composition codes are proposed to enable ML detection without statistical or instantaneous CSI, and shown to outperform uncoded transmission with optimal coherent and noncoherent detection, when the ISI is neglected \cite{jamali2018constant}.

\textbf{Asynchronous Detection:} The synchronization between the communicating devices is another major challenge. However, in the previously discussed studies, synchronization is assumed to be perfect. To overcome this limitation, an asynchronous peak detection method is developed in \cite{noel2017asynchronous} for the demodulation of MC signals. Two variants has been proposed. First method is based on measuring the largest observation within a sampling interval. This SbS detection method is of moderate complexity and non-adaptive, comparing the maximum observation to a fixed threshold. The second  method is adaptive and equipped with decision feedback to remove the ISI contribution. In this scheme, the receiver takes multiple samples per bit and adjusts the threshold for each observation based on the expected ISI.  

\textbf{Detection for Mobile MC:} The majority of MC studies assumes that the positions of Tx and Rx are static during communication. The mobility problem of MC devices has just recently started to attract researchers' attention. For example, MC between a static transmitter and a mobile receiver is considered in \cite{chang2018adaptive}, where the authors propose to reconstruct the CIR in each symbol interval using the time-varying transmitter-receiver distance estimated based on the peak value of the sampled concentration. Two adaptive schemes, i.e., concentration-based adaptive threshold detection and peak-time-based adaptive detection, are developed based on the reconstructed CIR. In \cite{ahmadzadeh2017stochastic}, different mobility cases including mobile TX and RX, mobile TX and fixed RX, and mobile RX and fixed TX are considered to develop a stochastic channel model for diffusive mobile MC systems. The authors derive analytical expressions for the mean, PDF, and auto-correlation function (ACF) of the time-varying CIR, through an approximation of the CIR with a log-normal distribution. Based on this approximation, a simple model for outdated CSI is derived, and the detection performance of a single-sample threshold detector relying on the outdated CSI is evaluated. 

\textbf{Discussion on Other Detection Techniques:} MC detection problem is also addressed for molecule shift keying (MoSK) modulation. In \cite{shahmohammadian2012optimum}, an optimal ML sequence detector employing Viterbi algorithm is proposed assuming that a passive receiver can independently observe MC signals carried by different types of molecules. This assumption greatly simplifies the problem and enable the application of detection methods developed for CSK-modulated MC signals for MoSK signals as well. 

Diffusion-based molecular timing (DBMT) channels are also addressed from detection theoretical perspective. DBMT channels without flow are accompanied by a Levy distributed additive noise having a heavy algebraic tail in contrast to the exponential tail of inverse Gaussian distribution, which DBMT channel with flow follows \cite{murin2017optimal}. In \cite{murin2017time}, an optimal ML detector is derived for DBMT channels without flow; however, the complexity of the detector is shown to have exponential computational complexity. Therefore, they propose sub-optimal yet practical SbS and sequence detectors based on the random time of arrivals of the simultaneously released information molecules, and show that the performance of the sequence detector is close to the one of computationally expensive optimal ML detector. 

In DBMT channels without flow, linear filtering at the receiver results in a dispersion larger or equal to the dispersion of the original, i.e., unfiltered, sample, rendering the performance of releasing multiple particles worse than releasing a single particle. Based on this finding, the authors in \cite{murin2017optimal} develop a low-complexity detector, which is based on the first arrival (FA) time of simultaneously released particles by the TX. The method is based on the observation that the probability density of the FA gets concentrated around the transmission time when the number of released molecules $M$ increases. Neglecting ISI, it is shown in the same paper that the proposed FA-based detector performs very close to the optimal ML detectors for small values of $M$. However, the ML detection still performs significantly better than the FA for high values of $M$. The detection based on the order statistics has been extended in the same authors' later work \cite{murin2018exploiting}, where they consider also the detection based on the last arrival (LA) time. Defining a system diversity gain as the asymptotic exponential decrease rate of error probability with the increased number of released particles, they showed that the diversity gain of LA detector approaches to that of computationally expensive ML detector. 

\subsubsection{\textbf{MC Detection with Reactive Receivers}}
\label{sec:RxDetectionReactive}
Another type of receiver considered for MC is reactive receiver, which samples the molecular concentration of incoming messages through a set of reactions it performs via specialized receptor proteins or enzymes, as shown in Fig. \ref{fig:receiverAssumptions}. The reactive receiver approach is more realistic in the sense that natural cells, e.g., bacteria and neurons, sense molecular communication signals through their receptors on the cell membrane, and many types of artificial biosensors, e.g., bioFETs, are functionalized with biological receptors for higher selectivity. Since synthetic biology, focusing on using and extending natural cell functionalities, and artificial biosensing are the two phenomena that are considered for practical implementation of MC receivers, studying MC detection with reactive receivers has more physical correspondence.  

Diffusion-based MC systems with reactive receivers, in most cases, can be considered as reaction-diffusion (RD) systems with finite reaction rates. Although RD systems, which are typically highly nonlinear, have been studied in the literature for a long time, they do not usually lend themselves to analytical solutions, especially when the spatio-temporal dynamics and correlations are not negligible. To be able to devise detection methods and evaluate their performance in the MC framework, researchers have come up with different modeling approaches, which will be reviewed next. For the sake of brevity, we focus our review on detection with receivers equipped with ligand receptors, which have only one binding site. 

Ligand-receptor binding reaction for a single receptor exposed to time-varying ligand concentration $c_L(t) $ can be schematically demonstrated as follows
\begin{equation}
\ce{U  <=>[{c_L(t) k_+}][{k_-}] B},
\end{equation}
where $k_+$ and $k_-$ are the ligand-receptor binding and unbinding rates, respectively; $U$ and $B$ denote the unbound and bound states of the receptor, respectively. When there are $N_R$ receptors, assuming that all of them are exposed to the same concentration of ligands, reaction rate equation (RRE) for the number of bound receptors can be written as \cite{pierobon2011noise}
\begin{align}
\frac{dn_B(t)}{dt} = k_+ c_L(t) \left(N_R - n_B(t)\right) - k_- n_B(t). \label{eq:rre}
\end{align}
As is clear, while the binding reaction is second-order depending on the concentrations of both ligands and available receptors, unbinding reaction is first-order and only depends on the number of bound receptors.  

Most of the time, the bandwidth of MC signals can be assumed to be low enough to drive the binding reaction to near equilibrium and allow applying quasi steady-state assumption for the overall system. In this case, time-varying concentration $c_L(t)$ can be treated constant, i.e., $c_L(t) = c_L$, and $dn_B(t)/dt = 0$, which results in the following expression for the mean number of bound receptors
\begin{align}
\E[n_B] = \frac{c_L}{c_L + K_D} N_R, \label{eq:meanBinomial}
\end{align}
where $K_D = k_- / k_ +$ is the dissociation constant, which is a measure of affinity between the specific type of ligand and receptor.  Even at equilibrium, the receptors randomly fluctuate between the bound and unbound state. The number of bound receptors $n_B$, at equilibrium, is a Binomial random variable with success probability $p_B = c_L / (c_L + K_D)$, and its variance can be given accordingly by
\begin{align}
\Var[n_B] = p_B (1-p_B) N_R.  \label{eq:varBinomial}
\end{align}

More insight can be gained by examining the continuous history of binding and unbinding events over receptors. The likelihood of observing a series of n binding-unbinding events at equilibrium can be given by 
\begin{align}
\p(\{\tau^B,\tau^U \}_{n}) = \frac{1}{Z} e^{T_U  \sum_{i=1}^{M} k_i^+ c_i}  \prod_{j=1}^{n}  \sum_{i=1}^{M} k_i^+ c_i k_i^- e^{-k_i^- \tau_j^B   } ,
\label{eq:likelihood}
\end{align}
where $Z$ is the normalization factor, $T_U = \sum_{j = 1}^n \tau_j^U$ is the total unbound time, $ \tau_j^U$ and  $\tau_j^B$ are the $j^{th}$ unbound and bound time intervals, respectively, $c_i$, $k_i^+$ and $k_i^-$ are the concentration, binding rate, and unbinding rate of $i^{th}$ type of ligand, respectively, $M$ is the number of ligand types present in the channel \cite{mora2015physical, kuscu2018maximum}. Note that the likelihood is equally valid for the cases of single receptor and multiple receptors, as long as the collected $n$ samples of unbound and bound time intervals are independent. These observable characteristics of the ligand-receptor binding reactions have been exploited to infer the incoming messages to different extents, as will be reviewed next. 

\paragraph{\textbf{Received Signal Models}}
The nonlinearities arising from the interaction of time-varying MC signals with receptors have led to different approaches for modeling MC systems with reactive receivers compromising on different aspects to develop detection techniques and make the performance analyses tractable. A brief review of these modeling approaches are provided as follows. 

\textbf{Reaction-Diffusion Models with Time-varying Input:} One of the first attempts to model the ligand-receptor binding reactions from an MC theoretical perspective is provided in \cite{pierobon2011noise}, where the authors develop a noise model for the fluctuations in the number of bound receptors of a receiver exposed to time-varying ligand concentrations as MC signals. The model is based on the assumption of a spherical receiver, in which ligand receptors and information-carrying ligands are homogeneously distributed. For an analytically tractable analysis, the concentration of incoming ligands is assumed to be constant between two sampling times, i.e., during a sampling interval, and the ligand-receptor binding reaction is assumed to be at equilibrium at the beginning of each sampling interval. In light of these assumptions, the authors obtain the time-varying variance and mean of the number of bound receptors, which are valid only for the corresponding sampling interval. A more general approach without the equilibrium assumption to obtain the mean number of bound receptors with time-varying input signals, i.e., ligand concentration, is contributed by \cite{ahmadzadeh2016comprehensive} and \cite{deng2015modeling}, through solving the system of differential equations governing the overall diffusion-reaction MC system. The authors of the both studies consider a spherical receiver with ligand receptors on its surface and a point transmitter, which can be anywhere on a virtual sphere centered at the same point as the receiver but larger than that,  to obtain a spherical symmetry to simplify the overall problem. As a result, the transmitter location cannot be exactly specified in the problem. In \cite{deng2015modeling}, the authors consider that the spherical receiver is capable of binding ligands at any point on its surface, which is exactly equal to the assumption of infinite number of receptors. On the other hand, \cite{ahmadzadeh2016comprehensive} considers finite number of receptors uniformly distributed on the receiver surface, and addresses this challenge through boundary homogenization. However, boundary homogenization for finite number of receptors does not take into account the negative feedback of the bound receptors on the second-order binding reaction (see \eqref{eq:rre}), and thus, the developed analytical model is not able to capture the indirect effects of finite number of receptors, e.g., receptor saturation. This is clear from their analysis, such that the discrepancy between the analytical model and the particle-based simulation results is getting larger with increasing ligand concentration.

\textbf{Frequency Domain Model:} Another modeling approach is provided in \cite{shahmohammadian2013modelling}, where the authors, assuming that the probability of a receptor to be in the bound state is very low, take the number of available, i.e., unbound, receptors equal to the total number of receptors at all time points. The completely first order characteristics of the resulting RRE enables them to carry out a frequency domain analysis, through which they show the ligand-receptor binding reaction manifests low-pass filter characteristics. However, this approximate model is relevant only when the probability of receptor-ligand binding is very low.

\textbf{Discrete Model based on Reaction-Diffusion Master Equation (RDME):} To capture the stochasticity of the reaction-diffusion MC, another approach is introduced in \cite{chou2013extended}, where the authors develop a voxel-based model based on RDME, with the diffusion and reactions at the receiver modeled as Markov processes. The three-dimensional MC system is discretized and divided into equal-size cubic voxels, in each of which molecules are assumed to be uniformly distributed, and allowed to move only to the neighboring voxels. In the voxel accommodating the Tx, the molecules are generated according to a modulation scheme, and the Rx voxel hosts the receptor molecules, where the ligands diffusing into the Rx voxel can react based on law of mass action. The jump of a ligand from one voxel to another is governed by a diffusion rate parameter, which is a function of the voxel size and the ligand diffusion coefficient. The number of ligands and bound receptors are stored in a system state vector, which is progressed with a given state transition rate vector storing the reaction, diffusion, and molecule generation rates. In the continuum limit, the model is able to provide closed-form analytical expressions for the mean and variance of the number of bound receptors for small-scale systems. However, for larger systems, with a high number of voxels, the efficiency of the model is highly questionable. 

\textbf{Steady-state Model:} In addition to the above approaches considering time-varying signals, some researchers prefer using the assumption of steady-state ligand-receptor binding reaction with stationary input signals at the time of sampling, based on fact that the bandwidth of incoming MC signals is typically low because the diffusion channel shows low-pass filter characteristics and the reaction rates are generally higher than the diffusion rate of molecules. This assumption enables the separation of the overall system into two; a deterministic microscale diffusion channel and the stochastic ligand-receptor binding reaction at the interface between the receiver and the channel. Accordingly, at the sampling time, the ligand concentration around the receptors assumes different constant values corresponding to different symbols. The only fluctuations are resulting from the binding reaction, where the random number of bound receptors follows Binomial distribution, whose mean and variance are given in \eqref{eq:meanBinomial} and \eqref{eq:varBinomial}, respectively. The steady-state assumption is applied in \cite{einolghozati2011capacity, aminian2015capacity}, where the authors derive reaction-diffusion channel capacity for different settings. 

\textbf{Convection-Diffusion-Reaction System Model:} Microfluidic MC systems with reactive receivers are studied by a few researchers. In \cite{kuscu2016modeling}, authors develop a one-dimensional analytical model, assuming that the propagation occurs through convection and diffusion, and a reactive receiver, which is assumed to be a SiNW bioFET receiver with ligand receptors on its surface, is placed at the bottom of the channel. The interplay between convection, diffusion and reaction is taken into account by defining a transport-modified reaction rate, tailored for the hemicylindrical surface of the SiNW bioFET receiver.  However, the authors assume steady-state conditions for the reaction, to be able to derive a closed-form expression for the noise statistics. The molecular-to-electrical transduction properties of the bioFET are reflected to the output current of the receiver through modeling the capacitive effects arising from the liquid-semiconductor interface and the 1/f noise resulting from the defects of the SiNW transducer channel. In \cite{kuscu2018modeling}, the authors considered a two-dimensional convection-reaction-diffusion system, which does not lend itself to closed-form analytical expressions for the received signal. Authors develop a heuristic model using a two-compartmental modeling approach, which divides the channel into compartments, in each of which either transport or reaction occurs, and derive an analytical expression for the time-course of the number of bound receptors over a planar receiver surface placed at the bottom of the channel. The model well captures the nonlinearities, such as Taylor-Aris dispersion in the channel, depletion region above the receiver surface, and saturation effects resulting from finite number of receptors, as validated through finite element simulations of the system in COMSOL. However, the model assumes the channel and the receiver is empty at the beginning of the transmission; therefore, does not allow an ISI analysis. 

\paragraph{\textbf{Detection Methods}}
The literature on detection methods for MC with reactive receivers is relatively scarce, and the reason can be attributed partly to the lack of analytical models that can capture the nonlinear ligand-receptor binding reaction kinetics and resulting noise and ISI. Nevertheless, the existing methods can be divided into three categories depending on the type of assumptions made and considered receiver architectures. 
  
\begin{figure}[!t]
	\centering
	\includegraphics[width=\linewidth]{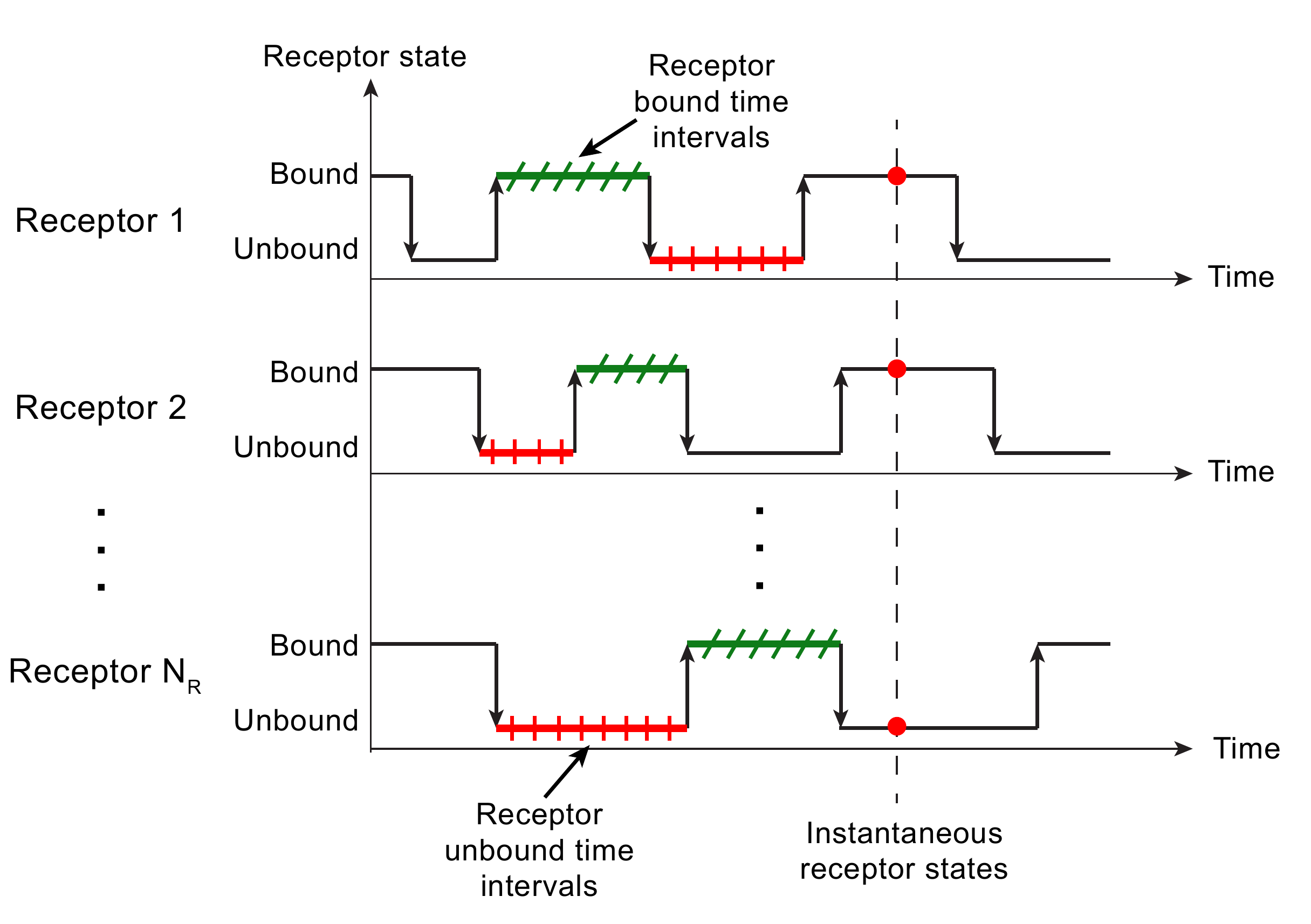}
	\caption{Different types of sampling of receptor states in reactive MC receivers.}
	\label{fig:receptorState}
\end{figure}

\textbf{Detection based on Instantaneous Receptor States:} The first detection approach is based on sampling the instantaneous number of bound receptors at a prespecified time, as shown in Fig. \ref{fig:receptorState}, and comparing it to a threshold. In \cite{deng2015modeling}, the authors study a threshold-based detection for OOK modulated ligand concentrations, using the difference between the number of bound molecules at the start and end of a bit interval. In \cite{shahmohammadian2013modelling}, converting the ligand-receptor binding reaction to a completely first-order reaction with the assumption that all of the receptors are always available for binding, the authors manage to transform the problem into the frequency domain. For the modulation, they consider MoSK with different receptors corresponding to different ligands; therefore, the problem basically reduces to a detection problem of the concentration-encoded signals for each ligand-receptor pair. To reduce the amount of noise, they propose to apply a whitening filter to the sensed signal in the form of number of bound receptors, and then utilize the same detection technique they proposed in \cite{shahmohammadian2012optimum}. 
An energy-based detection scheme is proposed in \cite{mahfuz2014strength}, where the test statistics is the total number of binding events that occur within a symbol duration. They propose a variable threshold-detection scheme with varying memory length. The article also takes into account the ISI; however, the model assumes that all the receptors are always available for binding, and completely neglects the unbinding of ligands from the receptors, making the reaction irreversible. In \cite{kuscu2018maximum}, the authors study the saturation problem in MC-Rxs with ligand receptors. As part of their analysis, they investigate the performance of an adaptive threshold ML detection using the instantaneous number of bound receptors. The effect of the receiver memory that stores the previously decoded bits, is also investigated, and their analysis reveal that the performance of MC detection based on number of bound receptors severely decreases when the receptors get saturated, which occurs when they are exposed to a high concentration of ligands as a result of strong ISI. This is because near saturation, it becomes harder for the receiver to discriminate between two levels of number of bound receptors corresponding to different bits.

\textbf{Detection based on Continuous History of Receptor States:} The second detection approach is based on exploiting the continuous history of binding and unbinding events occurring at receptors, or the independent samples of time intervals that the receptors stay bound and unbound, as demonstrated in Fig. \ref{fig:receptorState}. As we see in the likelihood function in \eqref{eq:likelihood}, the unbound time intervals are informative of the total ligand concentration, whereas the bound time intervals are informative of the type of bound molecules. This is exploited in \cite{kuscu2018maximum}, where the authors tackle the saturation problem in reactive receivers. For a receiver with ligand receptors and with a memory storing a finite number of previously decoded bits, they propose to exploit the amount of time the individual receptors stay unbound. Taking the ligand concentration stationary around the sampling time, and assuming steady-state conditions for the ligand-receptor binding reaction, they developed an ML detection scheme for OOK concentration-encoded signals, which outperforms the detection based on number of bound receptors in the saturation case.  A simple intracellular reaction network to perform the transduction of unbound time intervals into the concentration of certain type of molecules inside the cell is also designed. 
In a similar manner, in \cite{chou2015markovian, chou2015maximum}, using a voxel-based MC system model introduced in \cite{chou2013extended}, and by neglecting the ISI, the authors developed an optimal MAP demodulator scheme based on the continuous history of receptor binding events. Different from \cite{kuscu2018maximum}, the authors assume time-varying input signal. The resulting demodulator is an analog filter, which requires the biochemical implementation of mathematical operations, such as logarithm, multiplication, and integration. The demodulator also needs to count the number of binding events. In \cite{chou2015markovian}, they provide an extension of the demodulator for the ISI case by incorporating a decision feedback, and show the performance improvement with increasing receiver memory. 
 
One of the challenges of reactive receivers is their selectivity towards the messenger molecules, which is not perfect in practice. It is highly probable, especially in physiologically relevant environments, that there are similar ligands in the channel, which can also bind the same receptors, even though their unbinding rate is higher than that of the correct, i.e., messenger, ligands. This causes a molecular interference, which impedes the detection performance of the receiver especially when the concentration of interferer molecules is not known to the receiver. In \cite{muzio2018selective}, the authors evaluate the performance of the one-shot ML detection schemes based on number of bound receptors and unbound time intervals in the presence of interference from a similar ligand available in the environment, number of which in the reception space follows Poisson distribution, with a known mean. They proposed a new ML detection scheme based on estimating the ratio of messenger ligands to the inferring ligands using the bound time intervals sampled from each receptor. It is shown that the proposed method substantially outperforms the others, especially when the concentration of interferer molecules is very high. Note that the above detection techniques are built on MC models that neglect the receiver geometry by treating it as a transparent receiver, inside which ligands and receptors are homogeneously distributed, and they require the receiver to store an internal model, i.e., CIR, of the reaction-diffusion channel. 

\textbf{Detection for Biosensor-based MC Receivers:} The third set of detection methods deal with the biosensor-based receivers, where the binding events are transduced into electrical signals. In bioFET-based receivers the concentration of bound charge-carrying ligands are converted into electrical signals that are contaminated with additional noise. It is not possible to observe individual receptors states; therefore, the detection based on continuous history of binding events is not applicable for these receivers. Accounting for the 1/f noise and binding fluctuations at steady-state conditions, in \cite{kuscu2016modeling}, authors develop an optimal ML detection scheme for CSK in the absence of ISI. Approximating the binding and 1/f noise with a Gaussian distribution, they reduce the overall problem to a fixed-threshold detection problem and provide closed-form analytical expressions for the optimal thresholds and corresponding symbol error rates. The performance evaluation reveals that the 1/f noise, which is resulting from the defects of the semiconductor FET channel, surpasses binding noise resulting from the fluctuations of the receptor states, especially at low frequencies, and severely degrades the detection performance. 

\section{Challenges and Future Research Directions}
\label{sec:Challenges}
After providing a comprehensive survey of existing studies on different aspects of MC, in this section, we discuss the most important challenges toward realization of micro/nanoscale MC setups and evaluation of their ICT-based performance. In this direction, we highlight both the required theoretical studies and the experimental investigations to validate the theoretical models. Physical architecture of MC-Tx/Rx is one of the least studied topics in the MC literature; however, it significantly affects the accuracy of theoretical studies. Thus, we provide an in-depth discussion on the required investigations of MC Tx/Rx architecture. The second shortcoming of MC literature is use of non-realistic assumptions in the existing MC-Tx/Rx and channel models, which has led to imprecise performance estimations not applicable to real scenarios. Hence, we explain the requirements of realistic modeling, such as taking into account the impact of stochastic molecule generation process, Tx/Rx geometry, and stochastic noise dynamics. We also describe open research problems in developed modulation, coding and detection techniques for MC and emphasize important factors that must be considered in future investigations. 

\subsection{Challenges for Physical Design and Implementation of MC-Tx}
Theoretical modeling of MC-Tx in the literature generally relies on unrealistic simplifying assumptions due to the lack of experimental studies on micro/nanoscale implementation of MC-Tx. Therefore, MC-Tx is often assumed as an ideal point source capable of perfectly transmitting molecular messages, which completely neglects important factors in the IM release process such as the stochasticity in the molecule generation process, the effect of Tx geometry and channel feedback. As modeling micro/nanoscale MC-Tx without any experimental data seems not possible, requirement for empirical transmitter models is the most pressing challenge for MC as end-to-end channel models are required to consider the effects of both MC-Tx and MC-Rx. Based on the empirical channel models, the communication theoretical investigation of MC, i.e., analysis of capacity, modulation, coding and detection, needs to be revisited as communication parameters show strong dependence on the channel model. Other challenges regarding the practical implementation of a micro/nanoscale MC-Tx are listed as follows

\begin{itemize}
	\item \textbf{Energy:} Micro/nanoscale MC-Tx and MC-Rx are required to work as standalone devices with their own energy sources. Since battery-powered devices have limited lifetime, EH techniques are promising to develop energy neutral devices such that all operations of the device can be powered via harvested energy from various sources such as solar, mechanical and chemical. To design such systems, one needs to first calculate the harvestable energy budget of an MC-Tx and MC-Rx, i.e., the amount of energy harvestable from the surrounding based on application scenarios and medium. Then, both transmitter and receiver operations, e.g., modulation, coding, detection, are required to be developed accordingly not to exceed the available energy budget. Considering miniaturized MC-Tx and MC-Rx, the harvestable energy can be quite limited so that complex algorithms may not be feasible in a realistic MC scenario. 
	
	\item \textbf{Data Rate:} MC is mostly promising in applications without high data rate requirements as MC suffers from slow propagation channels. This problem can be tackled by encoding large amount of data in DNA/RNA strands, which is named as NSK as discussed in Section \ref{sec:TxModulation}. This way, the amount of transmitted information can be increased up to 100s of MB per IM such that MC can achieve data rates on the order of Mbps. Although sequencing of DNA/RNA strands can be performed via standalone devices utilizing nanopores, there is yet any practical low-cost system to write DNA sequences with a micro/nanoscale device. 
	
	\item \textbf{Molecule Leakage:} In the case of nanomaterial-based MC-Txs, there are two significant design challenges: molecule leakage and molecule reservoir/generation. Molecule leakage, while transmitting low logic or no information, is unavoidable for practical MC-Tx designs. The leaking molecules increase ISI in the channel, and also contribute to an additional problem in molecule reservoir/generation by lowering the molecule budget of MC-Tx. In order to tackle this problem, we have proposed some solutions based on molecule wax as an IM container and hybrid MC-Tx designs, where genetically engineered bacteria can be utilized to replenish IM sources. However, these solutions have not been implemented and the feasibility of such systems as MC-Tx still stands as an important open research issue. 

	\item \textbf{Biological Complexity:} In the case of biological MC-Tx architectures, the main difficulty in design stems from complexity of involved biological elements. Understanding of molecular basis underlying cellular mechanisms, i.e., cellular biology, and development of techniques to manipulate them, i.e., synthetic biology, are essential to unlock the reliable use of engineered biological entities for MC. Moreover, as in many applications data propagated in nanonetworks eventually needs to be connected to electronic devices, any biological network architecture must be interfaced with an electronic architecture. Thus, the development of MC architectures within this space requires an extremely high interdisciplinary engagement between the fields of ICT, biology and synthetic biology, as well as materials science and nanofabrication. Another issue, which again is caused by the inherent complexity of biological organisms, is the fact that, genetically engineered cells are not the best survivors, and complications in their engineered metabolisms cause accelerated death of the cell.
\end{itemize}

\subsection{Challenges for Physical Design and Implementation of MC-Rx}
Despite existing theoretical studies on the performance of MC \cite{meng2014receiver,kilinc2013receiver,noel2014optimal,mosayebi2014receivers,yilmaz20143,damrath2017equivalent,pierobon2011noise,aminian2015capacity,deng2015modeling} and few macroscale experimental setups \cite{farsad2013tabletop,koo2016molecular,farsad2017novel,unterweger2018experimental,wicke2018magnetic}, the literature lacks comprehensive investigation of the physical design of micro/nano-scale MC-Rx structures. This leads to critical open challenges needed to be tackled for realization of the MC promising applications.
\begin{itemize}
	\item \textbf{Ligand-Receptor Selection:} As mentioned in Section \ref{sec:RxPhysicalDesign}, both genetic circuit based and artificial architectures use ligand-receptor reactions to sense the concentration of target molecule by the MC antenna. This calls for careful selection of appropriate ligand-receptor pairs for MC paradigm. In this regards, the binding and dissociation, i.e., unbinding, rates of these reactions are among the important parameters that must be taken into consideration. The binding rate controls sensitivity, selectivity and the response time of the device. While high dissociation rates reduces the re-usability of the device, very low values are also not desired as they decrease the sensitivity of the receiver. The existing interferer molecules in the application environment and their affinities with the receptors are the next important design parameter that must be studied to minimize the background noise. 
	\item \textbf{Realistic ICT-based modeling of artificial structures:} For artificial biosensor-based MC-Rxs, the literature is lacking analytical models that can capture transient dynamics, as the available models developed from sensor application perspective are mostly based on equilibrium assumption. However, in MC applications, since the concentration signals are time-varying, equilibrium models may not be realistic. Therefore, devising MC detection methods for artificial receivers requires developing more complex models that can also capture the stochastic noise dynamics without steady-state assumption for different types of biosensors based on nanomaterials, e.g., graphene bioFET. The models should also include the effects of operating voltages for bioFETs, receiver geometry and gate configuration, channel ionic concentration determining the strength of Debye screening. The models should be validated through wet-lab experiments, and microfluidic platforms stand as a promising option for implementing testbeds for MC systems with artificial MC-Rxs. 
	\item\textbf{Biological Circuits Complexity:} In the case of implementing MC-Rxs with genetic circuits, the information transmission is through molecules and biochemical reactions, which results in nonlinear input-output behaviors with system-evolution-dependent stochastic effects. This makes the analytically studying the performance metrics difficult, if not impossible. Moreover, the existing noise in genetic circuits must be comprehensively studied and methods to mitigate this noise must be derived as it significantly reduces the achievable mutual information of the MC \cite{harper2018estimating}.
	\item \textbf{Bio-Cyber Interface:} While MC expands the functionality of nanomachines by connecting them to each other and making nanonetworks \cite{akyildiz2012nanonetworks}, connection of these nanonetworks to the cyber-networks, i.e., making IoBNT, further extend the applications of these nanomachines \cite{akyildiz2015internet}. Continuous health monitoring and bacterial sensor-actor networks inside human body are two promising examples of these applications. To this aim, implementation of micro/nanoscale bio-cyber interfaces are required. The bio-cyber interface needs to decode the molecular messages, process it and send the decoded information to a macroscale network node through a wireless link. In bioFET-based MC-Rxs, thanks to the molecular-to-electrical transducer unit, the electrical signal generated by the receiver can be sent to the cyber-networks through EM wireless communications. However, connection of biological MC-Rxs to the cyber-networks remains as an open issue.
\end{itemize}
Lastly, it is worthwhile to mention that the physical architecture of the device is mainly dictated by the application requirements. As an example, the biological MC-Rxs are only promising and feasible for \textit{in vivo} applications. On the other hand, utilizing artificial structures for \textit{in vivo} applications necessitates comprehensive investigation of the receiver biocompatibility. Moreover, the physiological
conditions imply solutions with high ionic concentrations, abundance of interferers and contaminants, and existence of disruptive flows and fluctuating temperature, which may degrade the receiver’s performance
in several aspects. First, high ionic concentration creates strong screening effect reducing the Debye length, thus, impedes the sensitivity of the receiver \cite{ rajan2013performance}. Moreover, contaminants and disruptive flows may alter the binding kinetics, impede the
stability of the receptors, even separate them from the dielectric layer. These call for comprehensive investigation of these factors, their impact on the performance of the device and methods to control their effects. To control screening problem, using highly charged ligands and very small size receptors like aptamers are theoretically efficient. However, frequency domain technique promises for much more realistic solutions. In \cite{zheng2010frequency}, the authors reveal that frequency domain detection outperforms the conventional time domain technique in terms of sensitivity in highly ionic solutions. An alternative solution to overcome the Debye screening limitations in detection is proposed in \cite{kulkarni2012detection}, where it is shown that applying a high-frequency alternating current between the source and the drain electrodes, instead of a DC current, weakens the double layer capacitance generated by the solution ions. They show that the effective charges of the ligands become inversely dependent on the Debye length instead of the exponential dependence. As a result, the screening effect on the bound ligands is significantly reduced, and the FET-based biosensing becomes feasible even for physiological conditions. Similar approaches are taken by others to overcome the Debye screening with radio-frequency operation of graphene bioFETs, which are summarized in \cite{fu2017sensing}.

\subsection{Challenges for Developing MC Modulation Techniques}
There are various modulation schemes that are proposed for MC by utilizing concentration, molecule type/order/ratio, and release timing. However, these studies mostly utilize simplified channel models based on MC-Tx, which is an ideal point source, and MC-Rx, which is capable of perfectly detecting multiple molecules selectively. Therefore, the performance of the proposed schemes under realistic conditions are still unknown, and this can be tackled by implementing the proposed schemes in an experimental MC setup. In addition, some of the modulation schemes require synchronization, which is hard to achieve considering the error-prone diffusive channel and low complexity MC devices at micro/nanoscale. Furthermore, energy efficiency is another important challenge for MC-Tx designs as being powered via limited energy harvested from the medium. Therefore, energy-efficient and low complexity modulation schemes for MC without strict synchronization requirements still stand as a significant research problem.

\subsection{Challenges for Developing MC Channel Coding Techniques}
Even though there is some literature on MC channel coding techniques, the field is still very new and open for research. Some of the most pressing directions are highlighted below.
	\begin{itemize}
		\item \textbf{Adaptive Coding:} The propagation time through MC channel and detection probabilities at the receiver are affected by various environmental parameters such as temperature, diffusion speed, channel contents, reaction rates and distance between nodes, which calls for adaptive channel coding techniques for error compensation \cite{atakan2008channel}. In this respect, none of the coding schemes investigated even considers to probe the channel for its characteristics in order to adapt itself.
		\item \textbf{Irregular Signaling:} All of the coding schemes discussed have been evaluated under the regular time-slotted transmission assumption, which in any realistic MC scenario will never be the case. For instance, fluctuations in the inter-symbol duration caused by irregular transmission would have a significant effect on suffered ISI, the primary source of BER in MC, and therefore needs to be considered by channel codes.
		\item \textbf{Simplicity of Models:} More channel codes for MC need to be invented. The only works that do so are \cite{shih2012channel,shih2013channel,ko2012new}, and they have very simple transmission and channel models, i.e., in all works the transmitter releases only a single molecule per transmission period, and the channel is one-dimensional.
	\end{itemize}

\subsection{Challenges for Developing MC Detection Techniques}

As reviewed in Section \ref{sec:RxDetection}, the MC detection problem has been widely addressed from several aspects for different channel and receiver configurations. However, the proposed solutions are still far from being feasible for envisioned MC devices. The main reason behind this discrepancy between the theoretical solutions and the real practice, which is also revealed by the preliminary experimental airborne MC studies performed with macroscale off-the-shelf components \cite{farsad2013tabletop, farsad2014channel}, is that there is no micro/nanoscale MC testbed that can be used as a validation framework to optimize the devised methods. In parallel to this general problem regarding MC technologies, major challenges for MC detection can be further detailed as follows:

\begin{itemize}
\item \textbf{Synchronization:} The majority of the proposed detection techniques assume a perfect synchronization between the transmitter and the receiver. In fact, synchronization is essential for the proper operation of the proposed solutions. There are many studies proposing different synchronization methods, e.g., using quorum sensing to globally synchronize the actions of nanomachines in a nanonetwork \cite{abadal2011bio}, blind synchronization and clock synchronization based on the ML estimation of the channel delay \cite{shahmohammadian2013blind, lin2016clock}, peak observation-time-based and threshold-trigger-based symbol synchronization schemes employing a different type of molecule for the purpose of synchronization \cite{jamali2017symbol}. However, these methods are either too complex for the limited capabilities for the nanomachines, or they rely on stable CSI, which is not the case for time-varying MC channels. Moreover, the effect of these non-ideal synchronization techniques on the performance of the proposed detection methods has not been revealed. Another potential solution to the synchronization problem could be to develop asynchronous detection techniques that obviate the need for synchronization. As reviewed in Section \ref{sec:RxDetectionPassive}, there are a few promising solutions for asynchronous MC detection, e.g., peak-detection and threshold-based detection methods \cite{noel2017asynchronous}. However, they are mostly built on simplifying assumptions for receiver and channel geometry and properties, which may result in unexpected performance in real applications. 

\item \textbf{Physical Properties of the Receiver:} Although there is little physical correspondence for passive and absorbing receivers, many of the detection schemes are built on these assumptions, as they enable a mathematically tractable analysis. As reviewed in Section  \ref{sec:RxDetectionReactive}, although they consider the effect of receptor reactions, the initial studies on reactive receivers also follow similar assumptions on the device architecture and geometry to simplify the analyses. However, for practical systems, the physical properties of the realistic receiver architectures and their impact on the molecular propagation in the MC channel should be taken into consideration to the most possible extent. Finite element simulations on microfluidic MC channel with bioFET receivers and macroscale MC experiments with alcohol sensors clearly reveal the effect of the coupling between the MC receiver and the channel \cite{kuscu2018modeling, farsad2014channel}. The coupling is highly nonlinear, and in most of the cases, it is not analytically tractable; therefore, beyond the available analytical tools, researchers may need to focus on stochastic simulations and experiments to validate the performance of the proposed detection techniques in realistic scenarios.

\item \textbf{Physical Properties of the Channel:} Most of the MC detection studies assume free diffusion, or diffusion plus uniform flow, for molecular propagation in an unbounded 3D environment. However, in practice MC channel will be bounded with varying boundary conditions, and can include nonuniform and disruptive flows, obstacles, temperature fluctuations, and charged molecules affecting the diffusion coefficient, and particles leading to channel crowding. Some of these aspects have recently started to be addressed through channel modeling studies, e.g., sub-diffusive MC channel due to molecular crowding \cite{mahfuz2016concentration}, diffusion-based MC in multilayered biological media \cite{mustam2017multilayer}. However, these simplified models are not able to provide enough insight into the detection problem in realistic channels. The highly time-varying properties of the MC channel have also been addressed by researchers through channel estimation techniques \cite{jamali2016channel, noel2015joint} and noncoherent detection techniques \cite{li2016low, li2016local, jamali2018non}, which are reviewed in Section \ref{sec:RxDetectionPassive}. These studies are built on simplifying assumptions on channel and receiver architecture.

\item \textbf{Reactive Receivers:} Although reactive receiver concept provides a more realistic approach to the detection problem, the research in this direction is still at its infancy, and the developed received signal models are not complex enough to reflect many intricacies. For example, most of the previous studies assume independent receptors being exposed to the same ligand concentration; however, this is not always the case, as the binding of one receptor can affect the binding of neighboring ones \cite{singh2016effects}, and receptors can form cooperative clusters to control sensitivity by exploiting spatial heterogeneity \cite{bray1998receptor, mugler2013spatial}. The interplay between diffusion and reaction is also often neglected in these studies by assuming the timescales of both processes are separated sufficiently. However, in most practical cases, reaction and diffusion rates are close to each other, and reactions are correlated with the transport process, which depends on the channel properties \cite{berezhkovskii2013effect, kaizu2014berg}. These problems are crucial to analyze the spatio-temporal correlations among the receptors and the coupling between the diffusion channel and reactive receiver. Moreover, except for a few recent studies \cite{chou2018molecular, chou2017chemical, marcone2018parity}, the design of intracellular reaction networks to implement the proposed detectors are usually neglected. The additional noise and delay stemming from these reaction networks should be taken into account while evaluating the performance of the overall detection. Furthermore, a proper analysis of the trade-off between energy, detection accuracy and detection speed is required to develop an optimization framework for the detector design in reactive receivers.

\item \textbf{Receiver Saturation:} Another challenge associated with reactive receivers is the saturation of receptors in the case of strong ISI or external interference, which can severely limit the receiver dynamic range, and hamper the ability of the receiver to discriminate different signal levels. The saturation problem has been recently addressed in \cite{kuscu2018maximum} through steady-state assumption for the ligand-receptor binding reactions; however, in general it is neglected assuming an infinite number of receptors on the receiver. The problem can also be alleviated by adaptive threshold detection techniques and adaptive transmission schemes as in detection with passive/absorbing receivers. 

\item \textbf{Receiver Selectivity:} Receiver selectivity is also a major issue for MC detection, and it has just started to be addressed from MC perspective. The physiologically relevant environments usually include many similar types of molecules, and the receptor-ligand coupling is not typically ideal, such that many different types of ligands present in the channel can bind the same receptors as the messenger ligands, causing molecular interference. Even if there are different types of receptor molecules for each ligand type in an MC system with MoSK, the cross-talk between receptors is unavoidable because of the non-ideal coupling between receptor and ligands. Therefore, there is a need for selective detection methods exploiting the properties of ligand-receptor binding reaction. The amount of time a receptor stays bound is informative of the ligand unbinding rate, which is directly linked with the affinity between a particular type of ligand-receptor pair. The bound time intervals are exploited in \cite{muzio2018selective} to detect the MC messages based on the ML estimation of the ratio of correct ligands to the interferers. However, this technique requires the receiver to know the type and probability distribution of the interferer molecules, and it is developed only for one type of interferer. To implement selective detection methods in engineered bacteria-based MC devices, the kinetic proofreading and adaptive sorting techniques implemented by reaction networks of the T cells in the immune system can be exploited \cite{lalanne2015chemodetection, mora2015physical, franccois2016case}. Additionally, MC spectrum sensing methods can be developed based on the information inferred from the receptor bound time intervals to apply cognitive radio techniques in crowded MC nanonetworks where many nanomachines communicate using same type of molecules.  
\end{itemize}

\section{Conclusion}
\label{sec:Conclusion}
MC has attracted significant research attention over the last decade. Although ICT aspects of MC, i.e., information theoretical models of MC channels, modulation and detection techniques for MC and system-theoretical modeling of MC applications, are well-studied, these research efforts mostly rely on unrealistic assumptions isolating the MC channel from the physical processes regarding transmitting and receiving operations. The reason being that although there are some proposals for MC-Tx/Rx based on synthetic biology and nanomaterials, there is no implementation of any artificial micro/nanoscale MC system to date. As a result, feasibility and performance of ICT techniques proposed for MC could not be validated. 

In this paper, we provide a comprehensive survey on the recent proposals for the physical design of MC-Tx/Rx and the state-of-the-art in the theoretical MC research covering modulation, coding and detection techniques. We first investigate the fundamental requirements for the physical design of micro/nanoscale MC-Tx/Rx in terms of energy and molecule consumption, and operating conditions. In light of these requirements, the state-of-the-art design approaches as well as novel MC-Tx/Rx architectures, i.e., artificial Tx/Rx designs enabled by the nanomaterials, e.g., graphene, and biological Tx/Rx designs enabled by synthetic biology, are covered. In addition, we highlight the opportunities and challenges regarding the implementation of Tx/Rx and corresponding ICT techniques which are to be built on these devices. 

The guidelines on the physical design of MC-Tx/Rxs that are provided in this paper will help the researchers to design experimental MC setups and develop realistic channel models considering the transceiving processes. In this way, the long-standing discrepancy between theory and practice in MC can be overcome towards unleashing the huge societal and economic impact of MC by paving the way for ICT-based early diagnosis and treatment of diseases, such as IoNT-based continuous health monitoring, smart drug delivery, artificial organs, and lab-on-a-chips.

\bibliographystyle{ieeetran}
\bibliography{references}

\end{document}